\documentclass{aa}  
\usepackage{graphicx}
\usepackage{txfonts}
\usepackage{fixmath}
\usepackage{tcolorbox}
\usepackage[g]{esvect}
\usepackage{natbib,twoopt}
\usepackage{amsmath,amsfonts,bm}
\usepackage{ulem}
\bibpunct{(}{)}{;}{a}{}{,} 

\newcommand{\vect}[1]{\mathbf{#1}}
\newcommand*\dif{\mathop{}\!\mathrm{d}}

\begin{document} 
   \title{Influence of stellar structure, evolution and rotation \\ on the tidal damping of exoplanetary spin-orbit angles}
   \author{C. Damiani
          \inst{1,2}
          \and 
          S. Mathis \inst{3,4}
          }
   \institute{Universit{\'e} Paris-Sud, CNRS, Institut d'Astrophysique Spatiale, UMR8617, 91405 Orsay Cedex, France
            \and
         Max Planck Institute for Solar System Research, Justus-von-Liebig-Weg 3, 37077 Göttingen, Germany ;   
          \email{damiani@mps.mpg.de}
          \and
          IRFU, CEA, Universit{\'e} Paris-Saclay, F-91191 Gif-sur-Yvette, France
          \and
          Universit{\'e} Paris Diderot, AIM, Sorbonne Paris Cit{\'e}, CEA, CNRS, F-91191 Gif-sur-Yvette, France
             }
             
\titlerunning{Influence of stellar evolution on the tidal damping of obliquities}

   \date{Received xxx, 2017; accepted xxx}
  \abstract
  {It is debated whether close-in giant planets can form in-situ and if not, which mechanisms are responsible for their migration. One of the observable tests for migration theories is the current value of the obliquity, i.e. the angle between the stellar equatorial plane and the orbital plane. But after the main migration mechanism has ended, the obliquity and the semi-major axis keep on evolving due to the combined effects of tides and magnetic braking. The observed correlation between effective temperature and measured projected obliquity in well characterised systems has been taken as evidence of such mechanisms being at play.}
 {Our aim is to produce an improved model for the tidal evolution of the obliquity, including all the components of the dynamical tide for circular misaligned systems. This model is developed to take into account the strong variations in structure and rotation of stars during their evolution, and their consequences for the efficiency of tidal dissipation.}
{Our model uses an analytical formulation for the frequency-averaged dissipation in convective layers for each mode, depending only on global stellar parameters and rotation. It also includes the effect of magnetic braking in the framework of a double zone stellar model.}
{For typical hot-Jupiters orbital configurations, the obliquity is generally damped on a much shorter timescale than the semi-major axis. The final outcome of tidal evolution is also very sensitive to the initial conditions, with Jupiter-mass planets being either quickly destroyed or brought on more distant orbit, depending on the initial ratio of planetary orbital momentum to stellar spin momentum. However we find that everything else being the same, the evolution of the obliquity around low-mass stars with a thin convective zone is not slower than around those with a thicker convective zone. On the contrary, we find that more massive stars, remaining faster rotator throughout their main-sequence, produce more efficient dissipation.}
{} 


   \maketitle
%

\section{Introduction}

	Over the last twenty years, tremendous observational efforts have resulted in the complete characterisation and determination of orbital properties of a few hundreds of exoplanetary systems \citep{Winn2015}. For about a hundred of them, this includes measurements of the Rossiter-MacLaughlin effect \citep{Rossiter1924,McLaughlin1924}, a signature in radial velocities produced during a transit that contains information about the obliquity $\Theta$ of the system, i.e. the angle between the stellar spin axis and the orbital normal. Most of the time, what is observed is the projection on the sky of this angle, usually called $\lambda$. The constraint obtained on $\Theta$ through the observation of $\lambda$ has a complex nature, but when $\lambda$ is observed to be large, then it is more likely that $\Theta$ is also large. Moreover, when the orbit is seen edge-on, the value of $\lambda$ can distinguish wether the orbit is prograde or retrograde \citep{Fabrycky2009}. Those measurements are especially valuable, because the obliquity of exoplanetary systems is one of the main observables allowing to put constraints on their formation and evolution.

	In the case of close-in planets however, which constitute the vast majority of systems with known obliquities, the orbital elements keep evolving after the end of the main migration mechanism under the influence of tides. According to \citet{Winn2010} and \citet{Albrecht2012}, observations seem to show that hot stars with hot Jupiters have high obliquities. They interpret this as evidence of tidal evolution, acting upon a population of initially randomly distributed obliquities. Indeed, tidal dissipation is expected to be more efficient in stars having important convective envelopes, because of the turbulent friction acting on the tidal flows \citep{Zahn1966a,Goldreich1977a,Ogilvie2012,Mathis2016}. Thus cooler star's spin would realign with the orbit more rapidly. Because the mass of the outer convection zone on the main sequence decreases rapidly with increasing stellar mass after 1.2 M$\odot$, which corresponds to an effective temperature of $T_{\rm eff} \approx 6250$ K, this would naturally explain the sharp increase in $\lambda$ that occurs around this effective temperature. This interpretation would favour a single migration mechanism, capable of producing randomly distributed obliquities.

	There is a major flaw in this reasoning: if tides are efficient at damping the obliquity of hot-Jupiters, they would also lead to significant orbital decay, and the planet would be destroyed before re-alignment. This result is obtained with the classical theory of the hydrostatic equilibrium tides \citep[see e.g. ][]{Zahn1966a,Remus2012,Ogilvie2013}, an approximation that seems to be incompatible with other orbital properties of hot Jupiters \citep{Ogilvie2007}. Corrections to the equilibrium tide are provided by the dynamical tide, i.e. the response of a rotating fluid to the harmonic forcing of a time-varying perturbing potential. This includes excitation and damping of inertial waves due to the Coriolis acceleration in convective regions and excitation and damping of gravity waves due to buoyancy modified by rotation (then called gravito-inertial waves) in radiative regions \citep{Ogilvie2004, Ogilvie2007}. 

	In the framework of the dynamical tide, \citet{Lai2012} suggested that tidal dissipation efficiency may not be the same for different processes, e.g. spin-orbit alignment and orbital decay. \citet[][appendix A]{Barker2009}  pointed out that in the case of a misaligned system, the tidal potential has a finite number of components, corresponding to all combinations of the order $m$ and degree $l$ ($-l\leq m \leq l$) of the spherical harmonics decomposition, and that in general each of the distinct components will generate a tidal perturbation with its own dissipation efficiency. The efficiency of tidal dissipation is usually parameterised by a dimensionless quality factor $Q$ proportional to the ratio of the total kinetic energy of the tidal distortion to the energy dissipated in one tidal period \citep[e.g.][]{Zahn2008}. In the theory, for the quadrupolar term, $Q$ always appears in the combination $Q' \equiv (3/2)(Q/k_2)$, where $k_2$ is the static Love number of the star that measures its density stratification. Therefore, $Q'$ is the most widely used parameter to estimate tidal dissipation efficiency. The lower the value of $Q'$ the stronger the tidal dissipation. In general, $Q'$ depends on $l$, $m$, the tidal frequency (a linear combination of stellar rotation frequency and orbital frequency), and the structure of the body under consideration. Thus a rigorous treatment of the tidal dissipation should consider the sum of the effects associated with the different tidal components each having its specific $Q'$ \citep{Ogilvie2014}.  

	Using the tidal prescription proposed by \citet{Lai2012}, \citet{Valsecchi2014} have computed the coupled evolution of the orbital elements and stellar spin of five representative systems, taking into account the combined effects of tides, stellar wind, mass loss, magnetic braking and stellar evolution. Their results show that, accounting for all the relevant physical mechanisms, the current properties of the systems they consider can indeed be naturally explained. However, an important limitation of this study is the absence of a reliable estimation for the tidal quality factor $Q'_{\rm 10}$ (in the notations introduced by \citet{Lai2012}) of the obliquity tide (corresponding to the component associated with a torque but with no dissipated power). In their study, $Q'_{\rm 10}$ is a free parameter which is allowed to vary within a broad range of values, instead of being imposed by stellar properties, to reproduce the observations. While this approach validates \citeauthor{Lai2012}'s basic idea, it is not enough to accurately constrain the initial distribution of the obliquity of the population of known hot Jupiters.

	Another fundamental limitation of their approach, inherited from \citeauthor{Lai2012}'s formulation, is that they only consider one of the components of the dynamical tide to have a possibly different tidal quality factor than the others. Indeed, this parametrisation may be reasonable for the \emph{current} observed orbital state of known hot-Jupiters, for which the orbital frequency is usually much greater than the stellar rotation frequency. Then in this case, there is only one mode that is capable of exciting inertial waves in the star. However, when the orbital frequency is shorter than the rotation frequency, which can in general be the case in early stages, other components of the tide may fall in the frequency range of inertial waves, and the resulting dissipation can also contribute to the evolution of the semi-major axis. As already shown by other studies, this clearly stresses the need to take into account the rotational history of the star to compute the tidal evolution of the system \citep{Bolmont2012, Damiani2015}.
	
	In this paper, we propose a tidal evolution model that follows Lai's theory and take into account the dynamical tide. We expand on this and \citet{Valsecchi2014} works by including all the seven components of the quadrupolar term of the tidal potential into the orbital evolution. Moreover, in a similar approach as \citet{Mathis2015}, we also include an analytical formulation of the tidal dissipation efficiency associated with each mode, as a function of stellar structure and rotation, by using the frequency-averaged formulation for dissipation in convective zones derived by \citet{Ogilvie2013}. In a series of paper \citep{Mathis2015, Lanza2016, Bolmont2016, Gallet2017, Bolmont2017}, this approach has already shown the importance of the evolution of stellar structure and rotation, for the frequency-averaged dissipation efficiency. This compels the use of a double-zone model, one zone representing the radiative core and the other the convective envelope, accounting for the evolution of stellar structure in time. We also use a double-zone semi-empirical model for magnetic braking \citep{Gallet2013,Penev2014}, to account for the evolution of stellar rotation and the internal redistribution of angular momentum within the star.

The paper is organised as follows. In Section \ref{sec:tidaleq}, we formulate the double-zone tidal model for the evolution of the obliquity in exoplanetary systems. We also describe and motivate our approximations. In Section \ref{sec:dissip}, we compute the frequency averaged-dissipation for different modes involved in the evolution of the semi-major axis and obliquity, and we discuss its dependence  with mass and rotation. In Section \ref{sec:compute}, we compute the coupled evolution of semi-major axis, stellar rotation and obliquity for a sample of initially misaligned systems and discuss the different characteristic time-scales. In Section \ref{sec:discus}, we discuss the main results and perspectives of this work.
	

\section{The double-zone weak friction model for tidal interaction}\label{sec:tidaleq}
We consider a star of mass $M_\star$ orbited by a planet of mass $M_{\rm p}$.  The star is modelled as a deformable core of homogenous density $\rho_{\rm c}$ and mean radius $R_{\rm c}$ in uniform rotation, surrounded by an envelope of homogeneous lesser density $\rho_{\rm e}$ and mean radius $R_{\star}$ also rotating uniformly at the angular frequency $\Omega_{\rm e}$, but not necessarily the same as that of the core $\Omega_{\rm c}$. We only consider the tides raised by the planet on the star. As shown by other authors \citep[e.g.][]{Leconte2010}, for typical hot-Jupiters, tides raised on the planet result in the evolution of the planetary spin magnitude and direction towards their equilibrium value within a time scale of about $10^5$ yr. This time is sufficiently short compared to the typical stellar evolution timescales to safely neglect the effects of tidal dissipation in the planet on the overall evolution in the case of a circular orbit. We also neglect the mutual interaction of the tidal bulges and consider that the tide-generating potential of the planet is that of a mass-point. This hypothesis should be relaxed only if the semi-major axis $a < 5 R_\star$ and if the gravitational quadrupolar moment $J_{2} > 10^{-2}$ for the planet \citep{MLP2009}. The planet generates a time-varying tidal potential $\Psi$ that changes the shape and as a result, the exterior potential $\Phi$ of the star. In polar coordinates, the time-varying tidal potential produced by the point-like planet can be expanded in terms of spherical harmonics $Y_{l}^{m}$ of degree $l$ and order $m$. For an eccentric orbit, $\Psi$ must be expanded for an infinite number of terms. But for a circular orbit, not necessarily in the equatorial plane of the reference frame, there is a finite number of tidal components for any given value of $l$. Here we are interested in the evolution of the obliquity of the orbit resulting from the dissipation of the tides in the star, so we limit our study to the case of circular orbits, and produce a tidal evolution model that is valid for arbitrary values of the obliquity.

\subsection{Dynamical tidal potential and response in the circular case}\label{sec:dynpot}
 We use spherical polar coordinates $(r, \theta, \phi)$ , where $r$ is the radial distance from the center of the star, $\theta$ is the colatitude, and $\phi$ is the azimuthal angle. We adopt an inertial coordinate system with polar axis aligned with the star's angular momentum $\mathbold{L}_\star$. We note $\Theta$, the angle between the orbital angular momentum $\mathbold{L_{\rm o}}$ and $\mathbold{L}_\star$, $a$, the semi-major axis of the relative orbit of the planet and $\Omega_0$, its mean orbital angular velocity. We chose the y-axis in the direction $\mathbold{L}_\star \times \mathbold{L_{\rm o}}$. Neglecting precession, the tide-raising potential of the planet can be written as
\begin{multline}\label{quadtide}
\Psi(r, \theta,\phi, t) ={\rm Re} \left[\sum_{l = 2}^{\infty} \sum_{m=-l}^{l} \sum_{p=0}^{l}  \frac{G M_{\rm p} R_\star^l}{a^{l+1}} \right. \\
\left. W_l^{l-2p} d^{(l)}_{m, l-2p}(\Theta) e^{-i(l-2p) \Omega_0 t} \left(\frac{r}{R_\star}\right)^l {Y}{_{l}^{m}}(\theta, \phi) \right],
\end{multline}
where $G$ is the gravitational constant, $d^{(l)}_{m, l-2p}(\Theta)$ are the Wigner d-matrix elements that only depend on $\Theta$, and 
\begin{equation} 
W_l^{l-2p} = \frac{4\pi}{2l+1} Y_l^{l-2p}(\pi/2, 0).
\end{equation}
Each tidal component rotates with the angular velocity $\omega$ given by
\begin{equation}
\omega = (l-2p)\Omega_0,
\end{equation}
where $p$ is an integer with $0\leq p\leq l$ that arises from the inclination of the orbital plane. This defines the tidal frequency in the inertial frame. In a frame that rotates with the spin angular velocity $\Omega_{\rm e}$, the tidal frequency must be Doppler shifted and we thus define $\hat{\omega}$, the tidal frequency in the fluid frame
\begin{equation}\label{tdff}
\hat{\omega} = (l-2p)\Omega_0 - m \Omega_{\rm e}.
\end{equation}
While the tidal frequency can be positive or negative, the physical forcing frequency is positive and is thus simply defined as $\chi = |\hat{\omega}|$. Since we have $\hat{\omega}_{l, m, p}=-\hat{\omega}_{l, -m, l-p}$, the $(m, p)$ component is physically identical to the  $(-m, l-p)$ component.

When the amplitude of the tidal disturbance is small, as this is generally the case for close-in exoplanetary systems, and for axisymmetric bodies, the tidal response of the perturbed body can be determined by treating each component of the tidal potential independently, and considering that the total response is simply the sum of each component. Taking the Fourier transform of the equations of motion, mass conservation and gravitational potential leads to a purely spatial problem in which the frequency $\omega$ appears as a parameter. 

Considering a tidal potential of the form ${\Psi=\sum \Psi_l^m(t) (r/R_\star)^l {Y}_{l}^{m} (\theta, \phi)}$, we write the perturbed stellar gravitational potential in the form ${\Phi=\sum \Phi_l^m(t) (r/R_\star)^{-(l+1)} {Y}_{l}^{m}(\theta, \phi)}$. We denote the temporal Fourier transforms with a tilde and we associate to each component of the perturbed gravitational potential an array of potential Love numbers $k_{l, l'}^{m, m'}(\omega)$, that quantify the frequency-dependent response of the star to tidal forcing : 
\begin{equation}
\tilde{\Phi}_l^m(\omega)= \sum_{l', m'}k_{l, l'}^{m, m'}(\omega) \tilde{\Psi}^{l'}_{m'}(\omega).
\end{equation}
For an axisymmetric body, $k_{l, l'}^{m, m'}(\omega)=0$ unless $m'=m$. The part of the response that has the same form as the applied potential is thus quantified by the potential Love numbers $k_l^m(\omega)$ :
\begin{equation}
k_l^m(\omega) = k_{l, l}^{m, m}(\omega).
\end{equation}

\subsection{Tidal dissipation} 
 Stars are fluid bodies that cannot be considered as purely frictionless. Dissipation of mechanical energy occurs even in adiabatically stratified fluid layers through turbulent friction or partial wave reflection at the interface with stably stratified zones for example. The fluid dynamics of tidally forced bodies is a very active field of research \citep[see e.g.,][]{Ogilvie2014}, but whatever the mechanism actually responsible for dissipation, it suffices to say here that the response of a fluid body to tidal excitation is not frictionless and some energy is transferred to the perturbed body. In the presence of dissipation, the perturbed external gravitational potential of the star involves complex Love numbers and the imaginary part of the Love numbers ${\rm Im}[k_l^m(\omega)]$ quantifies the component of the response that is out of phase with the tidal forcing, and is associated with transfers of energy and angular momentum. 

The rates of transfer of energy and of the axial component of angular momentum from the orbit to the body, measured in an inertial frame, define the tidal power $P$, and the tidal torque $T$, respectively. So for each component of the tidal potential of the form of Eq.~\eqref{quadtide}, averaged over azimuth in the case $m=0$, it can be shown \citep{Ogilvie2013} that $P=\omega \mathcal{T}$ and $T=m\mathcal{T}$, where $\mathcal{T}$ depends on the amplitude of the tidal component and on ${\rm Im}[k_l^m(\omega)]$. To compute the evolution of orbital elements associated with these transfers of energy and angular momentum, we use an approximation similar to the one made in the equilibrium theory of tides. We consider that the tidal response resembles the hydrostatic one in the absence of dissipation, and that dissipation introduces a small frequency-dependent phase lag in the response of the body. However, contrary to the usual equilibrium theory of tides, we do not impose that all components have the same phase-lag. So for each component of the tidal potential, we assume that dissipation introduces separate frequency-dependent phase-lags $\Delta_l^m(\omega)$ which can be related to the imaginary part of the Love numbers \citep{Efroimsky2013,Ogilvie2013} with 
\begin{equation}
{\rm Im}[k_l^m(\omega)] =  |k_l^m(\omega)| \sin \Delta_l^m(\omega)
\end{equation}
where ${\rm Im}[k_l^m(\omega)]$ has the same sign as $\hat{\omega}$.
As detailed in \citet{Ogilvie2014}, typically ${\rm Re}[k_l^m(\omega)]$ is a quantity of order unity, only weakly dependent on $m$ and $\omega$, and can be well approximated by its hydrostatic value. The hydrostatic Love number $k_l$ is a real quantity and does not depend on $m$. Its evaluation for a fluid body is a classical problem involving the Clairaut's equation (see Appendix.~\ref{statLove}). In the weak friction approximation \citep{Alexander1973}, we thus consider that $|\Delta_l^m(\omega)| << 1$ so that 
\begin{equation}
|k_l^m(\omega)| \approx k_l
\end{equation}
and 
\begin{equation}
{\rm Im}[k_l^m(\omega)] \approx  k_l \Delta_l^m(\omega).
\end{equation}

In general, the tidal response involves resonances with stellar oscillations when the tidal frequency matches that of an appropriate mode. In convection zones, for solar-like slow rotating stars, the frequencies of tidal oscillations typically lie in the range of inertial modes, for which the Coriolis acceleration provides the restoring force. The eigenfrequencies of inertial modes in the frame that rotates with angular velocity $\Omega_{\rm e}$ are dense in the interval $[-2\Omega_{\rm e}, 2\Omega_{\rm e}]$, but do not exist outside of this range. The spatial structure of the inertial modes imply that any spherical harmonic potential resonates with only a finite number of inertial modes. As shown for example by \citet{Ogilvie2004, Ogilvie2007}, enhanced tidal dissipation rates are expected when the tidal potential can resonantly excite those inertial modes. Dissipation has then a complex dependence on the tidal frequency, and thus on both the rotational and orbital period. As dissipation depends on the internal structure of the body and on its rotation, the computation of ${\rm Im}[k_l^m(\omega)]$ becomes very heavy when one is interested in the secular effects of the tidal torque in the evolution of the orbital element as the star itself is evolving \citep{Mathis2016}. 

Here, we use a simplified model for tidal dissipation, which does not involve the complicated details of the frequency-dependence of the response functions, and neglect the enhancement of dissipation due to resonances. Indeed, as shown by \citet{Ogilvie2013}, the typical level of dissipation can be approximated using a simple analytical formulation of the frequency-averaged dissipation $\int^{\infty}_{-\infty} {\rm Im}[k_l^m(\omega)] {\rm d}\omega/\omega$, obtained by means of an impulse calculation. Their solution depends on the internal structure of the body, either homogeneous or double-layered, but makes no assumption on the details of dissipation mechanisms, so it is smooth and free of surface effects (like wave reflection at the interface between convective and radiative zones). Their derivations are done for arbitrary degree and order of the tidal components. Taking-up on their work, we use here $\int^{\infty}_{-\infty} {\rm Im}[k_l^m(\omega)] {\rm d}\omega/\omega$ as a measure of the tidal response in the low-frequency part of the spectrum where inertial waves are found, when the details of the frequency-dependent response curve are filtered out. Thus, we define an average phase-shift $\bar{\Delta}_{l}^{m}$ for the tidal component of degree $l$ and order $m$ obtained using the frequency-averaged formulation of \citet{Ogilvie2013}\footnote{We recall the reasoning behind the formulation of $\int^{\infty}_{-\infty} {\rm Im}[k_l^m(\omega)]  {\rm d}\omega/\omega$ and derive its expression in terms of global stellar parameters for arbitrary values of $l$ and $m$ in Appendix~\ref{hydrostat}.}
\begin{equation}\label{imkomega}
\int^{\infty}_{-\infty} {\rm Im}[k_l^m(\omega)]  \frac{{\rm d} \omega}{\omega} \approx k_l \int^{\infty}_{-\infty}\Delta_l^m(\omega)\frac{{\rm d} \omega}{\omega} \equiv k_l \bar{\Delta}_{l}^{m}.
\end{equation}

This averaged phase-shift $\bar{\Delta}_{l}^{m}$ can be related to an average positive-definite time-lag $\bar{\tau}_l^m$. They are  obtained by assuming that, in the fluid frame, components of different $p$ but same $l$ and $m$ have the same time-lags, which is equivalent to assume that the relevant tidal frequency is $\hat{\omega} = m \Omega_{\rm e}$. Since the tidal dissipation rate is a positive-definite quantity, by construction  $\bar{\Delta}_{l}^{m}$ is also positive-definite. Because $\hat{\omega}_{l, m, p}=-\hat{\omega}_{l, -m ,l-p}$,  we must have $\bar{\tau}_{l}^{-m} = \bar{\tau}_{l}^{m}$, and thus
\begin{equation}\label{eqdeltalm}
k_l  \bar{\tau}_l^m = \frac{k_l \bar{\Delta}_{l}^{m}}{|m| \Omega_{\rm e}}.
\end{equation}
Subsequently, we compute the tidal power and torque associated to every component of the tidal potential, but instead of using a frequency dependent time-lag associated to ${{\rm Im}[k_l^m(\omega)] = k_l \tau_{l, m, p} \hat{\omega}}$, we approximate it to ${{\rm Im}[k_l^m(\omega)] \approx k_l \bar{\tau}_{l}^{m} \hat{\omega}}$. For the axisymmetric $m=0$ component, the axial component of the tidal torque vanishes, but not the tidal power averaged over azimuths, because tidal deformation modulates the moment of inertia of the star. Hence, we compute the average time-lag for $m=0$ considering that the relevant frequency is the orbital frequency
\begin{equation}\label{eqdeltal0}
k_l \bar{\tau}_l^0  = \frac{k_l \bar{\Delta}_{l}^{0}}{\Omega_0}.
\end{equation}
The frequency averaged formulation derived from \citeauthor{Ogilvie2013}'s work corresponds to the dissipation of tidally excited inertial waves. Yet those waves exist only when $|\hat{\omega}| < 2 \Omega_{\rm e}$. Outside of this range, the dissipation efficiency corresponds to the one of the equilibrium tide. Hence, we introduce a frequency dependence for the value of $\bar{\tau}_{l}^{m}$ by considering
\begin{equation}\label{eq:cases}
k_l \bar{\tau}_{l}^{m, p}=
\begin{cases}
k_l \bar{\tau}_{l}^{m},\quad \text{if } |\hat{\omega}| <2 \Omega_{\rm e} \\
\frac{3}{2 \Omega_0 Q_{\rm eq}'},\quad \text{otherwise}
\end{cases}
\end{equation}
where the value of $Q_{\rm eq}'$ remains uncertain and we perform calculations considering a range of value to study the sensitivity of the model (see Sec.~\ref{sec:compute}).

Following \citet{Ogilvie2013}, we thus write the average energy transfer $\langle \dot{E}\rangle$ associated to the component of degree $l$ as
\begin{multline}\label{trasenergy}
\langle \dot{E}\rangle = \Omega_0 \sum_{m, p} (l-2p)\frac{(2l+1)}{8 \pi }\frac{G  M_{\rm p}^2 R_\star^{2l+1}}{a^{2l+2}} \\
\left(W_l^{l-2p} d^{(l)}_{m, l-2p}(\Theta)\right)^2 k_l  \bar{\tau}_{l}^{m, p} \hat{\omega}_{l, m, p},
\end{multline}
and the corresponding axial component of the tidal torque $T_{z}$
\begin{multline}\label{axtorque}
T_{z} = \sum_{m, p} m \frac{(2l+1)}{8 \pi } \frac{G  M_{\rm p}^2 R_\star^{2l+1}}{a^{2l+2}} \\
\left(W_l^{l-2p} d^{(l)}_{m, l-2p}(\Theta)\right)^2 k_l  \bar{\tau}_{l}^{m, p} \hat{\omega}_{l, m, p} ,
\end{multline}
where we calculate the value of $\bar{\tau}_{l}^{m, p}$ using Eqs.~\eqref{imkomega}, \eqref{eqdeltalm}, \eqref{eqdeltal0} and \eqref{eq:cases}.

Up to this point, all our derivations are valid for arbitrary value of $l$ and $m$, and in principle, they allow the computation of the tidal response for any spherical harmonic of the multipole expansion of the tidal potential. As can be seen in Eq.~\eqref{quadtide}, the contribution of the spherical harmonic of degree $l$ is proportional to $r^l$, so usually only the quadrupolar terms are retained in the formulation of the temporal evolution of orbital elements. Explicitly then, for the quadrupolar components,  Eq.~\eqref{trasenergy} yields
\begin{multline}\label{trasenergyquad}
\langle \dot{E}\rangle = \frac{3}{64} \frac{G M_{\rm p}^2 R_\star^5}{a^6} \Omega_0  \left[ 6 \sin^4 \Theta (2\Omega_0)   k_2  \bar{\tau}_{2}^{0,2} \right.\\
+4 \sin^2 \Theta \left(1+\cos \Theta \right)^2 (2\Omega_0-\Omega_{\rm e}) k_2 \bar{\tau}_{2}^{1,0}  \\
- 4 \sin^2 \Theta \left(1-\cos \Theta \right)^2  (-2\Omega_0-\Omega_{\rm e})  k_2 \bar{\tau}_{2}^{1,2} \\
+ \left(1+\cos \Theta \right)^4 (2\Omega_0-2\Omega_{\rm e})  k_2 \bar{\tau}_{2}^{2,0} \\
\left. - \left(1 - \cos \Theta\right)^4 (-2\Omega_0-2\Omega_{\rm e})  k_2 \bar{\tau}_{2}^{2,2} \right],
\end{multline}
and for Eq.~\eqref{axtorque} we have
\begin{multline}\label{axtorque2}
T_{z} = \frac{3}{64} \frac{G M_{\rm p}^2 R_\star^5}{a^6}   \left[ 2 \sin^2 \Theta (1+\cos \Theta)^2  (2\Omega_0 - \Omega_{\rm e})  k_2   \bar{\tau}_{2}^{1,0} \right. \\
+ 8 \sin^2 \Theta \cos^2 \Theta (-\Omega_{\rm e})  k_2   \bar{\tau}_{2}^{1,1} + 2 \sin^2 \Theta  (1-\cos \Theta)^2(-2\Omega_0 - \Omega_{\rm e})  k_2   \bar{\tau}_{2}^{1,2}  \\ 
+  \left(1+\cos \Theta\right)^4  (2\Omega_0 - 2\Omega_{\rm e}) k_2   \bar{\tau}_{2}^{2,0} +  4  \sin^4 \Theta k_2   \bar{\tau}_{2}^{2,1}  (- 2\Omega_{\rm e}) \\
\left.+  \left(1-\cos \Theta \right)^4  k_2   \bar{\tau}_{2}^{2,2}  (-2\Omega_0 - 2\Omega_{\rm e}) \right]
\end{multline}

The y-component only contributes to precession that we are neglecting here. The perpendicular component of the torque is computed following the formulation of \citet{Lai2012}\footnote{The computations of the tidal power in \citet{Lai2012}  (Eq. 25) and \citet{Ogilvie2013}  (Eq. 41) are equivalent to a factor $(2l+1)/8\pi$ with the notations we have introduced.}
\begin{multline}\label{transtorque}
T_x =   \frac{3}{64}\frac{G M_{\rm p}^2 R_\star^5}{a^6}   \sin \Theta \Big[ 6 \sin^2 \Theta (2\Omega_0) k_2 \bar{\tau}_{2}^{0,2} \\
+  2 (1+\cos \Theta)^2 (2-\cos \Theta) (2\Omega_0 - \Omega_{\rm e}) k_2  \bar{\tau}_{2}^{1,0} - 8  \cos^3 \Theta (-\Omega_{\rm e}) k_2  \bar{\tau}_{2}^{1,1} \\
- 2 (1-\cos \Theta)^2 (2+\cos \Theta) (- 2\Omega_0 - \Omega_{\rm e}) k_2  \bar{\tau}_{2}^{1,2}\\
+ (1+\cos \Theta)^3 (2\Omega_0 - 2\Omega_{\rm e}) k_2  \bar{\tau}_{2}^{2,0} -  4 \sin^2 \Theta  \cos \Theta (- 2\Omega_{\rm e}) k_2  \bar{\tau}_{2}^{2,1}\\
\left. - ( 1-\cos \Theta)^3 (-2\Omega_0 - 2 \Omega_{\rm e})  k_2 \bar{\tau}_{2}^{2,2}\right]
\end{multline}

\subsection{Stellar angular momentum modeling}\label{sec:magbrak}
The angular momentum of the star $\mathbold{L}_\star = L_\star \mathbold{\hat{L}}_\star $ is modelled with core-envelope decoupling under the assumptions of the double-zone model introduced by \citet{MacGregor1991};
\begin{equation}
L_\star = L_{\rm c} + L_{\rm e} = I_{\rm c} \Omega_{\rm c} + I_{\rm e} \Omega_{\rm e}
\end{equation}
where $I_{\rm c}$ and $I_{\rm e}$ are the moment of inertia for the core and the envelope respectively. The core and the envelope are assumed to rotate as solid bodies with different angular velocities. As noted in \citet{Ogilvie2013}, in a differentially rotating body there is no simple global relation between the tidal dissipation rate and the rates of energy and angular momentum transfer as computed from Eqs.~\eqref{trasenergy}, \eqref{axtorque}, \eqref{transtorque}. However, those quantities depend on stellar parameters only through the mean stellar radius at the surface, the static Love number and the average time-lag. The static (real) Love numbers link the interior structure of the star to its external potential through the surface value of the logarithmic derivative of the deformation \citep{Kopal1959}. Here we neglect the effect of the centrifugal force on the shape of the star, so the internal discontinuity in rotation speed is irrelevant. The impulsing forcing approach used to estimate the average time-lag also neglects the effect the Coriolis acceleration on the non-wavelike part of the displacement, so the boundary condition specifying the solution remains the same even if we consider that the core rotates at a different angular velocity than the envelope. Moreover, as is clearly apparent in Eq.~(96), (97) and (98) of \citet{Ogilvie2013}, only the convective envelope of the star is involved in the tidal energy transfer, since we are only considering the dissipation of tidally-excited inertial wave. Here we suppose that the envelope rotates rigidly and as such our assumptions are fully compatible with the derivations of \citet{Ogilvie2013}, setting the angular velocity to the one of the envelope. Furthermore, we consider that the tidal torque is applied on the envelope of the star, but we allow angular momentum exchanges between the two zones. As a matter of fact, in the piece-wise homogenous rigidly rotating model considered in \citet{Ogilvie2013}, the non-barotropic nature of the density-jump results in a singularity of vorticity at the interface. In the double-zone model considered here, considering our simplifying assumptions regarding rotation, and other overlooked mechanisms that may be involved, we simply parametrise the transfer of angular momentum between the zones by a quantity $\Delta L$ defined as
\begin{equation}
\Delta L = \frac{I_{\rm c} I_{\rm e} }{I_{\rm c}+ I_{\rm e}} \left(\Omega_{\rm c} - \Omega_{\rm e}\right)
\end{equation}
at a rate determined by a coupling time-scale $\tau_{c}$. For the dependence of $\tau_c$ on stellar rotation, we follow \citet{Spada2011} and consider a two-valued function with
\begin{equation}
\tau_c =
\begin{cases}
10 \text{ Myr }, \quad \text{if } \Omega_{\rm e}(t_0) \geq \Omega_{\rm crit}, \\
128 \text{ Myr } ,\quad \text{otherwise}.
\end{cases}
\end{equation}

We follow the evolution of the star from its formation until the end of its main sequence, i.e. before the core starts contracting during the subgiant phase. Initially the star is fully convective and is assumed to rotate rigidly. We consider a phase of disc-locking, where the net effect of interactions with the disc is that of keeping the surface angular velocity of the star constant for the disc lifetime, i.e. 
\begin{equation}
\frac{\dif{\Omega_{\rm e}}}{\dif{t}} = 0,\quad \text{while} \quad t\leq \tau_{\rm disc}.
\end{equation}
The disc lifetime $\tau_{\rm disc}$ may vary, but it is in general shorter than the PMS, as observations show that most primordial discs have disappeared by the first $10$ Myr \citep{Ribas2014}. As the radiative core develops, a quantity of material contained in a thin shell at the base of the convective zone, with a velocity of $\Omega_{\rm e}$, becomes radiative, producing an angular momentum transfer towards the core
\begin{equation}
\left.\frac{\dif{L}}{\dif{t}}\right\vert_{\rm growth} = \left(\frac{2}{3} R^2_{\rm c} \frac{\dif{M_{\rm c}}}{\dif{t}}\right) \Omega_{\rm e}.
\end{equation}
From the moment the disc disappears, the star experiences angular momentum loss through its magnetised wind. We assume that because of wind braking, the envelope loses angular momentum at a rate given by the following parametric formula \citep{Kawaler1988}
\begin{equation}
\left.\frac{\dif{L_{\rm e}}}{\dif{t}}\right\vert_{\rm wind} = K_{\rm w} \left(\frac{R_\star}{{\rm R_\odot}}\right)^{\frac{1}{2}}\left(\frac{M_\star}{\rm M_\odot}\right)^{-\frac{1}{2}} {\rm min} \left(\Omega^3_{\rm e}, \Omega^2_{\rm sat}\Omega_{\rm e}\right),
\end{equation}
where $K_{\rm w}$ determine the braking intensity and $\Omega_{\rm sat}$ is the angular frequency threshold defining a saturated (when $\Omega_{\rm e} \geq \Omega_{\rm sat}$) regime of angular momentum loss.

In fine, the evolution of the stellar angular momentum is modelled with the following differential equations
\begin{align}
\frac{\dif{L_{\rm c}}}{\dif{t}} &= -\frac{\Delta L }{\tau_c} + \left.\frac{\dif{L}}{\dif{t}}\right\vert_{\rm growth} \\
\frac{\dif{L_{\rm e}}}{\dif{t}} &= + \frac{\Delta L }{\tau_c} - \left.\frac{\dif{L}}{\dif{t}}\right\vert_{\rm growth} -\left.\frac{\dif{L_{\rm e}}}{\dif{t}}\right\vert_{\rm wind} + T_z. \label{evolom}
\end{align}

\subsection{Tidal evolution of orbital elements}\label{sec:orbitevol}
 \begin{figure}
	\centering
	\includegraphics[width=0.9\columnwidth]{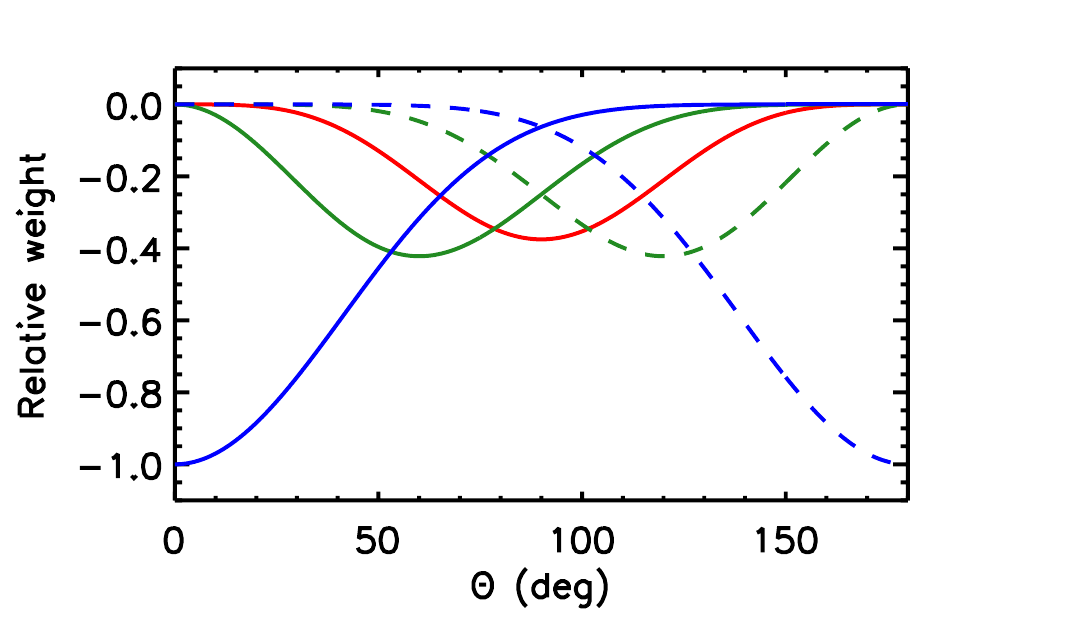}\\
	\includegraphics[width=0.9\columnwidth]{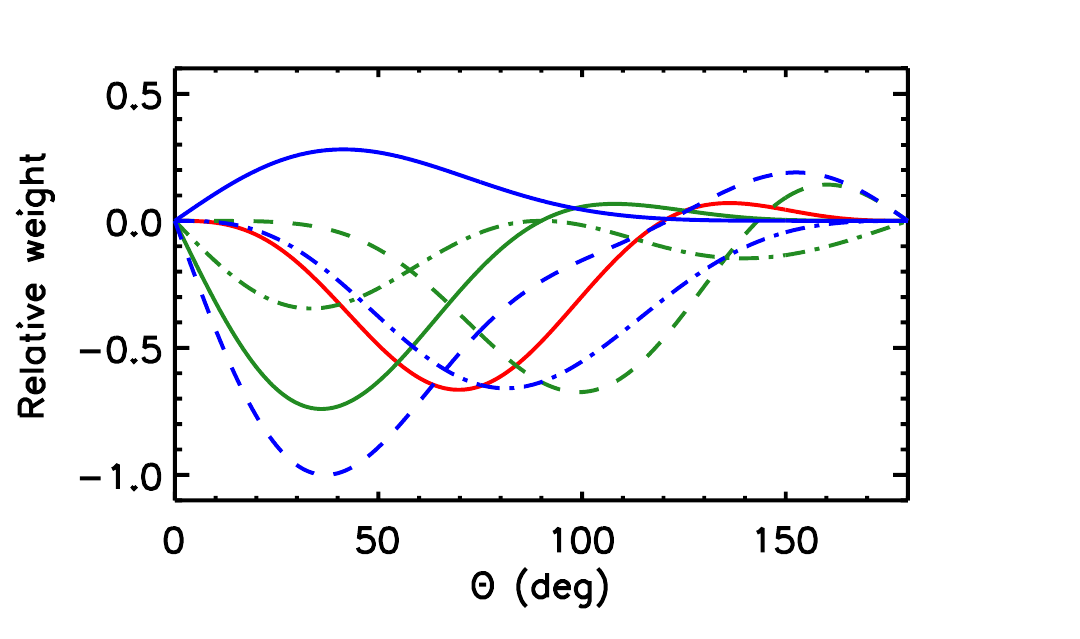}
	\caption{Relative geometrical contribution of each tidal component of degree $l=2$ in the evolution of the semi-major axis (Eq.~\ref{evola}) (top) and the obliquity (Eq.~\ref{evoli}) as a function of the instantaneous value of the obliquity for $m=0$ (red), $m=1$ (green) and $m=2$ (blue); $p=0$ (solid), $p=1$ (dot-dashed) and $p=2$ (dashed). For the obliquity (bottom panel), the relative contribution of each component depend on the orbital to spin angular momentum magnitude ratio, here taken for illustration purposes at 0.5. Those factors do not include the factor $k_2  \bar{\tau}_{l}^{m,p}$, so they do not contain information about the relative dissipation efficiencies of the different components for orbital evolution.}
	\label{fig:coeffs}
\end{figure}

The transfer of energy from the orbit to the rotation of the star due to tides results in the variation of the semi-major axis through the relationship
\begin{equation}\label{evola}
\langle \dot{E}\rangle = \frac{\dif}{\dif{t}} \frac{G M_\star M_{\rm p}}{2a} = - \frac{G M_\star M_{\rm p}}{2a^2} \frac{\dif{a}}{\dif{t}}.
\end{equation}
For hot-Jupiters, the quadrupolar components of the tidal potential is strongly dominant due to the small ratio between stellar radius and semi-major axis, even during the PMS. The evolution of the semi-major axis is straightforwardly obtained using Eq.~\eqref{trasenergyquad}. As can be seen in Eq.~\eqref{trasenergy}, tidal components with $p=1$ do not contribute to the transfer of energy. Moreover, for $m=0$, the components with $p=0$ and $p=2$ are physically identical. So there are only 5 components that contribute to the evolution of the semi-major axis. The importance of each of them depends on their respective tidal frequency, the average time-lag, and a weighting factor that is a function of the instantaneous value of $\Theta$. This quantity is shown in Fig.~\ref{fig:coeffs} (top panel). Note that for $\Theta = 0$, only the $l=2, m=2, p=0$ component contributes to the evolution of the semi-major axis.

The magnitude of the orbital angular momentum $L_{\rm o}= || \mathbold{L_{\rm o}} || $ is 
\begin{equation}
L_{\rm o}=\frac{M_\star M_{\rm p}}{M_\star + M_{\rm p}} \Omega_0 a^2.
\end{equation}
Because we consider here that only the envelope is involved in tidal energy transfer, we will consider the evolution of the orientation of the direction of the spin of the envelope so 
\begin{equation}
\cos \Theta = \frac{\mathbold{L}_{\rm e} . \mathbold{L_{\rm o}}}{L_{\rm e} L_{\rm o}}.
\end{equation}
By definition, $0 \leq \Theta \leq \pi$, the orbit is polar when $\Theta=\pi/2$ and retrograde when $ \Theta> \pi/2$. Supposing that in the absence of magnetic braking, the total angular momentum is conserved we get 
\begin{equation} \label{evoli}
\frac{\dif{\Theta }}{\dif{t}}= - \frac{T_x}{L_{\rm e}} -\frac{T_x}{L_{\rm o}} \cos \Theta + \frac{T_z}{L_{\rm o}} \sin \Theta,
\end{equation}
The differential equation governing the evolution of the obliquity can thus be obtained using Eq.~\eqref{axtorque2} and \eqref{transtorque}. Here, all the 7 physical components of the tidal forcing will contribute to the evolution of the obliquity. Their relative importance is a function of the instantaneous value of $\Theta$ and of $ L_{\rm o}/{L}_{\rm e}$. It is shown in Fig.~\ref{fig:coeffs} (bottom panel) for $ L_{\rm o}/{L}_{\rm e} = 0.5$.


\section{Evolution of tidal dissipation as a function of stellar mass and evolutionary stage}\label{sec:dissip}
        
As can be seen from Eq.~ \eqref{trasenergyquad}, \eqref{axtorque2}, \eqref{transtorque}, \eqref{evolom}, \eqref{evola} and \eqref{evoli}, the coupled evolution of the orbital elements depends on the dissipation efficiency of the tides raised by the three components of the quadrupole moment $l=2$ of the tidal potential : the $m=0$ zonal harmonic, the $m=1$ tesseral harmonic and the $m=2$ sectoral harmonic. The so-called obliquity-tide ($l=2,m=1,p=1$), which has the frequency $\hat{\omega}=-\Omega_{\rm e}$ regardless of the orbital frequency, contributes to the tidal torque but has no associated variation of mechanical energy. However, it is important to note that other terms associated with the zonal or tesseral harmonics participate to both angular momentum and energy transfer, but their contribution to the evolution of the semi-major axis is non-null only when $\sin \Theta \neq 0$, i.e. when the system is not aligned or anti-aligned. 

The effect of stellar structure and evolution on the frequency-averaged dissipation of the sectoral ($l =m = 2$) harmonic have already been computed in other studies \citep{Mathis2015, Lanza2016, Bolmont2016, Gallet2017, Bolmont2017}. In the impulsive forcing problem, the value of $m$ affects the energy dissipated by inertial waves through the coupling of the spheroidal and toroidal wavelike velocity components due to the Coriolis force.  The only difference between the energy transfer from the sectoral harmonic and the zonal and tesseral ones stems from the contribution of the toroidal part, related to the coupling coefficient $\tilde{q_l}$,
\begin{equation}\label{eqcoupling}
\tilde{q_l} = \frac{1}{l} \left(\frac{l^2-m^2}{4l^2 -1}\right)^{1/2}.
\end{equation}

\begin{figure}
	\centering
	\includegraphics[width=0.94\columnwidth]{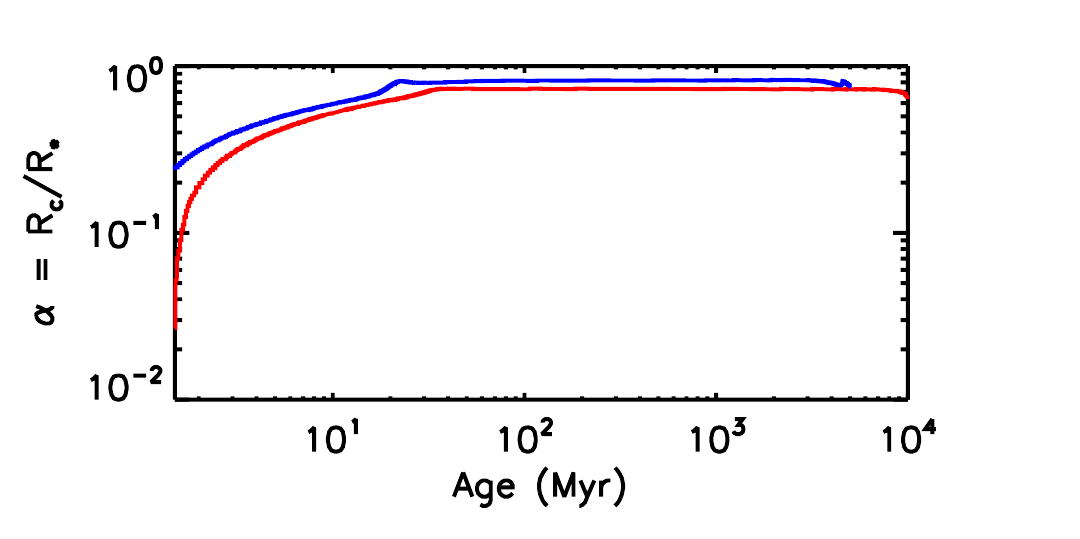}\\
	\includegraphics[width=0.94\columnwidth]{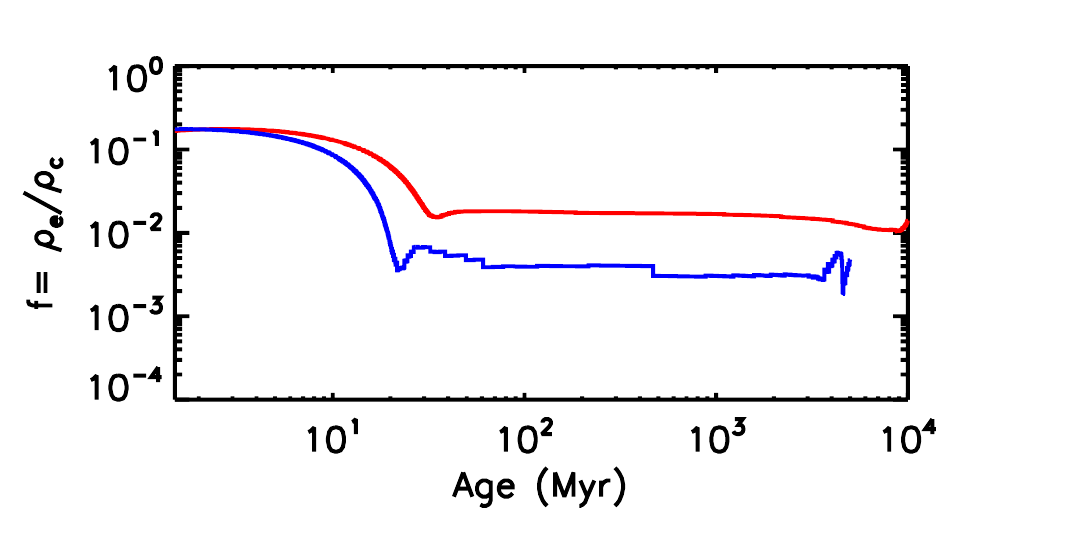}
	\caption{Temporal evolution of the core to envelope ratio (top panel) and envelope to core density ratio (bottom panel) for our models of a 1~M$_\odot$ star (in red) and a 1.2~M$_\odot$ star (in blue).}
	\label{fig:alphaandf}
\end{figure}
  
To compare the evolution of $\int^{\infty}_{-\infty} {\rm Im}[k_l^m(\omega)] {\rm d}\omega/\omega$ for the sectoral and tesseral harmonics, we computed it for models of a 1~M$_\odot$ and a 1.2~M$_\odot$ and two values of initial stellar rotation frequency $3 \Omega_\odot$ and $25 \Omega_\odot$. Those values correspond to rotation periods of $\sim 8.5$ and $1$~days respectively, and are representative of the range of surface rotation periods measured in clusters younger than $\sim 2$~Myr for stars in this mass range \citep[e.g.][]{Gallet2015}. The stellar structure quantities and their time derivatives are taken from evolutionary models by \citet{Baraffe1998}. To obtain the variables of our piecewise-homogeneous fluid model, we compute the mean density in each layer simply using the mass and thickness of the corresponding zone. We plot in Fig.~\ref{fig:alphaandf} the temporal evolution of $\alpha$ the core to envelope radius ratio (top panel), and $f$ the envelope to core density ratio ( bottom panel; see also Appendix \ref{hydrostat}) that we get from the stellar models for the two masses considered here.

 Our evolution begins at $t_0 \sim 1.5$ Myr, when stars are almost fully convective, justifying our assumption of solid-body rotation as our initial condition. After the disk-locking phase, the angular momentum of the star evolves as a consequence of both structural evolution and magnetic braking (see Sec.~\ref{sec:magbrak}). Here the parameters are taken from \citet{Spada2011} and summarised in Table~\ref{tab:params}. There is a lack of reliable calibration of the magnetic braking laws in the framework of the double-zone model for stars with $M_\star > 1.1$~M$_\odot$. Here, we simply assume the same values apply also to our  1.2~M$_\odot$ model for all parameters, except for $K_w$, that we divide by a factor 10 as in \citet{Barker2009}. 

   \begin{table}
      \caption[]{Values of the parameters used to plot Fig.~\ref{fig:dissp} for the 1~M$_\odot$ model \citep[from][]{Spada2011}. We use $\Omega_\odot = 89.94$ yr$^{-1}$}
         \label{tab:params}
         \centering
         \begin{tabular}{ll}
            \hline
            Parameter    &  Value  \\
            \hline
            $\tau_{\rm disk}$ & 5.8 Myr\\
            $\Omega_{\rm crit}$ & 3.89 $\Omega_\odot$\\
            $\Omega_{\rm sat}$ & 5.5 $\Omega_\odot$\\
            $K_w$ & $1.71\times 10^{33}$ kg.m$^2$.yr \\
            \hline
           \end{tabular}
          \end{table}

\begin{figure*}
	\centering
	\includegraphics[width=0.47\linewidth]{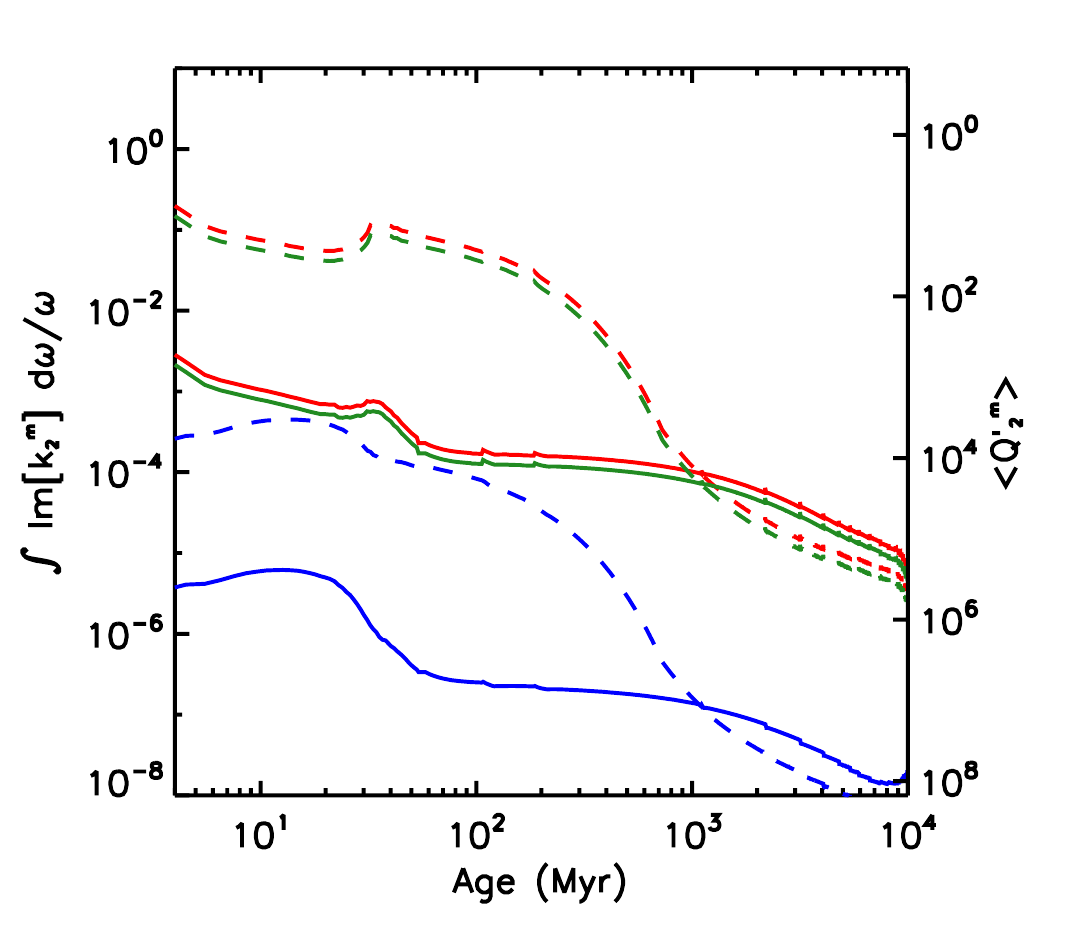}\quad \includegraphics[width=0.47\linewidth]{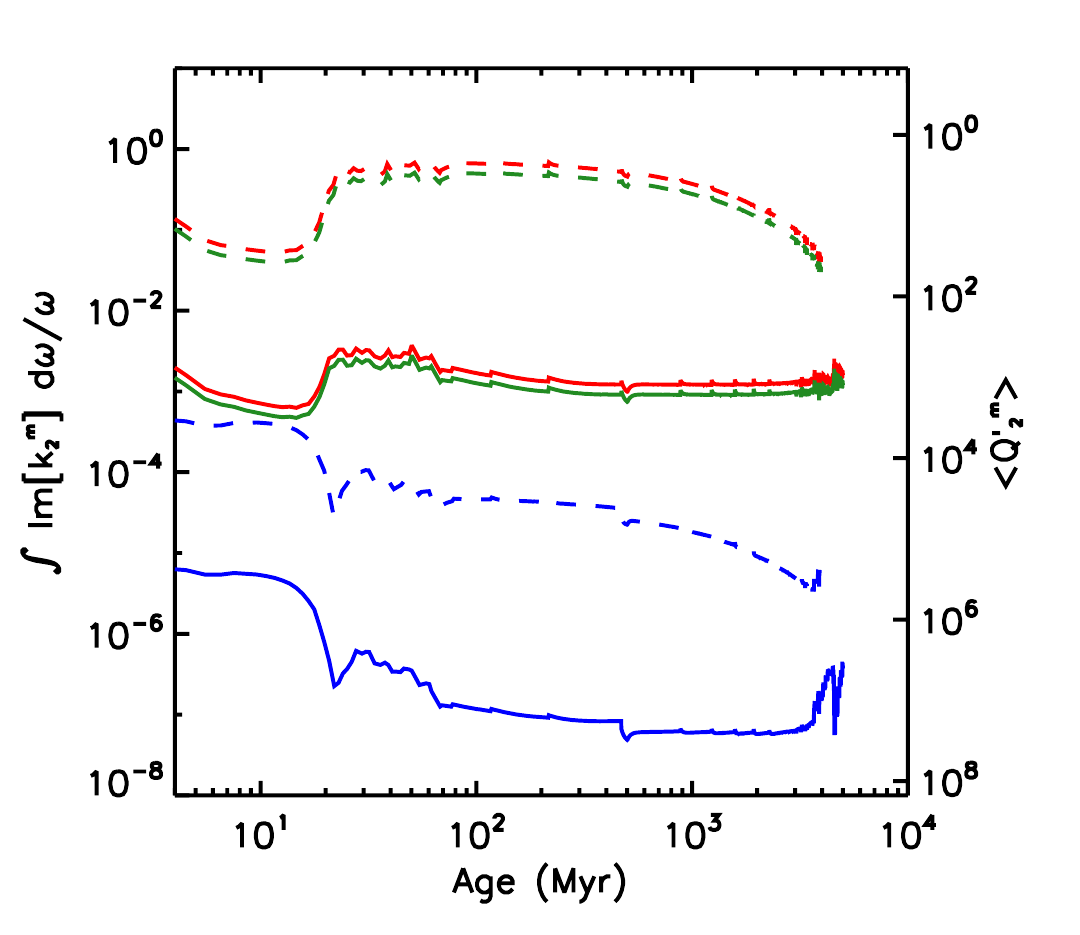}
	\caption{Left: Temporal evolution of the frequency-averaged imaginary part of the potential Love number corresponding to components of degree $l=2$ and $m=0$ (red), $m=1$ (green) and $m=2$ (blue) for models of a 1~M$_\odot$ (left) and 1.2~M$_\odot$ (right)  mass stars. The solid and dashed lines correspond to a star initially rotating with $\Omega(t_0) = 3 \Omega_\odot$, and $\Omega(t_0) = 25 \Omega_\odot$ respectively. A conversion to the modified tidal quality factor $Q'$ is given on the right $y$-axis.}
	\label{fig:dissp}
\end{figure*}

To allow easy comparison with the values found in the literature for the equilibrium tide, we also give the conversion of our estimation of the dissipation rate as the traditional equivalent modified tidal quality factor $Q'$, using the relationship
\begin{equation}
<Q_l^{'m}> = \frac{3}{2 k_l \bar{\Delta}_{l}^{m}}. 
\end{equation}
Here we get a value of $Q'$ that is not exactly frequency-averaged, but rather corresponds to a frequency average of the phase lag \citep[see][ for a discussion on this point]{Mathis2015sf2a}. The results are shown in Fig.~\ref{fig:dissp}. 

We see that the $m=0$ and $m=1$ components are almost identical due to similar contributions of the toroidal part to the impulse energy, and they are more than one order of magnitude greater than the $m=2$ component during the main-sequence. This clearly shows that assuming that different components have the same time-lag, as is done in the equilibrium tide theory, severely underestimates the participation of the tesseral harmonics of the tide, when the tidal frequency is within the range allowing the existence of inertial waves.  Overall, we find that the value of  $Q^{'0}_2$ and $Q^{'1}_2$ assumes value between 1 and $10^6$ throughout stellar evolution across slow and fast rotators. For 1.2~M$_\odot$ mass stars, fast rotators can have $Q^{'0}_2$ and $Q^{'1}_2$ smaller than 100 throughout their evolution. This is extremely efficient for fluid dissipation, and is comparable to the value of $Q'$ estimated for the rocky bodies of the solar system, estimated in the framework of the equilibrium tide \citep{Lainey2016}. 

We must stress that we are using here a simplified model which considers that the densities in the radiative core and the convective envelope are both homogeneous. For the sectoral harmonic $l=m=2$, \citet{Ogilvie2013} has shown, using polytropes of different indexes and in the presence of a solid core, that using various density profiles can change the frequency-averaged dissipation-rate by orders of magnitude depending on the core to envelope radius ratio $\alpha$ \citep[see Fig. 10 in][]{Ogilvie2013}. The general trend, for $\alpha \gtrsim 0.3 $, is that the dissipation becomes less efficient as the star gets more centrally condensed. In addition, \citet{LinOgilvie2017} have shown numerically that the dissipation rate of the obliquity tide, for a fluid body containing a solid core, is enhanced by 1 to 2 orders of magnitude in the homogenous case compared to the more realistic case of a polytropic density of index 1(we refer the reader to Figs. 4 \& 7 in their article). Here, we use a piecewise homogeneous fluid model. Therefore, we cannot exclude that modelling the convective envelope as homogeneous may artificially enhance the dissipation and this could potentially significantly modify our results regarding the timescales of tidal evolution. In addition, we know from previous works \citep[e.g.][]{Ogilvie2004,Ogilvie2007,Barker2009} that the dissipation of tidal inertial waves can vary over several orders of magnitude as a function of the tidal frequency. At a given frequency, it may thus differ from its frequency-averaged value. However, we expect this latter to provide us a reasonable order of magnitude of the dissipation value that we can use in planetary systems secular evolution models.

Dissipation is strongly affected by stellar rotation. In the magnetic braking model considered here, initially fast and slow rotators converge to a common rotation rate after the first Gyr of evolution, but they have very different behaviours during the firsts hundreds of Myr. During this phase, the angular momentum evolution is driven by the evolution of the structure of the star, dominated by its global contraction. Thus, the value of $Q'$ during this phase is very similar for the two stellar masses considered here, with $Q^{'0}_2$ and $Q^{'1}_2 \sim 10^1 - 10^3$ and  $Q^{'2}_2 \sim 10^3 - 10^5$ depending on the value of the initial rotation. Those values are maintained during the first 20-30 Myr of evolution. This implies that any hot-Jupiter that would reach a 3-day period orbit in the first tens of Myr would spiral into the star and be destructed by tides within a few $10^5$ years, if the tidal frequency falls within the range of inertial waves and if the rotation period of the star is greater than 3 days, regardless of the mass of the host. Yet, one should keep in mind that the values obtained here may be considered as upper bounds on the dissipation because of the hypothesis of homogeneous density in the convective envelope.

 On the main sequence, the stellar structure stops evolving and the evolution of the dissipation is mainly controlled by the rotation frequency of the envelope. For the 1~M$_\odot$ mass star, we see a steady decrease in dissipation efficiency, with $Q^{'0}_2$ and $Q^{'1}_2 \sim 10^4 - 10^5$ and  $Q^{'2}_2 \sim 10^7 - 10^9$. For the 1.2~M$_\odot$ mass star, dissipation efficiency appears to plateau around $Q^{'0}_2$ and $Q^{'1}_2 \sim 10^1 - 10^3$ and  $Q^{'2}_2 \sim 10^5 - 10^7$, because angular momentum loss is reduced and the star spins down less efficiently. After about 4 Gyrs, the mass of the convective envelope becomes negligible and our computation of $Q'$ suffers from numerical instabilities inherent to the stellar evolutionary track we are using. 
These order of magnitudes are in agreement with those obtained by \citet{Gallet2017}, who also computed the impact of the simultaneous structural and rotational evolution of low-mass stars on the dissipation of tidal inertial waves in their convective envelope (we refer the reader to their Figs. 4 and 5 - top left panel) for the $l=m=2$ components.
 
Interestingly here, unlike what is found in \citet{Gallet2017}, we see that despite the less important convective zone of the 1.2~M$_\odot$ mass star, the energy dissipated by the inertial waves for this stellar mass is about the same order of magnitude, if not greater, than for the 1~M$_\odot$ mass star, for a given set of $l, m$ and initial rotation frequency. 
 
\begin{figure*}
	\centering
	\includegraphics[width=0.47\linewidth]{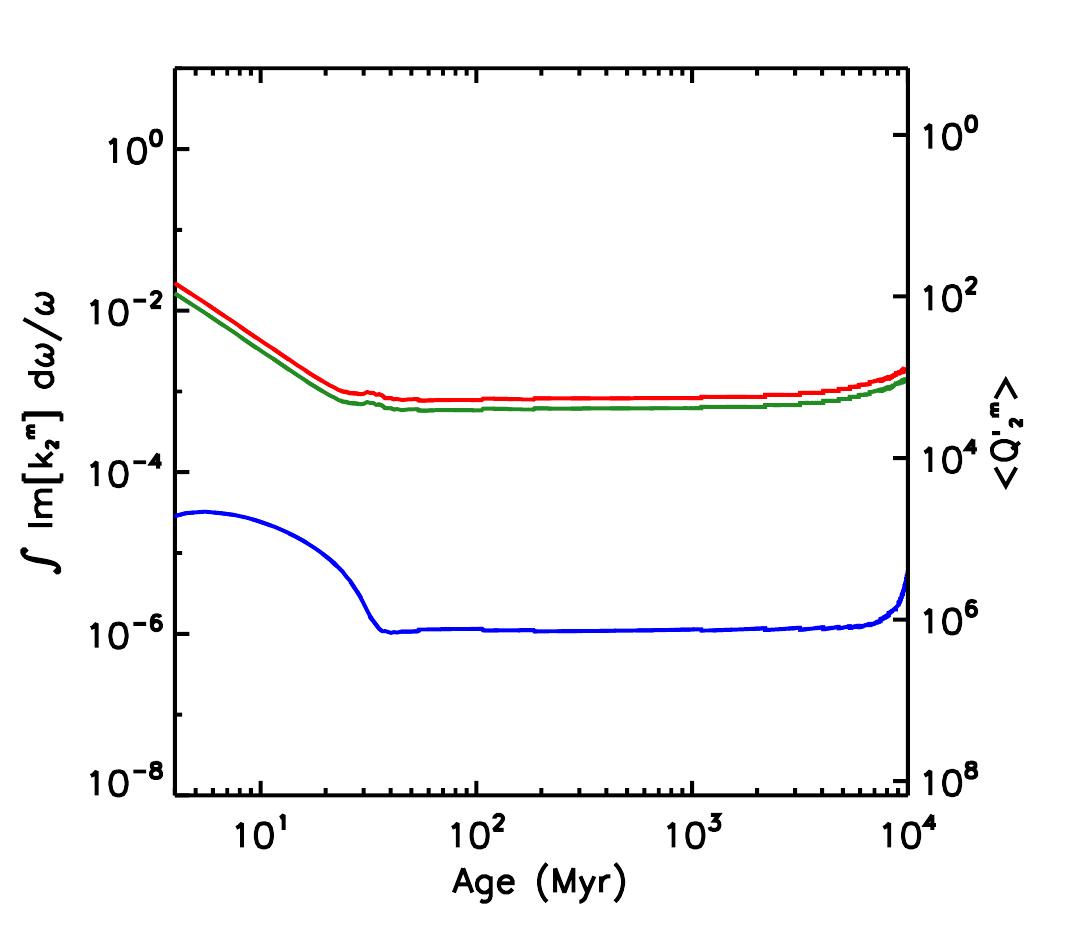}\quad \includegraphics[width=0.47\linewidth]{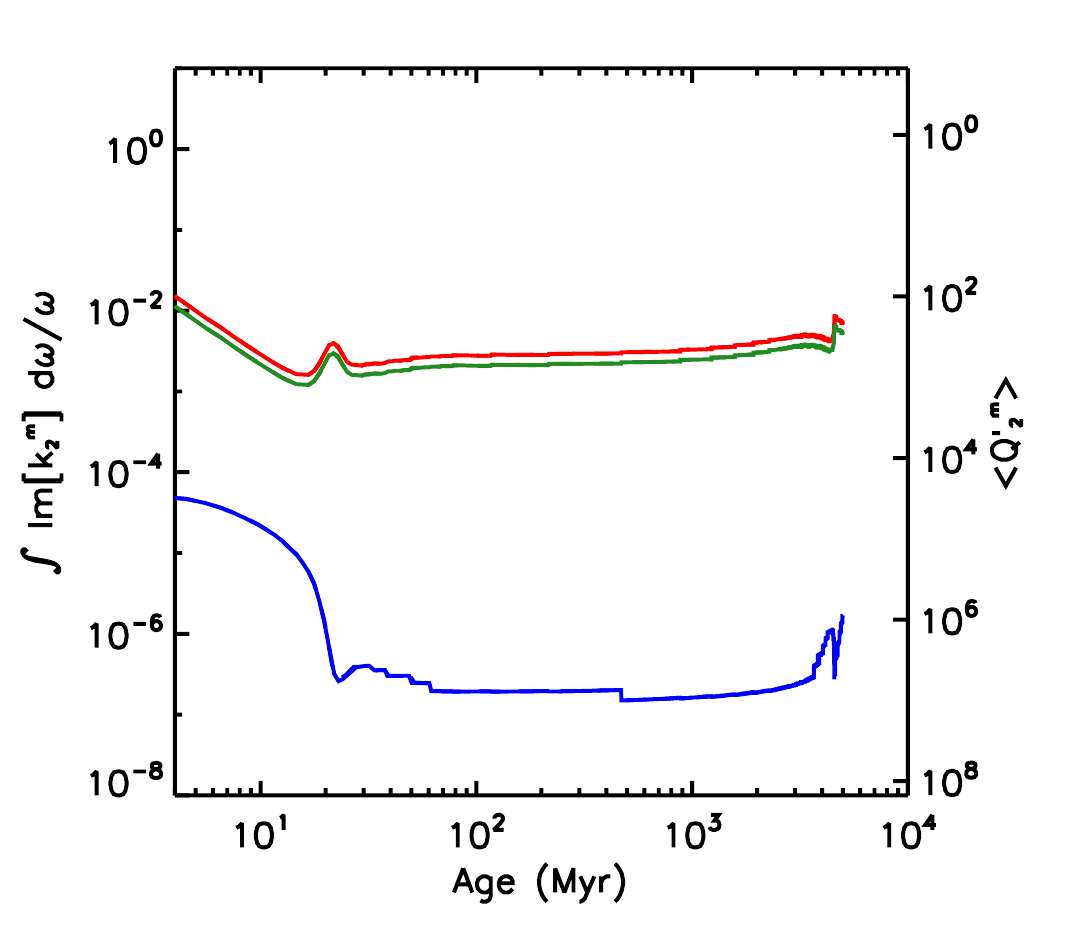}
	\caption{Left: Temporal evolution of the frequency-averaged imaginary part of the potential Love number corresponding to components of degree $l=2$ and $m=0$ (red), $m=1$ (green) and $m=2$ (blue) for models of a 1~M$_\odot$ (left) and 1.2~M$_\odot$ (right)  mass stars. The rotation rate of the star does not evolve in time and is set at all times at $\Omega = 8.5 \Omega_\odot$ for both models. A conversion to the modified tidal quality factor $Q'$ is given on the right $y$-axis.}
	\label{fig:imklm3d}
\end{figure*}

This may seem to be also in contradiction with previous numerical simulations which showed that this dissipation rate should scale with the extent of the convection zone \citep{Ogilvie2007, Barker2009}. But one must remember that those simulations were computed at a fixed rotation rate while here we compute the evolution of the rotation of the star as its internal structure evolves. To better compare our results with these numerical simulations, we show in Fig.~\ref{fig:imklm3d} the frequency-averaged dissipation obtained with our model along the evolution of the stellar structure, but fixing the rotation frequency of the star at $8.5\Omega_\odot$, which corresponds to the rotation period of 3 days, a value chosen arbitrarily by \cite{Ogilvie2007} and \cite{Barker2009} to illustrate their results.

Firstly, let us consider the case of the $l=m=2$ component. For this component, the frequency dependent dissipation rate of inertial waves for a solar mass star rotating with the 3-day period has been computed in  \citet{Ogilvie2007}. Besides the complex dependance upon tidal frequency, their equivalent $Q'$ ranges between $10^6$ and $10^8$. Using the same numerical set up, \citet{Barker2009} obtained, for a 1.2~M$_\odot$ mass star rotating at the same angular velocity, a significantly lower $Q'$, ranging between $10^8$ and $10^{10}$ for the same tidal component. Here, on the main sequence, we obtain $Q'\sim 10^6$ and $Q'\sim 10^7$ for the respective masses. Considering again that our model assumes piecewise homogenous density, which leads to upper bounds on the dissipation, our results are thus in reasonable agreement with the trends of these more realistic numerical simulations, albeit the difference in dissipation as a function of mass is less contrasted. Again, this is in broad agreement with \citet{Gallet2017} (compare their Fig. 4 top-left and bottom-left panels) where the difference in dissipation between a 1~M$_\odot$ and a 1.2~M$_\odot$ mass star is found to be greater when considering only the effect of structure, compared to the case when the evolution of both the structure and rotation is taken into account. Comparing Fig.~\ref{fig:dissp} and \ref{fig:imklm3d} (and also top-left and bottom-left panels of Fig. 4 in \citet{Gallet2017}, although in their case the 1.2~M$_\odot$ mass star always dissipates less), it is clear that the main factor affecting the difference in dissipation on the main sequence between the two stellar models considered here is thus the evolution of stellar rotation : the 1.2~M$_\odot$ mass star spins down less rapidly and thus maintains a dissipation rate throughout its main sequence similar to those of the 1~M$_\odot$ mass star independently from the thickness of their respective convective envelope.

\begin{figure}
	\centering
	\includegraphics[width=0.9\columnwidth]{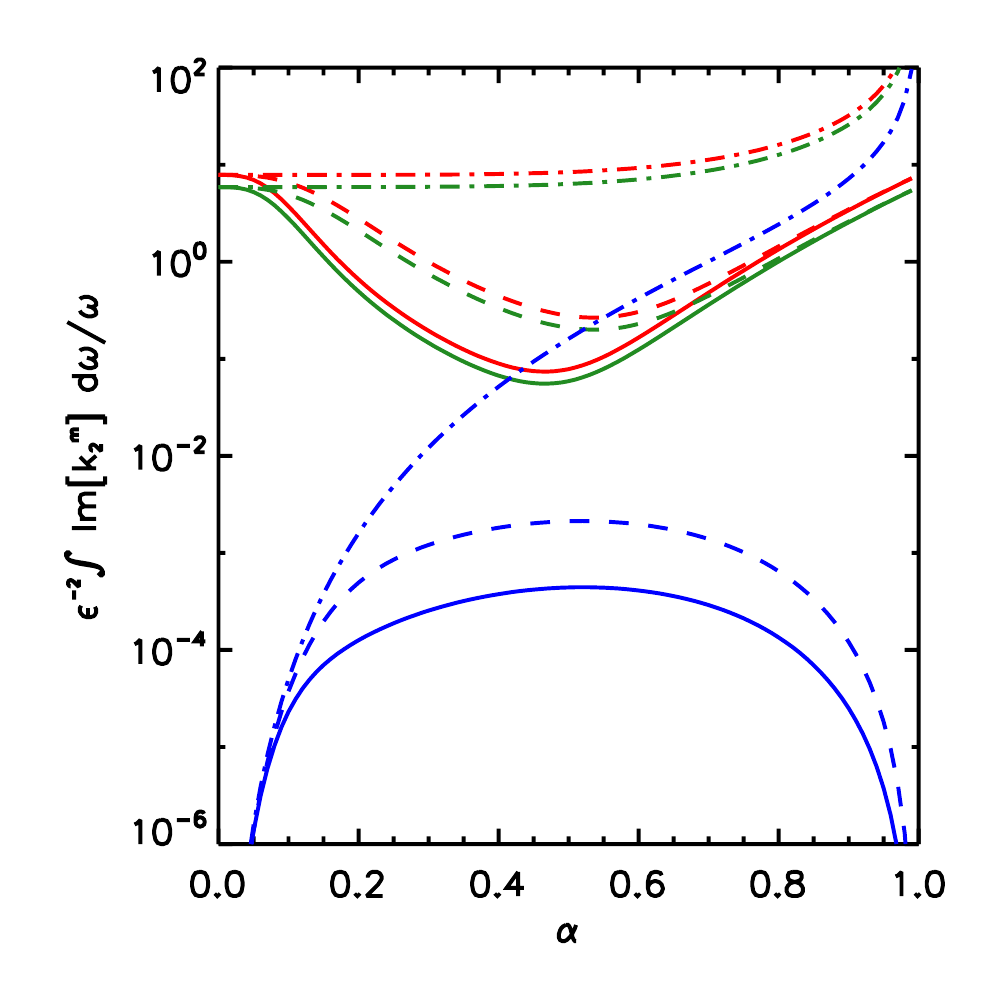}
	\caption{Frequency-integral of the imaginary part of the Love numbers, weighted by $1/\omega$ and divided by $\epsilon^2$, versus fractional core size for the quadrupolar component $l=2$. The colour of the lines correspond to different values of $m$ :  $m=0$ (red), $m=1$ (green) and $m=2$ (blue). The dashed-dotted line shows the analytical result of the impulse calculation for a homogeneous fluid body with a solid core. The dashed line shows the same but in the case of the piecewise homogeneous fluid model with a density ratio of $2\times10^{-2}$, adequate for a solar mass star on the main sequence. The solid line corresponds again to the piecewise homogeneous fluid model but with a density ratio of $4\times10^{-3}$, appropriate for a 1.2~M$_\odot$ mass star on the main sequence.}
	\label{fig:fig6}
\end{figure}

Nevertheless, for a given initial rotation rate, as previously noted, the dissipation of the $m=0$ and $m=1$ components is always orders of magnitude greater than for the $m=2$ component. This seems to disagree with previous results from numerical simulations computed in \citet{Barker2009}, who showed that the dissipation evaluated for different values of $m$ are similar in magnitude. Such a discrepancy is also observed between the frequency-averaged estimation and numerical results obtained by \citet{LinOgilvie2017} for the dissipation rate of the obliquity tide in the case of an homogeneous fluid containing a solid core. Indeed, they show in their Fig. 4 the dimensionless dissipation rate $\tilde{D}$ of the obliquity tide as defined in their Eq.~(69) for a homogeneous fluid. It ranges between $10^{-5}$ and $1$ depending on the value of the fractional core radius. However, Fig. 6 of \citet{Ogilvie2013} (second red line from the top, also partly reproduced in Fig. \ref{fig:fig6}) shows that the frequency-averaged, dimensionless, rotation normalised, equivalent dissipation for the same homogenous fluid is greater than $10$ and diverges as $\alpha \rightarrow 1$. The dissipation rate computed in \citet{LinOgilvie2017} is not exactly equivalent to the quantity obtained in the impulsive response formulation in \citet{Ogilvie2013}. However, the forcing frequency is independent of the orbital frequency of the planet for the obliquity tide, and both quantities, if we relate them through an equivalent quality factor $Q'$, should agree to a factor of the order of unity. This does not seem to be the case and may be related to the effect of precessional forces.

In addition, compared to the fluid case, the boundary conditions imposed by a solid core have large repercussions on the functional dependence of the frequency-averaged tidal dissipation with the core size. This can be appreciated on Fig.~\ref{fig:fig6}, where we plot the result of the frequency-averaged dissipation, divided by $\epsilon^2=\Omega^2 \left({G M_\star}/{R_\star}\right)$, in the cases of a homogenous fluid with a solid core and of the piecewise fluid model considered here, for all components of the quadrupolar tidal forcing. 

First, we can clearly see that the piecewise homogenous fluid has overall less efficient dissipation for all values of $m$. This is expected since the core in this model is not perfectly rigid but instead undergoes some tidal deformation. This produces slower horizontal flow than in the solid core case, and through the Coriolis force, a smaller wavelike velocity. 

Next, in the range of parameters corresponding to main sequence values (see Fig. \ref{fig:alphaandf}), the dissipation of the zonal and tesseral components clearly dominates that of the sectoral component in the case of the piecewise fluid model. This quantitative difference exists also in the solid-core case but is mainly important for thick convection zones. 

Here in the piecewise homogeneous fluid model, the integrated response to a tesseral harmonic is always much stronger than the one of the sectoral harmonic. This seems to be a robust result for the dissipation computed for an impulsive forcing. However, the question of the effect of the density profile on the relative dissipations of the different tidal components has not yet been investigated beyond the two simple models discussed here. This requires further theoretical investigations that are beyond the scope of this article. Let us note here that, in the present model, the difference in stellar structure between a 1~M$_\odot$ and a 1.2~M$_\odot$ mass star on the main sequence (mainly stemming from the density of the envelope) has little effect on the tidal dissipation for the $m=0$ and $m=1$ components, but has a greater impact on the $m=2$ component.
Again, the predominant factor here will be the evolution of stellar rotation. Since the 1.2~M$_\odot$ mass star remains a fast rotator throughout its main-sequence, it has overall a more efficient dissipation, if we consider that our impulsive response formulation provides a robust measure of the dissipative properties of the star.

This would tend to produce an anti-correlation of obliquity with stellar effective temperature, contrary to what is observed. Nevertheless, this result is very sensitive to the model chosen for angular momentum evolution. For example, \citet{Gallet2017} use a different prescription for the angular momentum evolution of the star. They use stellar evolution models that take into account rotation, with angular momentum transport operated through meridional circulation and diffusion by shear turbulence. This results in a much less efficient coupling between the radiative zone and convective envelope than the model used here \citep{Charbonnel2013}. Moreover, in their model, the magnetic braking law involves a parametrisation based on the Rossby number. Consequently, for the $l=m=2$ mode, they find that dissipation scales with the extent of the convective zone for stellar mass between 1 and 1.4~M$_\odot$ on the main sequence, with $<Q_2^{'2}>$ increasing by about an order of magnitude over this mass range. Here our results differ because we allow the coupling of the radiative and convective zones, which affects the rotation rate of the envelope and thus the value of the dissipation. We estimate tidal dissipation efficiency using a simplified model, valid in the convective zone only and which assumes rigid body rotation, but our computations show that even in this framework, the internal angular momentum transport throughout the star may have a strong impact on tidal dissipation efficiency.

To conclude, we find different dissipation efficiencies associated to different modes, and this supports the possibility of having different timescales for the tidal evolution of the semi-major axis and the obliquity. But the absolute value of the dissipation is also very sensitive to the evolution of the rotation of the star. With the magnetic braking law used here, this would tend to produce an anti-correlation of obliquity with stellar effective temperature, contrary to what is observed. In any cases, the value we find here would imply typical timescales for tidal evolution that could be much shorter than the main-sequence life-time of the star, depending on the orbital configuration. Thus the final state of Hot-Jupiter systems is not easily inferred from the relative values of $Q'^l_m$ alone, but must also consider the initial rotation period of the star and the coupled evolution of orbital elements.

\section{Coupled evolution of semi-major axis and obliquity}\label{sec:compute}
Using Eqs.~\eqref{trasenergyquad}, \eqref{axtorque2}, \eqref{transtorque},  \eqref{evolom}, \eqref{evola} and \eqref{evoli}, we compute the coupled temporal evolution of the orbital elements for different values of initial obliquity, semi-major axis and stellar rotation. Our evolution begins again at $t_0 \sim 1.5$~Myr, when stars are almost fully convective, and thus rotating as a solid-body. It stops at the end of the main-sequence phase (i.e. after 10 Gyr for the 1~M$_\odot$ star and 5 Gyr for the 1.2~M$_\odot$ star) or when the planet reaches the Roche limit ($a_{\rm R}=2.422 R_{\rm p}\sqrt[3]{M_\star/M_{\rm p}}$, with $R_{\rm p} = 1.2 R_{\rm Jup}$ for a Jupiter-mass planet), or alternatively when the star is spinning faster than the break-up velocity ($\Omega_{\rm B} = \sqrt{G M_\star/ R_\star^3}$). During the disk-locking phase, we consider that the planet is maintained on its orbit by its interaction with the disk. Since the rotational evolution of the star is clearly important for tidal evolution, and the distinction between initially slow and fast rotator may be paramount, we opted to relate the disk-lifetime $\tau_{\rm disk}$ and the coupling time-scale $\tau_{\rm c}$ to the initial rotation rate of the star following the prescription of \citet{Gallet2015}. However, keeping the prescription of \citet{Spada2011} for the other parameters as in the previous section resulted in problematic evolutions in the case of initially fast rotating 1.2~M$_\odot$ mass stars. For most of the initial parameter space explored, tidal evolution would lead either to the star spinning up to the break-up velocity or the planet reaching the Roche limit before the first few Myrs. This suggests that the chosen angular momentum loss model may not apply to this kind of host. Therefore, and in the lack of calibration for angular momentum loss rate in the framework of the double-zone model for 1.2~M$_\odot$ mass stars, we opted for a parametrisation akin to the one proposed by \citet{Matt2015} for the solid rotation case. Namely, we take the same value for the scaling factor $K_{\rm w}$ for all masses, and we include the mass dependence  through the value of the saturation threshold  $\Omega_{\rm sat}$. In this prescription,  $\Omega_{\rm sat}$  is inversely proportional to the convective turnover timescale in the star $\tau_{\rm cz}$
\begin{equation}
\Omega_{\rm sat} =  \Omega_{\rm sat_\odot} \frac{\tau_{\rm cz_\odot}}{\tau_{\rm cz}},
\end{equation}
where  $\Omega_{\rm sat_\odot}$ and ${\tau_{\rm cz_\odot}}$ are normalisation values and where $\tau_{\rm cz}$ is computed from the effective temperature on the ZAMS according to the prescription of \citet{Cranmer2011}.  Table~\ref{tab:param2} and \ref{tab:param3} lists the values of all adopted parameters for stellar angular momentum loss in the present section. This slightly different magnetic braking model produces a qualitatively similar evolution of dissipation as the one presented in Fig.~\ref{fig:dissp}, but now, most of the 1.2~M$_\odot$ hosts do not enter the saturated regime of the magnetic braking law. The peak of their rotation speed is not as high during the initial spin up phase, so even though they do not enter the saturated regime, their instantaneous rate of angular speed loss is weaker than for the 1~M$_\odot$ stars, during the first Gyr of evolution.
\begin{figure}
	\centering
	\includegraphics[width=0.94\linewidth]{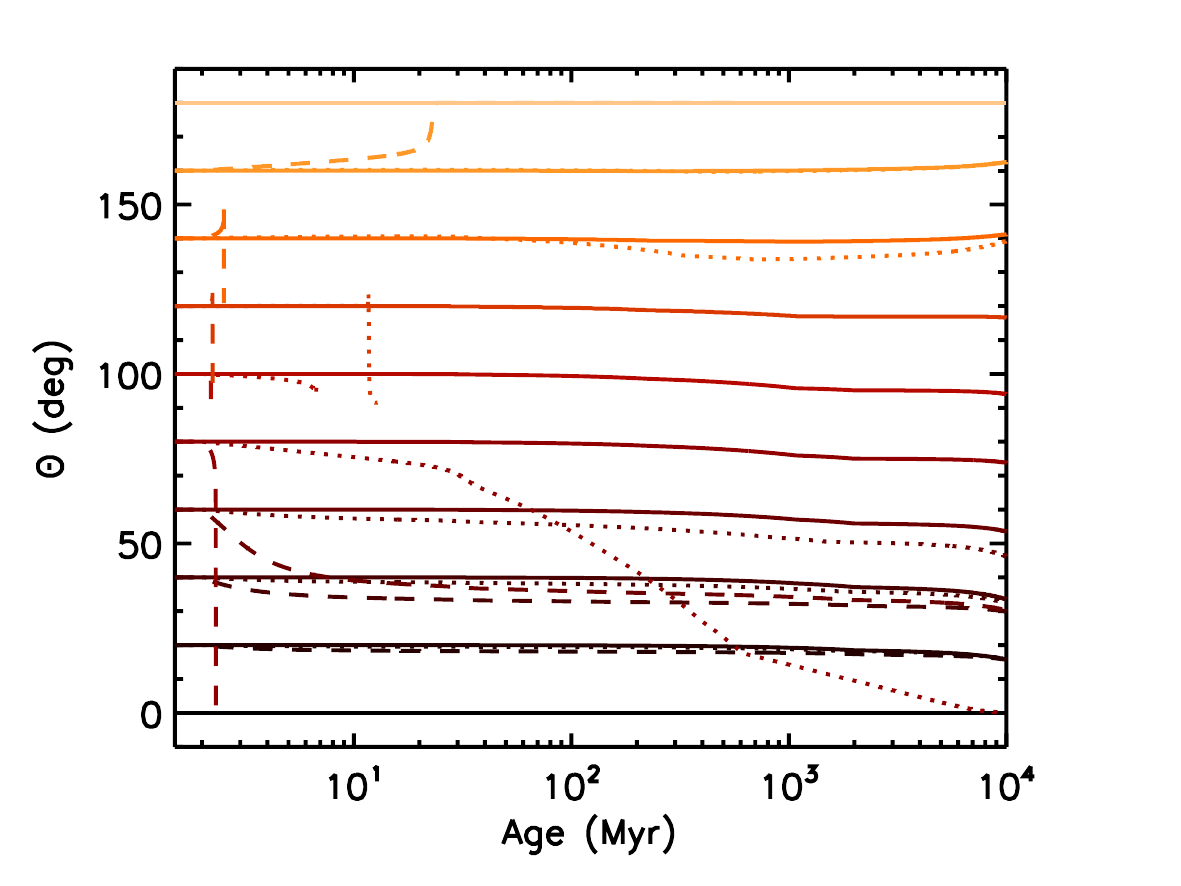}\\
	\includegraphics[width=0.94\linewidth]{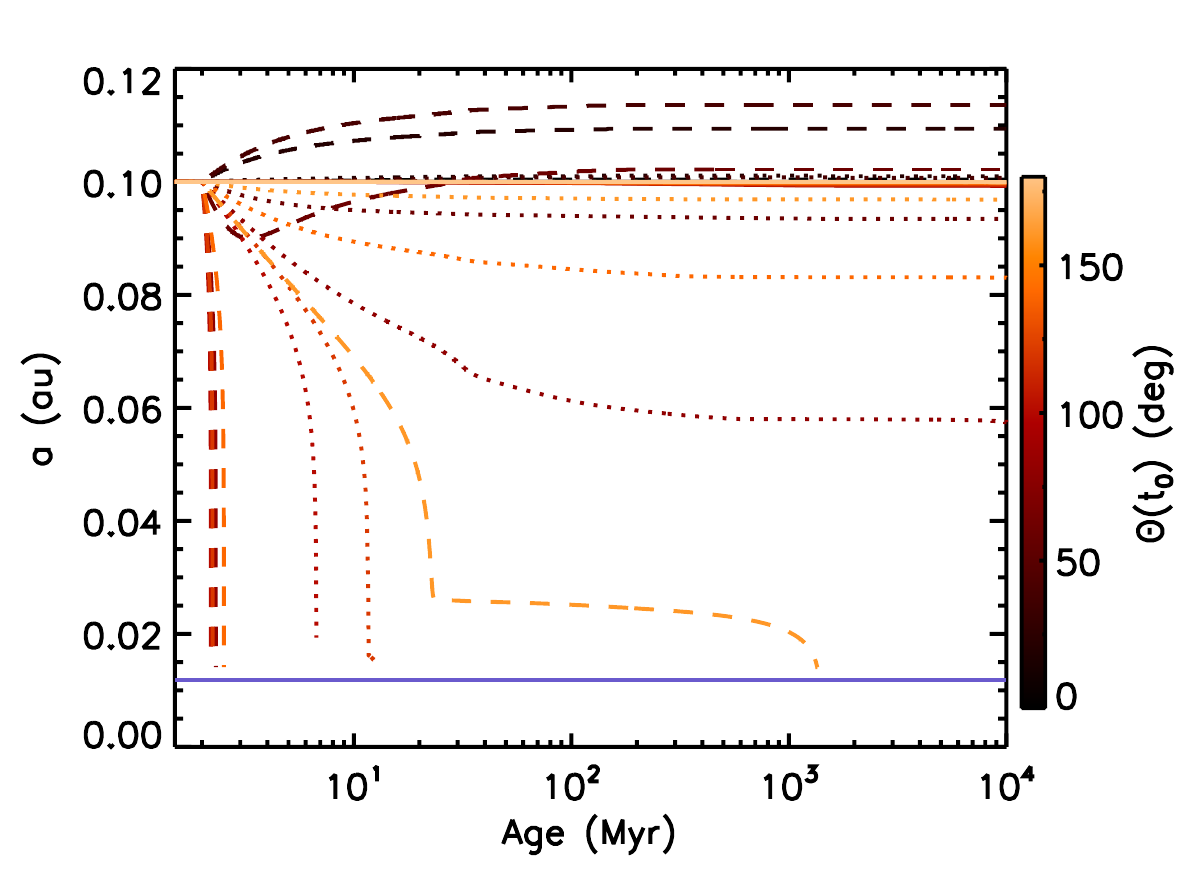}\\
	\includegraphics[width=0.94\linewidth]{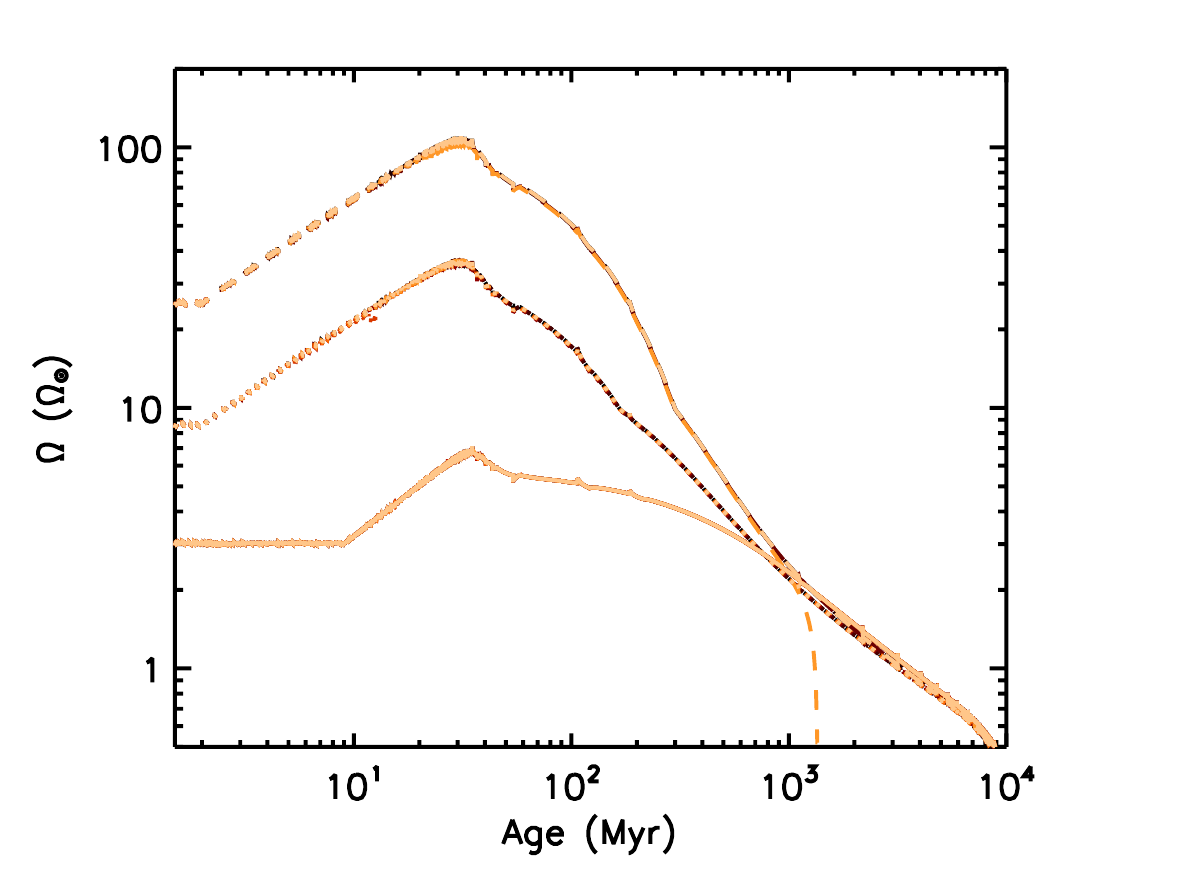}	
	\caption{Temporal evolution of the obliquity (top panel), semi-major axis (middle panel) and rotation frequency of the envelope (bottom panel) for a 1M$_{\rm Jup}$ planet orbiting a 1~M$_\odot$ star starting at $a(t_0)=0.1$ AU. The line colors indicate the value of the initial obliquity as indicated on the colour scale in the middle panels. The solid, dotted and dashed lines correspond to a star initially rotating with $\Omega(t_0) = 3, 8.5$ and $25\Omega_\odot$ respectively. The solid blue line on the middle panels indicates the Roche limit, where our evolution is stopped.}
	\label{fig:orbitevolout100}
\end{figure}
\begin{table}
	\caption[]{Values of the magnetic braking model parameters depending on the initial rotation frequency, used to compute the evolutions shown in Fig.~\ref{fig:orbitevolout100} to Fig.~\ref{fig:klmevolout120}, for all stellar masses. We use $\Omega_\odot = 89.94$ yr$^{-1}$ and $\Omega_{\rm crit} = 3.89 \Omega_\odot$. }
         \label{tab:param2}
         \centering
         \begin{tabular}{lll}
            \hline
              & $\Omega_{\rm e}(t_0) \geq \Omega_{\rm crit}$ & $\Omega_{\rm e}(t_0)  < \Omega_{\rm crit}$\\
            \hline
            $\tau_{\rm disk}$ & 2 Myr & 9 Myr \\
            $\tau_c$ & 10 Myr & 30 Myr \\
            \hline
           \end{tabular}
          \end{table}
   \begin{table}
      \caption[]{Values of the magnetic braking model parameters depending on stellar mass, used to compute the evolutions shown in Fig.~\ref{fig:orbitevolout100} to \ref{fig:klmevolout120}, for all initial rotation rates. We use $\Omega_\odot = 89.94$ yr$^{-1}$ and $\Omega_{\rm crit} = 3.89 \Omega_\odot$. }
         \label{tab:param3}
         \centering
         \begin{tabular}{lll}
            \hline
               & 1~M$_\odot$ & 1.2~M$_\odot$\\
            \hline
             $\Omega_{\rm sat}$ & 10 $\Omega_\odot$ & 42 $\Omega_\odot$ \\
            $K_w$ & $1.71\times 10^{33}$ kg.m$^2$.yr & $1.71\times 10^{33}$ kg.m$^2$.yr\\
            \hline
           \end{tabular}
          \end{table}
          
Finally, we performed orbital evolution calculations considering a range of the value of $Q_{\rm eq}'$ defined in Eq.~\eqref{eq:cases} (corresponding to the frequency range where inertial waves cannot exist). It turns out that in most case, the results do not change significantly for $Q_{\rm eq}'\simeq 10^6 - 10^{10}$ \citep[see e.g. ][]{Hansen2010, Hansen2012}. Thus we present here the results obtained with $Q_{\rm eq}'=10^7$. We integrate the set of equations presented in Section~\ref{sec:orbitevol} using a fourth-order Runge-Kutta integrator with an adaptive time-step routine \citep{Press1992}. Examples of evolutions are given in Fig.~\ref{fig:orbitevolout100} and for a 1~M$_\odot$ mass star and Fig.~\ref{fig:orbitevolout120}.

\begin{figure}[hbt]
	\centering
	\includegraphics[width=0.94\linewidth]{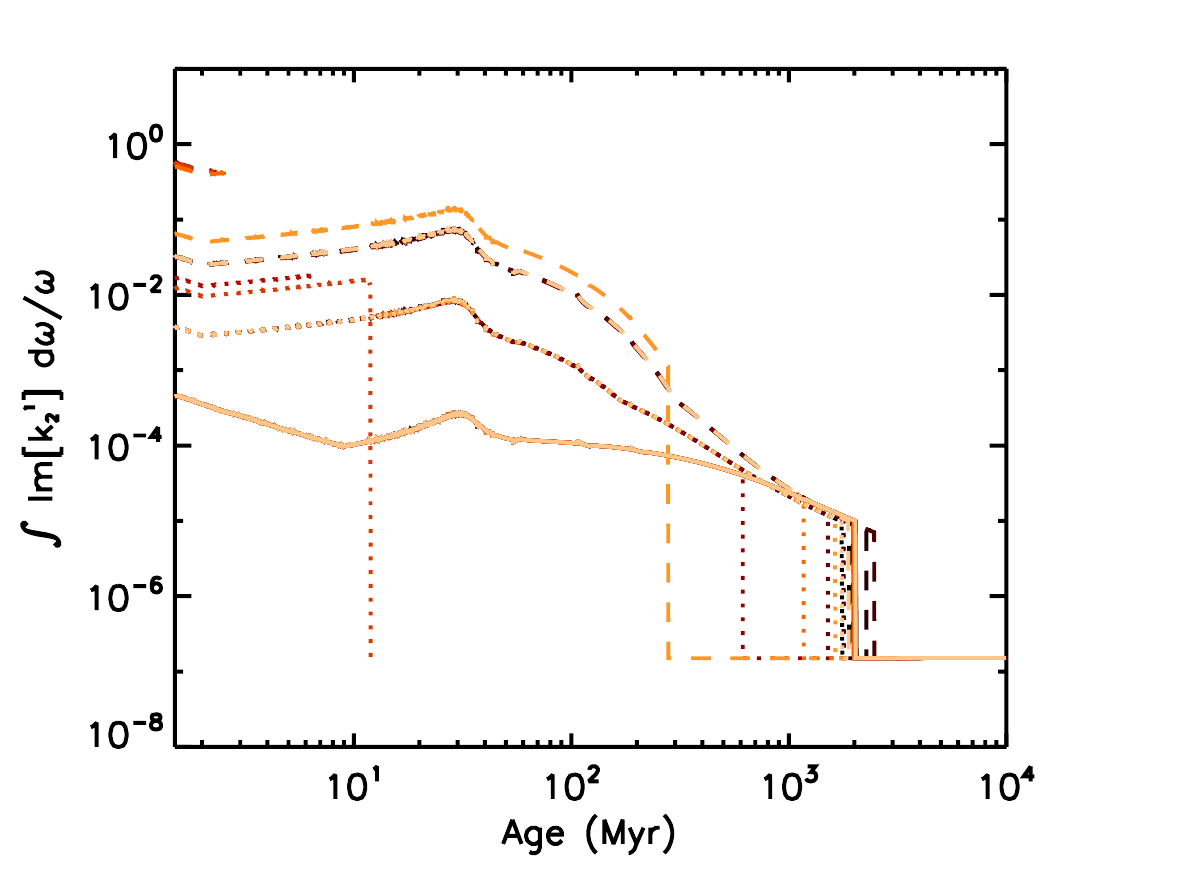}\\	\includegraphics[width=0.94\linewidth]{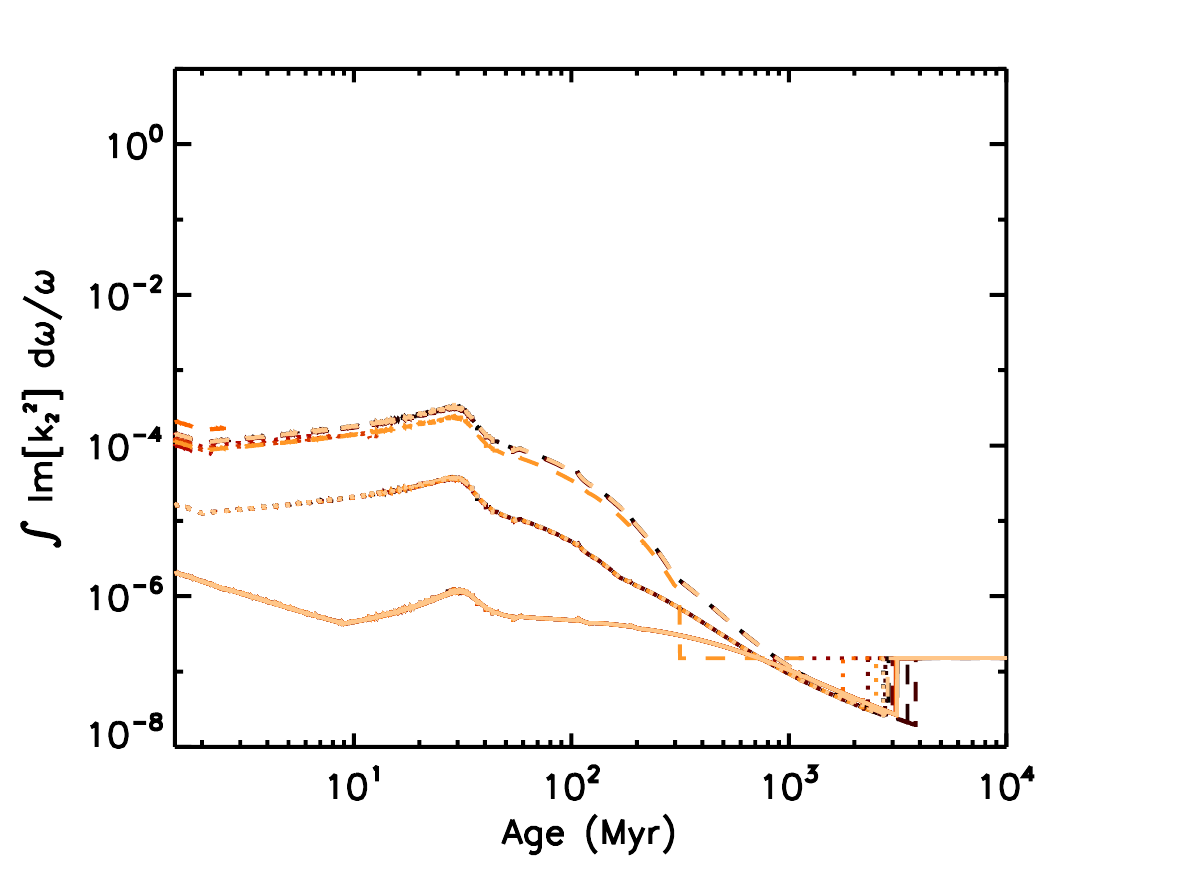}	
	\caption{Temporal evolution of the frequency-averaged imaginary part of the potential Love number corresponding to components of degree $l=2, m=1$ (top) and $l=m=2$ (bottom) along the tidal evolution of a the systems shown in Fig.~\ref{fig:orbitevolout100} of a a 1M$_{\rm Jup}$ planet orbiting 1~M$_\odot$ mass star starting at $a(t_0)=0.1$ AU. The line-styles and colors correspond match those of the tracks shown in Fig.~\ref{fig:orbitevolout100}. The dissipation starts with the value expected from the dissipation of inertial waves, and assumes the equilibrium tide constant Q' value only after the first Gyr for most orbital configurations.}
	\label{fig:klmevolout100}
\end{figure}

\begin{figure}
	\centering
	\includegraphics[width=0.94\columnwidth]{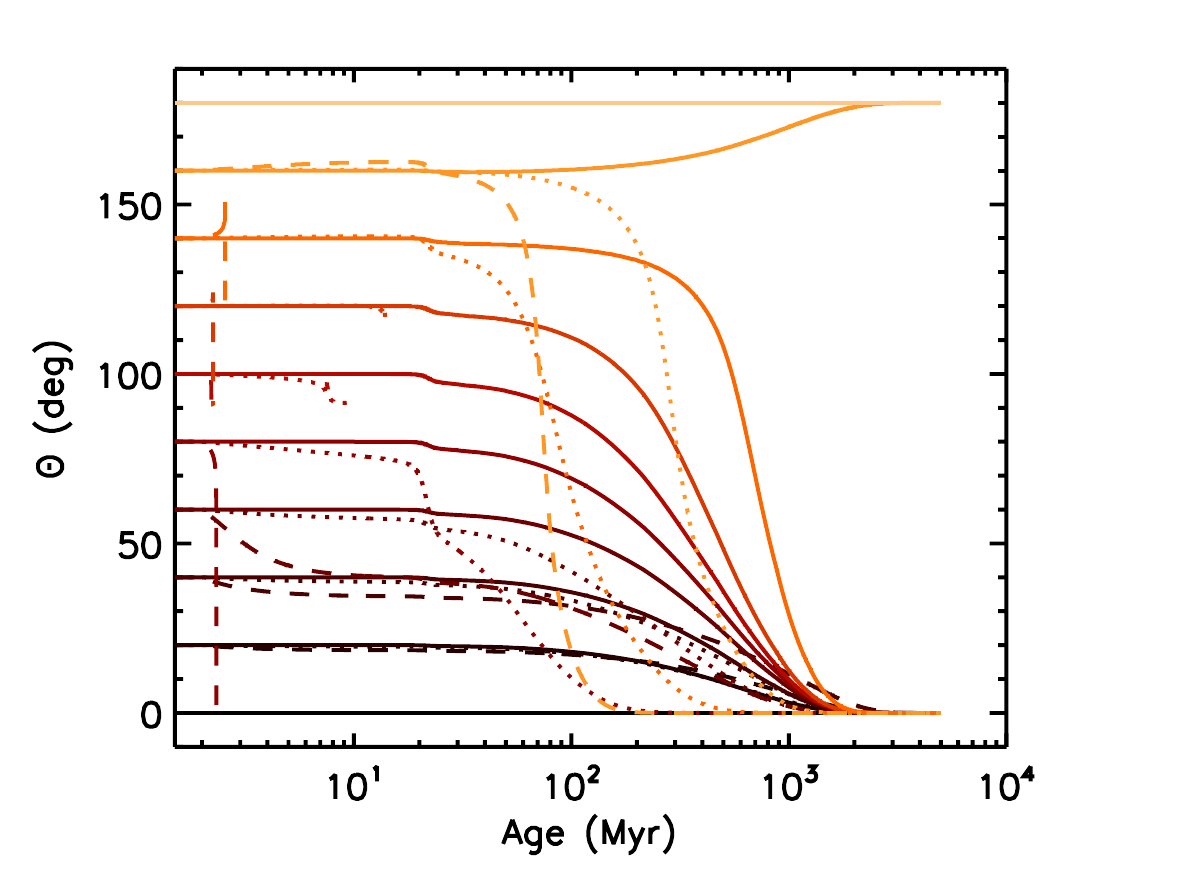}\\
	\includegraphics[width=0.94\columnwidth]{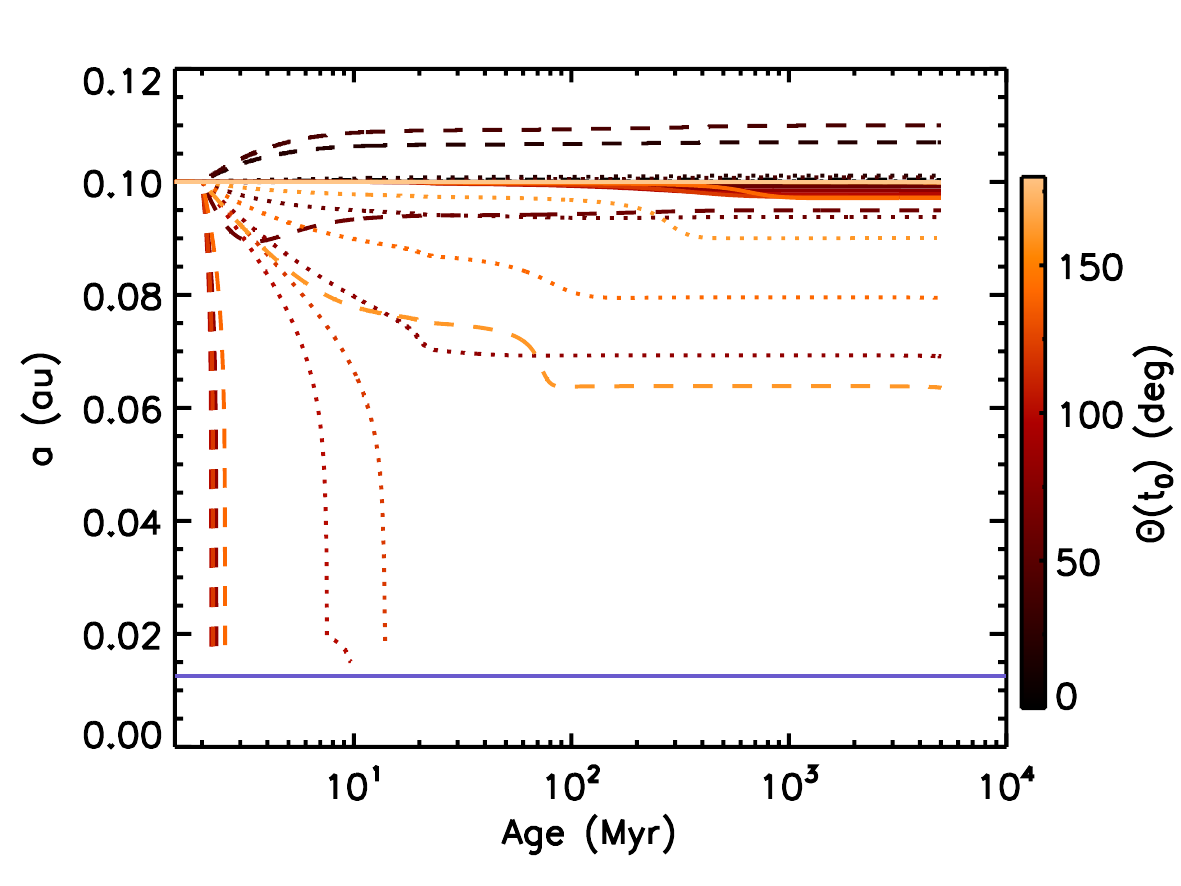}\\
	\includegraphics[width=0.94\columnwidth]{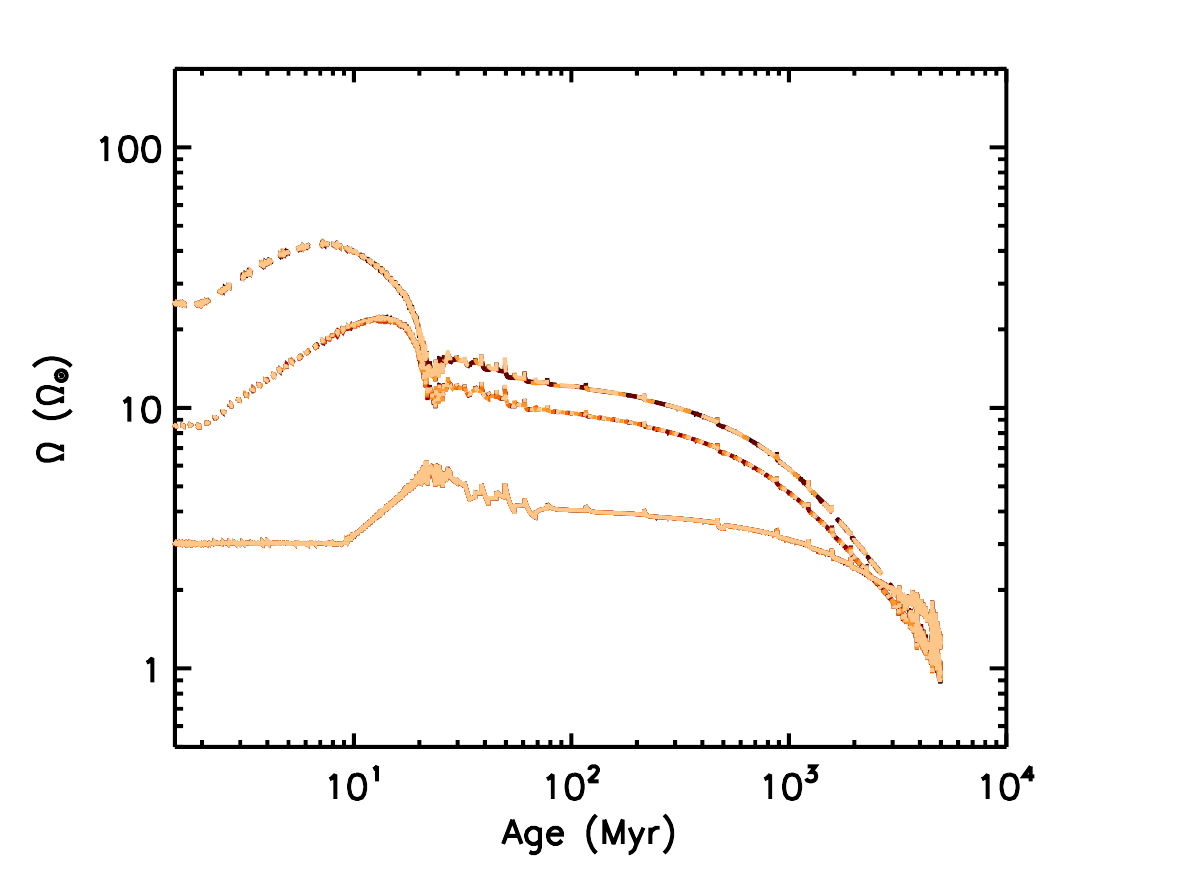}	
	\caption{Same as Fig.~\ref{fig:orbitevolout100} but for a 1.2~M$_\odot$ host. The line colors ans styles have the same meaning.}
	\label{fig:orbitevolout120}
\end{figure}

Let us first consider the case of a solar-mass star orbited by a Jupiter-mass planet at $0.1$ AU at the start of the evolution (Fig.~\ref{fig:orbitevolout100}). On the top panel, we see that the evolution of the obliquity depends mainly on the initial rotation frequency of the star, whatever the obliquity's initial value. This is expected in our model where the efficiency of tidal dissipation is directly related to the rotation frequency of the star, knowing that in this orbital configuration, inertial waves can be excited throughout the first Gyr of evolution, as can be seen on Fig.~\ref{fig:klmevolout100}. 

For initially slow rotators (solid lines), tidal dissipation is not efficient and this result in very little evolution of both the semi-major axis and the obliquity over the lifetime of the system. For initially moderately rotating stars (dotted-lines), tidal dissipation becomes significant and the value of $\sin \Theta$ has an important contribution to typical timescale of evolution of the semi-major axis. For both quasi-aligned prograde and retrograde system, $\sin \Theta \approx 1$ and both the obliquity and the semi-major axis do not change significantly in time. But for $80^\circ \lesssim\Theta \lesssim 120^\circ$,  $\sin \Theta << 1$, the semi-major axis decreases very rapidly and the planet reaches the Roche Lobe during the first 10 Myrs of evolution. However in this case, tidal evolution does not change significantly the value of the obliquity before the final engulfment. 

For initially fast rotators (dashed lines) the prograde and retrograde orbits do not behave in a mirrored way. For systems starting almost aligned ($\Theta < 60^\circ$), the typical timescale for evolution of the semi-major axis and the obliquity appear to be similar, and faster than for the initially moderately rotating star. Since at the beginning of the evolution, $\Omega_0 << \cos \Theta \Omega_{\rm e}$, the tides tend to push the planet on a wider orbit, and the obliquity is damped as the start contracts and $\Omega_{\rm e}$ increases. When  $\Omega_{\rm e}$ reaches its maximum, the star enters its phase of spin down in the saturated regime, and tidal dissipation becomes less efficient. Essentially, both the semi-major axis and the obliquity stop evolving and they keep the value they reached after the first $\sim 40$ Myrs of evolution. The almost polar orbit starting with $\Theta=80^\circ$ has a somewhat different behaviour starting with a faster evolution of the semi-major axis which, in this case, decreases in time, because $\cos \Theta \approx 0$ and thus $\Omega_0 >> \cos \Theta \Omega_{\rm e}$. When the star starts spinning down, the tidal regime changes and the evolution of the obliquity becomes faster than that of the semi-major axis. This changes again when the frequency ratio between the star and the orbit does not allow the excitation of the dynamical tide any more and tidal evolution resumes, damping the obliquity but with little effect on the semi-major axis. 

On the other hand, systems with fast rotators starting retrograde with $\Theta > 100^\circ$ are much more unstable and the planet is immediately engulfed as negative angular momentum is transferred (from the orbit to the star because $|\Omega_0| << |\cos \Theta \Omega_{\rm e}|$) into the axial component of the orbital angular momentum through an increasing value of $\Theta$ because the system is retrograde.

The evolution starting with $\Theta=160^\circ$ is a remarkable example. As the obliquity reaches perfect retrograde alignement ($\Theta=180^\circ$) the $m=0$ and $m=1$ components of the tides cease to dissipate energy, and the rate of evolution of the semi-major axis reduces dramatically. The system is now capable of maintaining a retrograde very close-in orbit for about two Gyrs, the star continuously being spun down to the point of reaching sub-solar rotation by the end of the evolution. Within the space of initial conditions considered here, this is the only track for which tidal evolution affect significantly the spin of the star on secular timescales.

Finally Fig.~\ref{fig:orbitevolout120} shows similar evolutionary track but this time for a 1.2~M$_\odot$ star. Qualitatively, we find again the same respective differences between the outcome of tidal evolution of initially slow, moderate and fast rotators. However, the star is now rotating faster throughout its main sequence, and consequently maintains a strong dissipation for most of its lifetime. This results in a faster evolution of the obliquity even around slowly rotating stars. In this set of simulations, all systems reach values of obliquity inferior to $10^\circ$ after the first Gyr of evolution, except some initially retrograde systems that settle on well-aligned retrograde orbits.

\begin{figure}
	\centering
	\includegraphics[width=0.93\columnwidth]{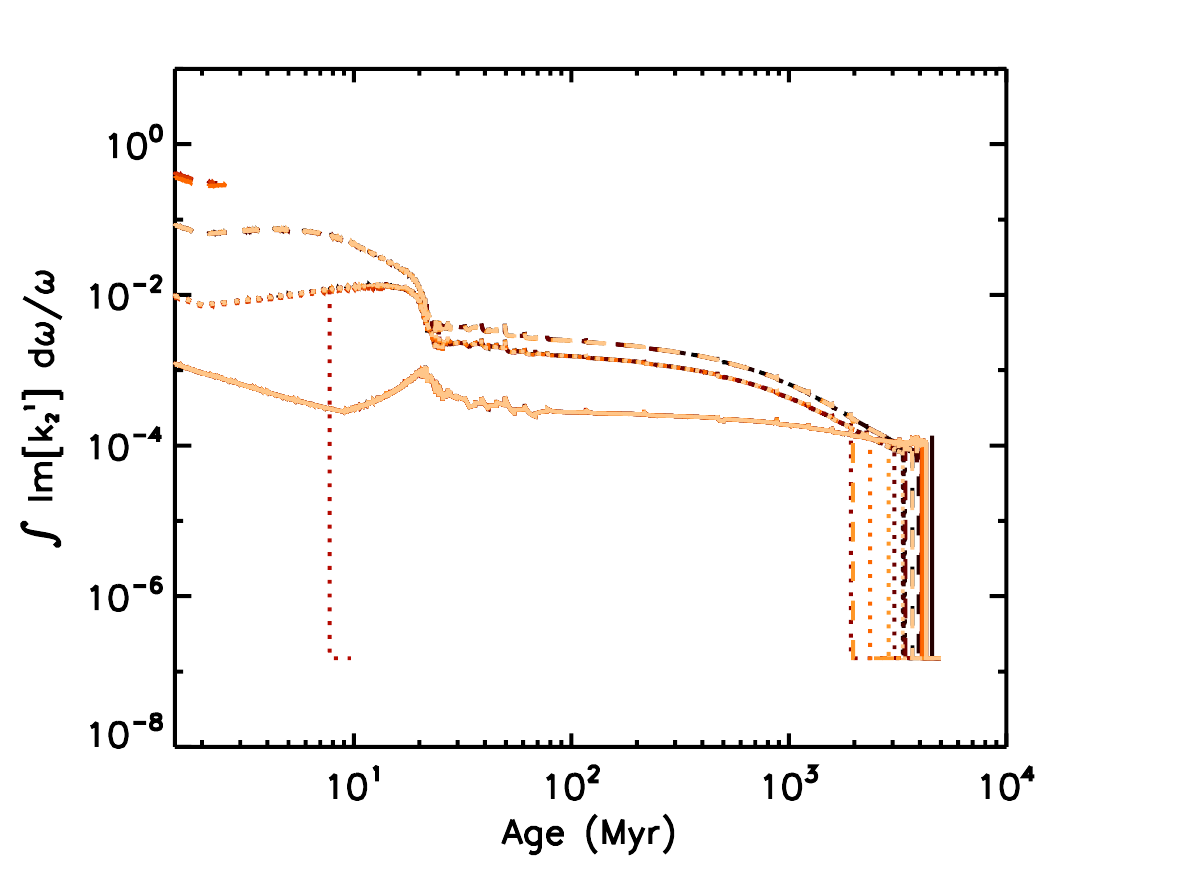}\\
	\includegraphics[width=0.93\columnwidth]{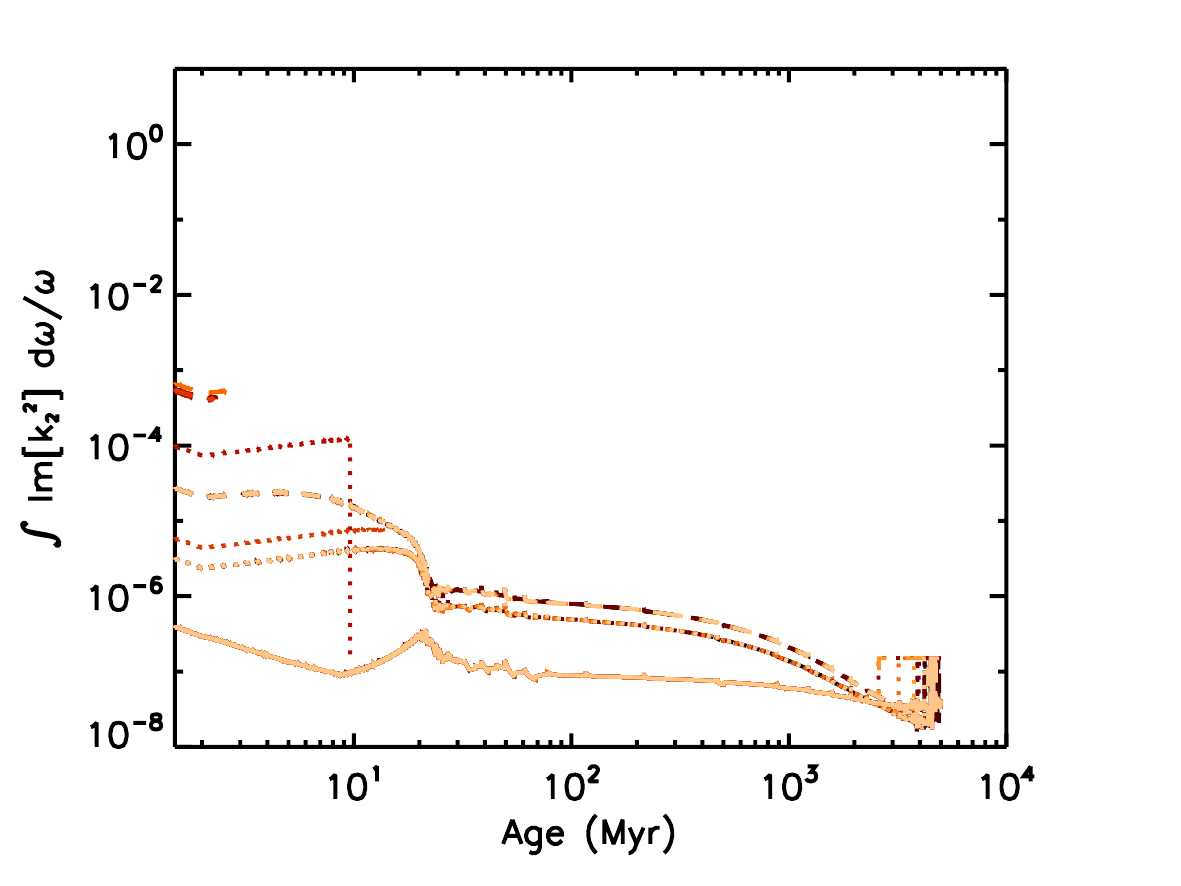}	
	\caption{Same as Fig.\ref{fig:klmevolout100} but for a 1.2~M$_\odot$ mass host. The line-styles and colors correspond match those of the tracks shown in Fig.~\ref{fig:orbitevolout120}. The dissipation starts with the value expected from the dissipation of inertial waves, and assumes the equilibrium tide constant Q' value only after the first Gyr for most orbital configurations.}
	\label{fig:klmevolout120}
\end{figure}

\begin{figure*}
	\centering
	\includegraphics[width=0.47\linewidth]{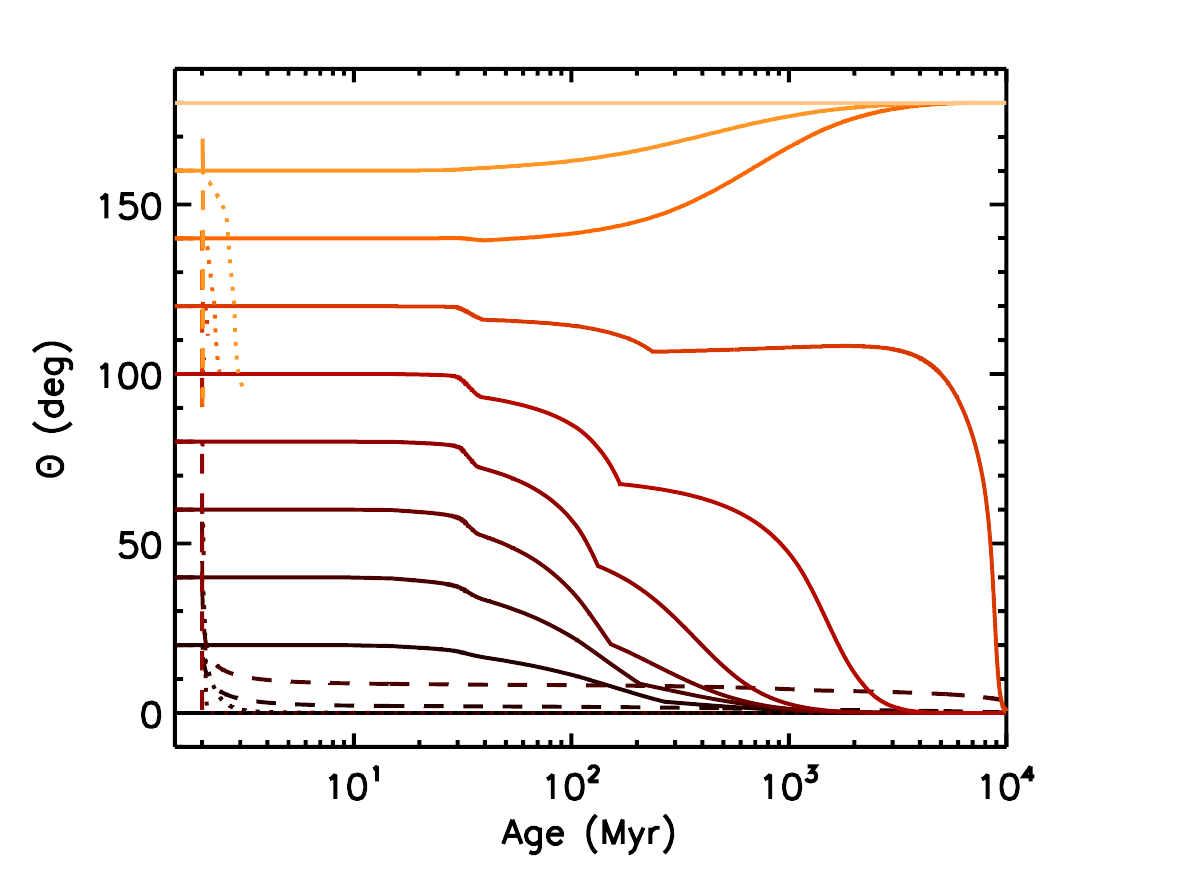}
	\quad\includegraphics[width=0.47\linewidth]{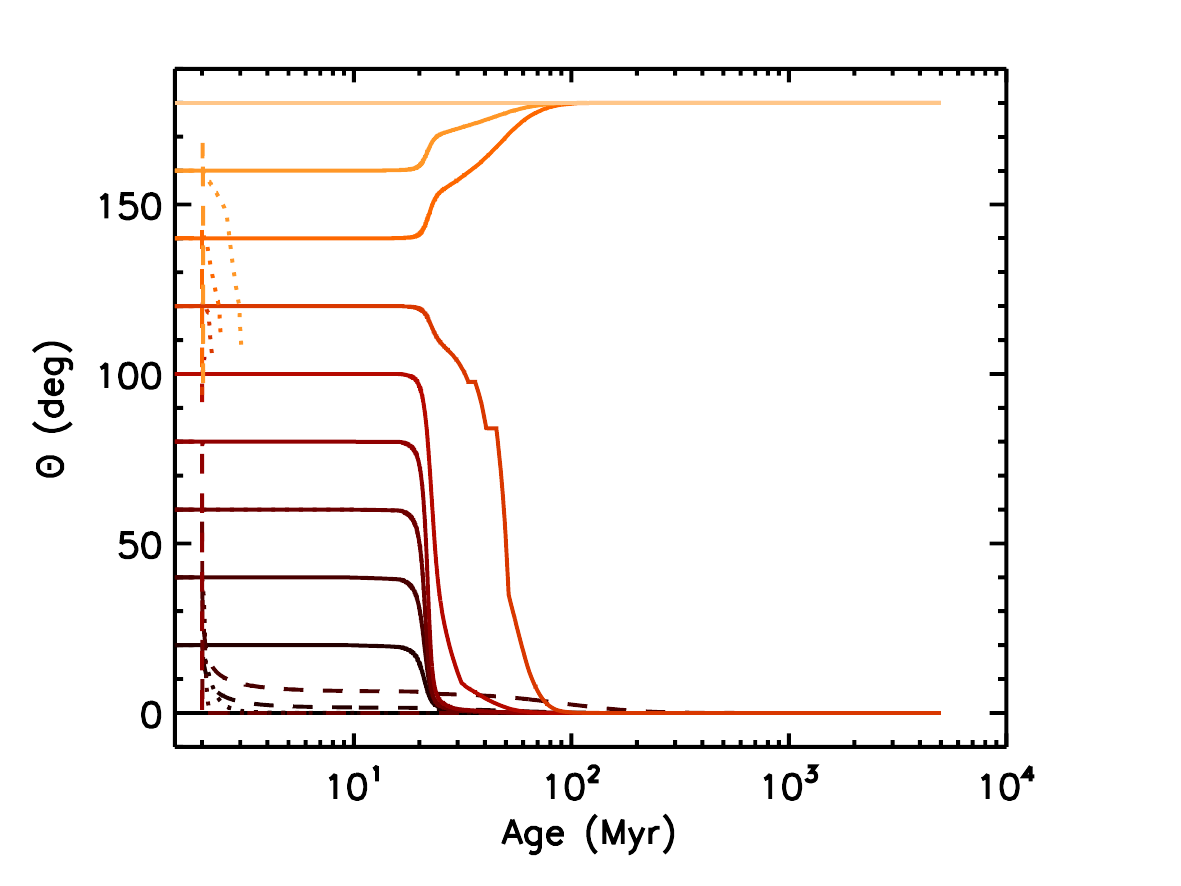}\\
	\includegraphics[width=0.47\linewidth]{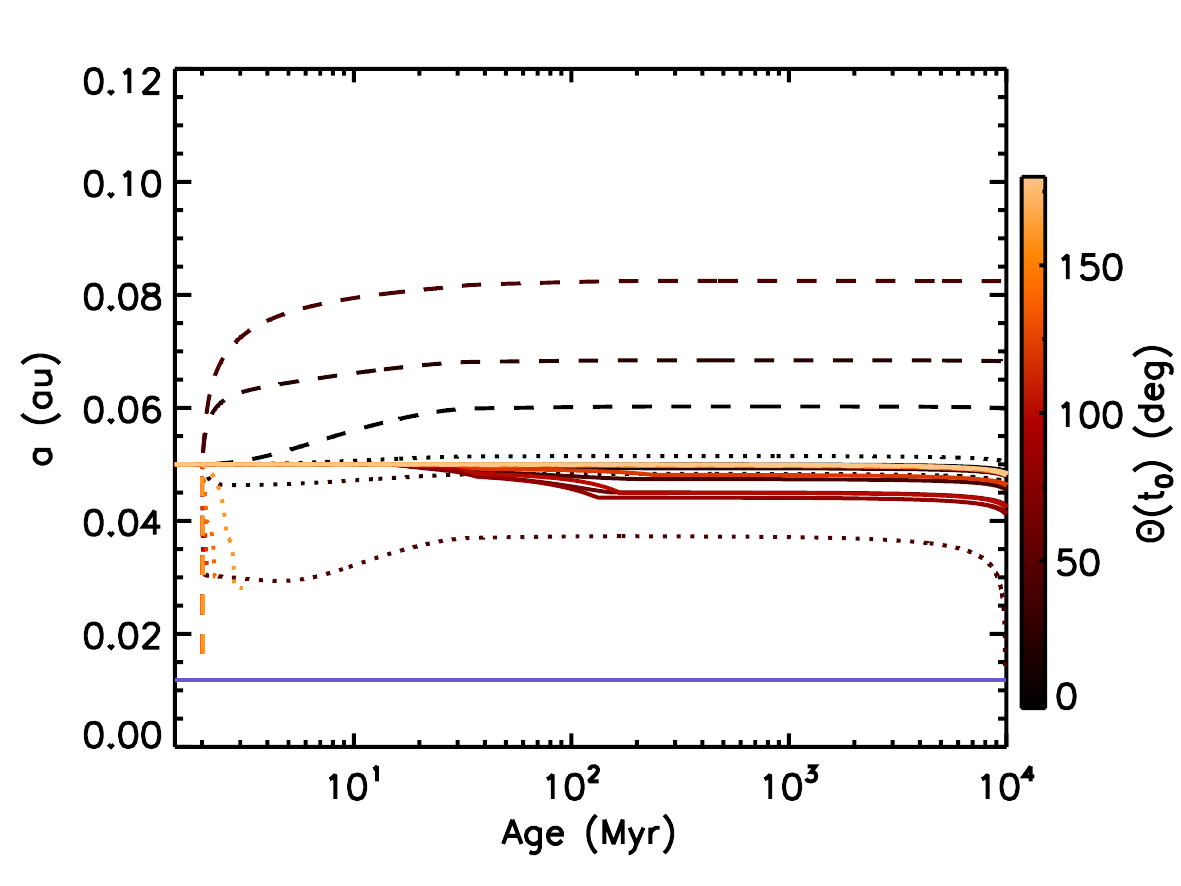}
	\quad\includegraphics[width=0.47\linewidth]{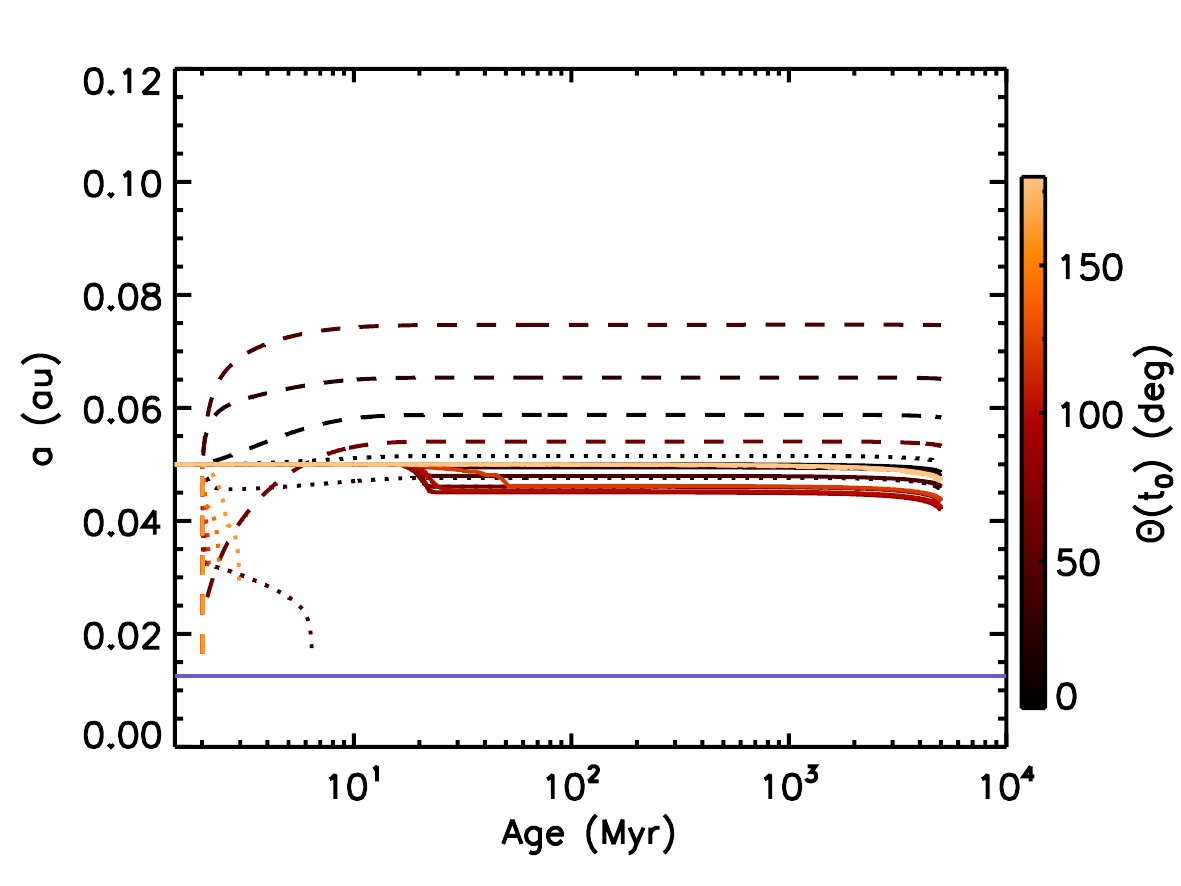}\\
	\includegraphics[width=0.47\linewidth]{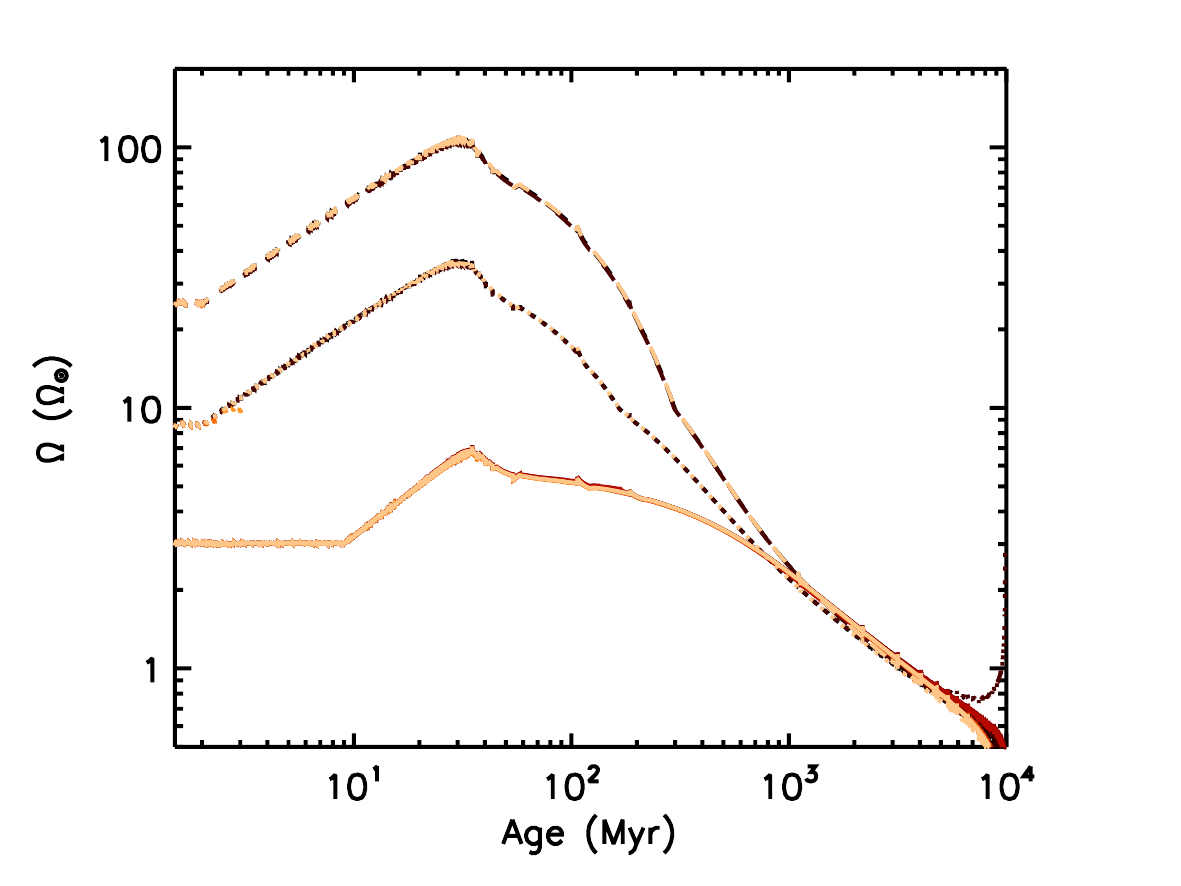}
	\quad\includegraphics[width=0.47\linewidth]{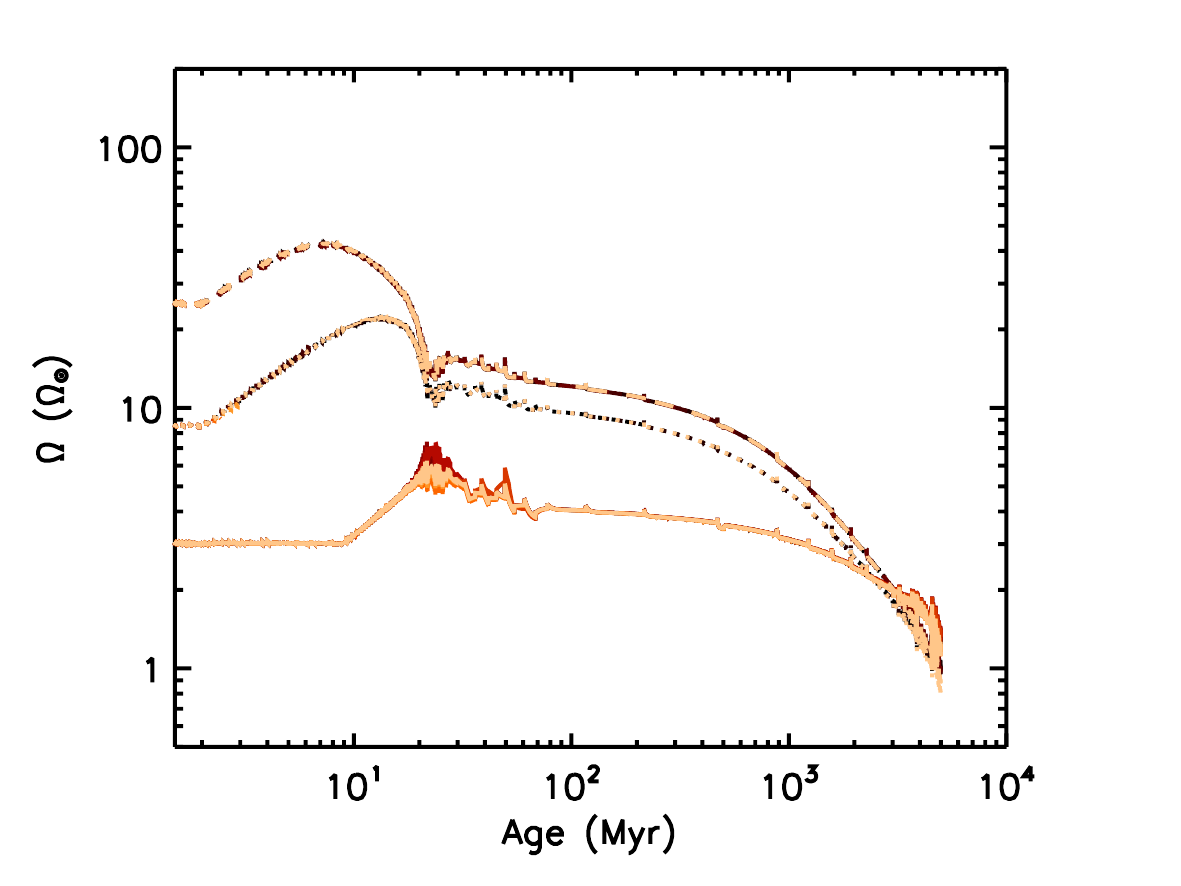}	
	\caption{Same as Fig.\ref{fig:orbitevolout100} (left) for a 1~M$_\odot$ mass host and Fig.\ref{fig:orbitevolout120} (right) for a 1.2~M$_\odot$ mass host, but with a planet orbiting at $a(t_0)=0.05$ AU when the evolution starts.}
	\label{fig:orbitevolin}
\end{figure*}

\begin{figure*}
	\centering
	\includegraphics[width=0.47\linewidth]{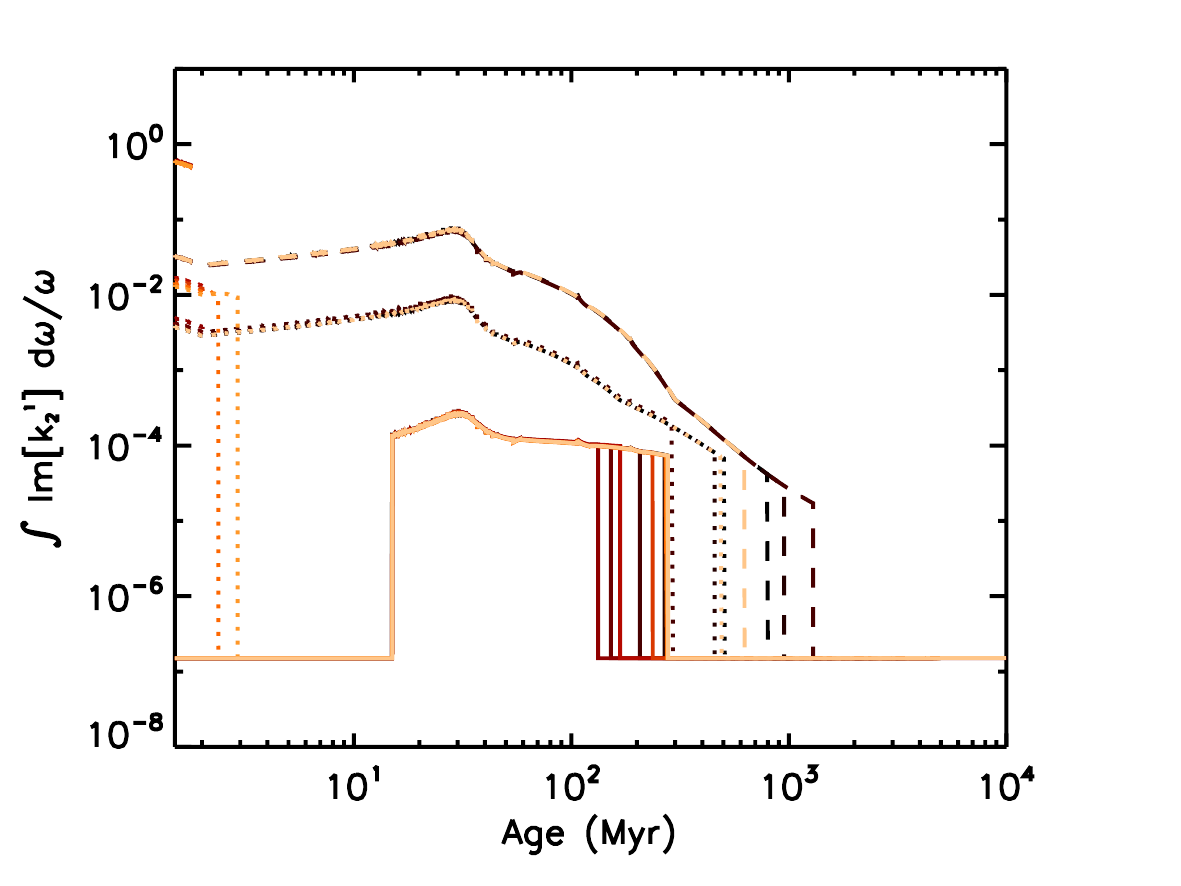}\quad \includegraphics[width=0.47\linewidth]{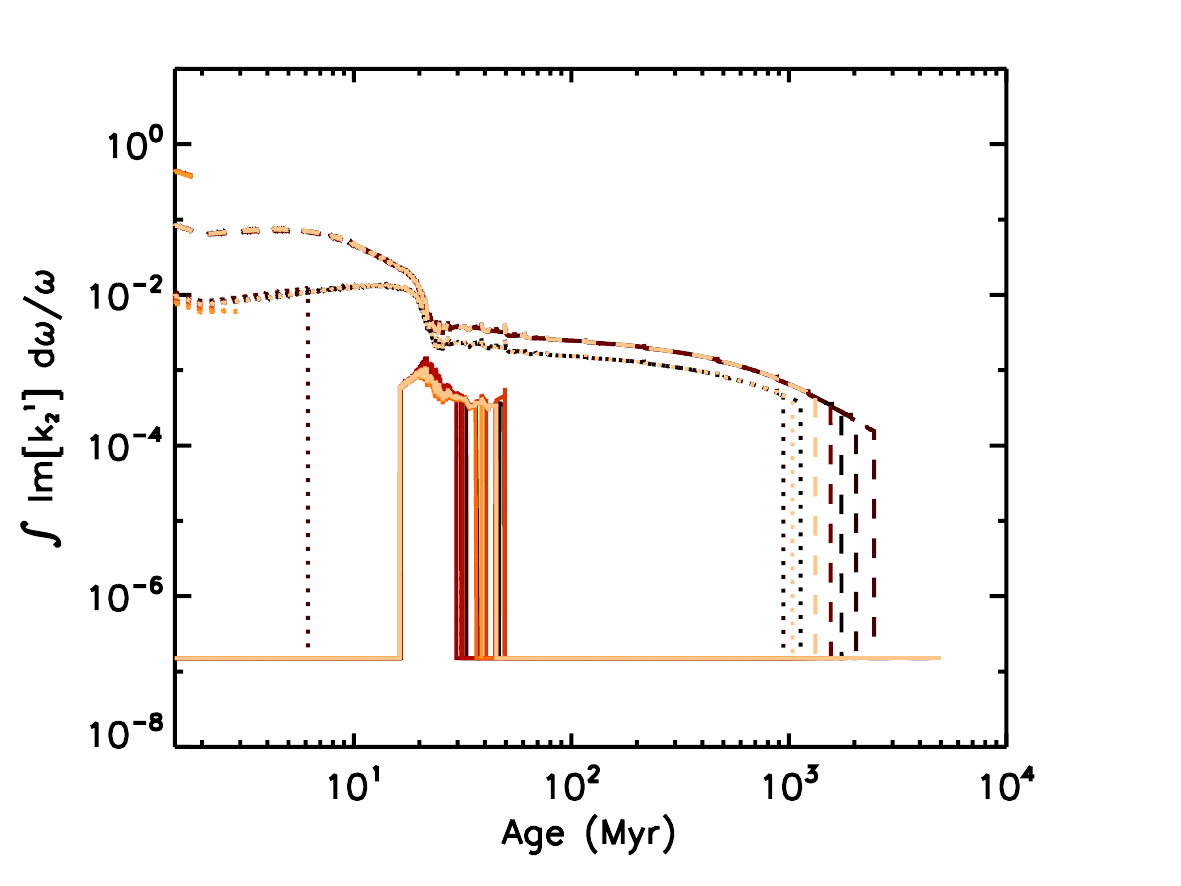}\\
	\includegraphics[width=0.47\linewidth]{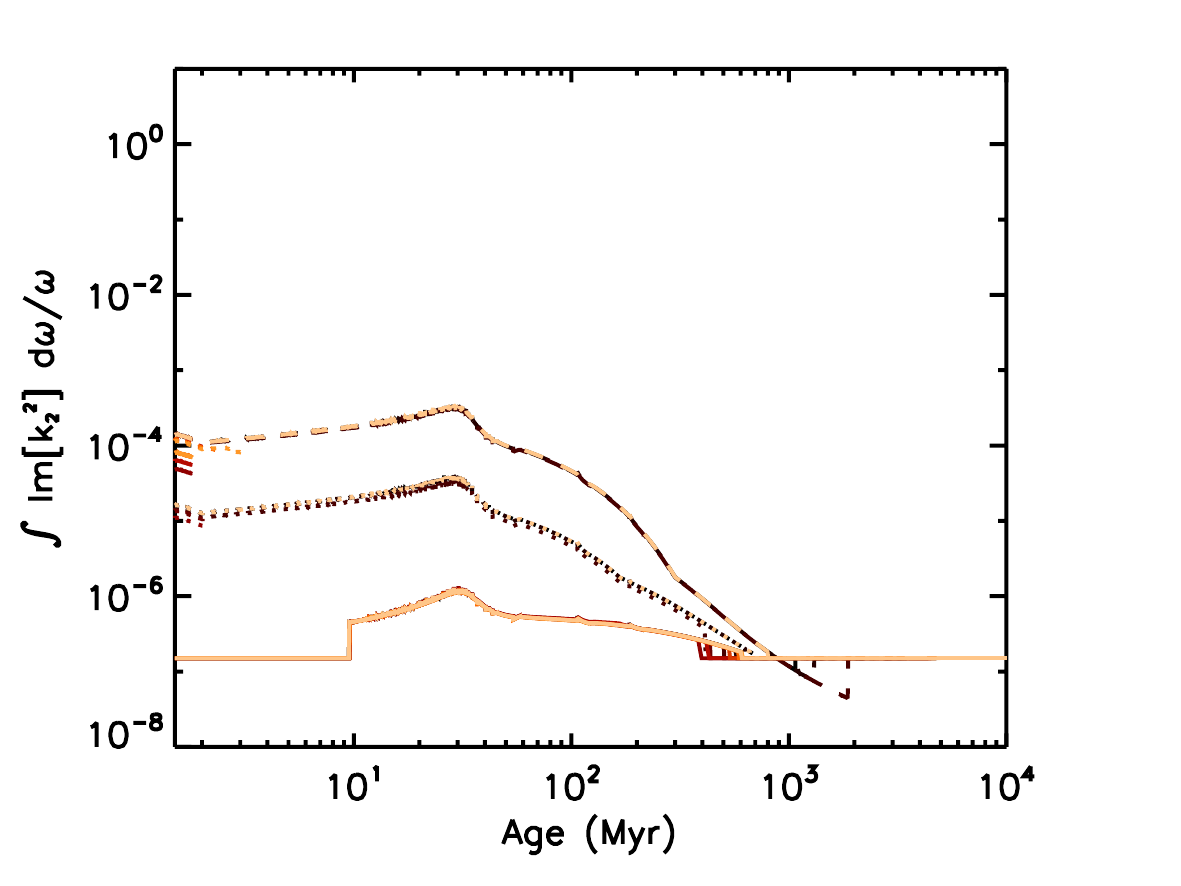}\quad \includegraphics[width=0.47\linewidth]{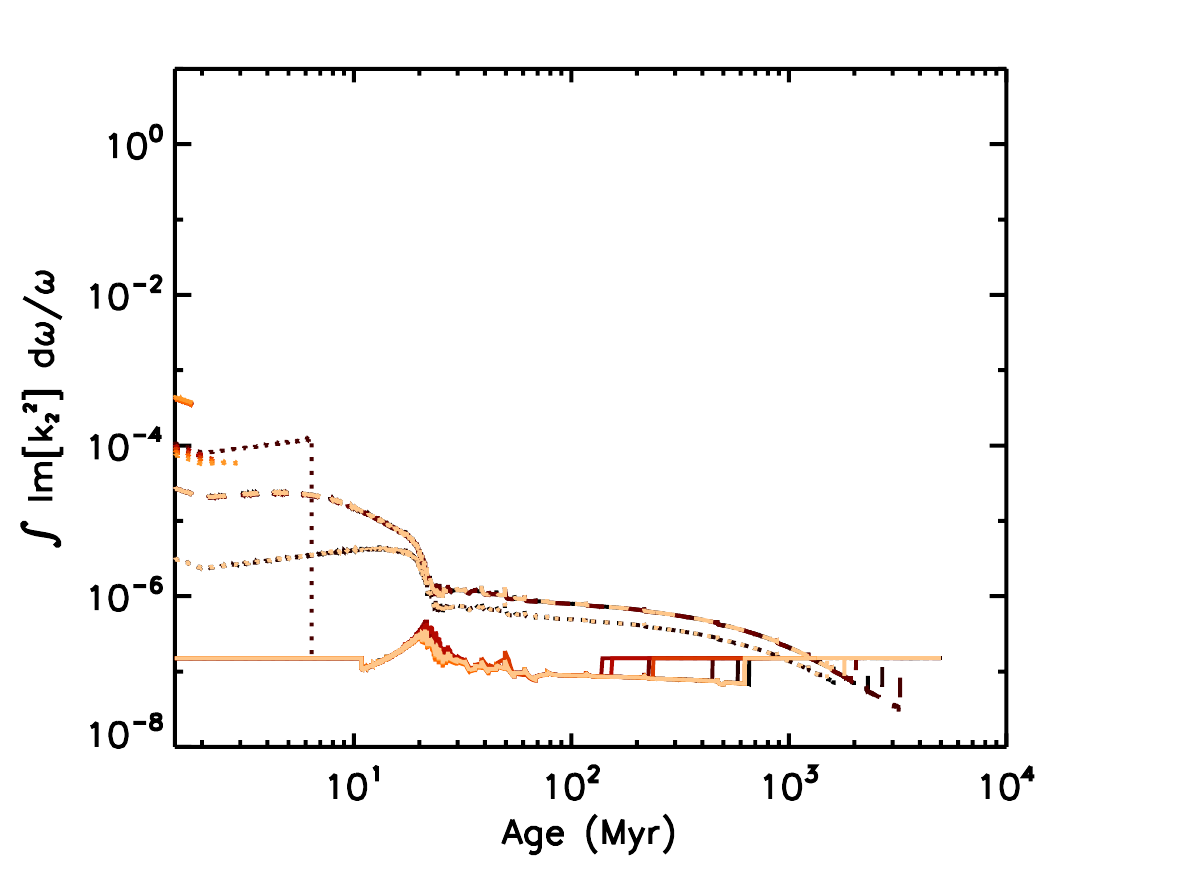}	
	\caption{ Same as Fig.~\ref{fig:klmevolout100} (left) and Fig.~\ref{fig:klmevolout120} (right), but for a 1M$_{\rm Jup}$ planet initially starting with  $a(t_0) = 0.05$ AU.}
	\label{fig:klmevolin}
\end{figure*}

The same general trends are observed if the planet is starting its evolution on a shorter orbit. Fig.\ref{fig:orbitevolin} show the same kind of simulations but with now a Jupiter-mass planet starting on a $0.05$ AU orbit, close to where hot-Jupiters are observed. Now in these orbital configurations, inertial waves can be only be excited episodically during the evolution, especially for initially slow rotators, as can be seen on Fig.~\ref{fig:klmevolin}. However, there is little qualitative difference compared to the previous case. Typical timescales of evolution are faster, but the obliquity tend to evolve faster than the semi-major axis, in all cases. However, the 1.2~M$_\odot$ star remains more efficient at damping any initial non-null obliquity. This suggests that tidal dissipation in the convective zone alone is not enough to interpret the observed correlation between effective temperature and obliquity. Clearly others physical mechanisms are at play.

\section{Conclusions}\label{sec:discus}
We have implemented for the first time a model of tidal evolution for the obliquity in circular exoplanetary systems that uses an ab-initio estimation of tidal dissipation based on the physical properties of the host star, and tied to a known mechanism, the excitation and damping of inertial waves in convection zones. Even though we don't apply a full frequency-dependent treatment of the dynamical tide, our formulation retains some dependence on tidal frequency, through the use of different tidal dissipation efficiency for different components of the tidal potential. Moreover, we also condition the value of dissipation on the range of tidal frequencies where inertial waves are excited \citep{Ogilvie2013, Mathis2015}. We use here a simplified piecewise homogeneous fluid body model. This may enhance the obtained value of the dissipation compared to a case where the realistic density profile is taken into account \citep{Ogilvie2013}. Moreover, the actual dissipation at a given frequency could differ from the frequency-averaged dissipation by several orders of magnitude \citep[e.g.][]{Ogilvie2004,Ogilvie2007}, so that we cannot exclude that real systems behave differently from this simplified model \citep[e.g.][]{WS2002,ADLPM2014}.

We find that the frequency-averaged tidal dissipation is very sensitive on the angular velocity of the convective envelope, as it scales as $\Omega_{\rm e}^2$ for all modes. We also find that for the quadrupole moment of the tidal potential, the dissipation due to the sectoral harmonic ($l = 2, m = 2$) is about one to two orders of magnitude less efficient than the dissipation due to both the zonal ($l = 2, m=0$) and the tesseral ($l = 2, m = 1$) harmonic, for both models of 1~M$_\odot$ and 1.2~M$_\odot$ mass stars. We predict that the dissipation in the 1.2~M$_\odot$ mass star is not significantly reduced compared to the one of the 1~M$_\odot$ mass star for all modes considered here, despite a smaller convective zone. This is due to the fact that the evolution of tidal dissipation in convective envelopes of low-mass stars is mainly driven by their rotational evolution on the main-sequence \citep{Gallet2017}. More massive stars remain fast rotators until the end of the main sequence, as a consequence of a less efficient magnetic braking. This explains why their dissipation is comparable to the one of less massive stars.

In this framework, typical timescales of evolution for the orbital elements cannot be simply formulated in terms of global parameters. However our simulations show that, for typical orbital configurations of hot-Jupiters, the damping timescale of the obliquity is generally shorter than that of the semi-major axis, except for the case of rapidly rotating stars with planets on very inclined orbits. This is a formal proof of concept of the conjecture put forward by \citet{Lai2012}, where the dynamical tide does indeed lead to different speed of evolution for the semi-major axis and the obliquity.

Nevertheless, our simulations fail to produce slower tidal evolution for more massive stars, in the range of mass corresponding to low-mass stars developing a radiative core and a convective envelope. The obliquity and/or the semi-major axis are damped before the star reaches the main-sequence for both stellar masses considered here, when the initial rotation period is shorter than $\sim 3$ days (i.e. $\Omega_{\rm e} \gtrsim 8.5 \Omega_{\odot}$) and the initial semi-major axis is smaller than $\sim 0.05$ AU (i.e. $P\sim 4$ days). For planets starting with $\Theta \gtrsim 150^\circ$, tidal dissipation can bring the planet on a meta-stable retrograde aligned orbit ($\Theta =180^\circ$), as found in other studies \citep{Lai2012,Valsecchi2014}. For initially slow rotators, an initial misalignment can be maintained throughout the life of the system, but in this case, the obliquity of planets orbiting more massive host is consistently damped on shorter time-scales.

This would thus fail to explain the observed correlation between effective temperature and obliquity. However, many factors are at play here, and our simulations show that the final outcome of tidal evolution is very sensitive to the initial conditions. In addition, the few first Myr of evolution are determinant for the final outcome. This is also where our model is more affected by the parameters of the magnetic braking law, such as the duration of the disc-locking phase, the coupling timescale for exchange of angular momentum between the radiative core and the convective envelope, or the saturation threshold.

One of the most severe limitations of magnetic braking laws using the double-zone model is the paucity of calibration points. To this day, there is only a handful of main sequence Sun-like stars for which constraints on internal differential rotation and age can be obtained using asteroseismology \citep{Nielsen2017,Benomar2015}. Precise measurements of internal differential rotation are only available for the Sun \citep{Thompson2003,Garcia2007,Fossat2017,Gehanetal2018}, where it is measured to be weak through most of the interior, except maybe in the innermost region of the core. There is also a general lack of stars of known age, rotation period and mass for masses smaller than 1 M$_\odot$ and ages greater than 600 Myr \citep{Gallet2015}.

Clearly those parameters play a role as important as the actual mechanism responsible for tidal dissipation in the global outcome of tidal evolution. A better understanding of the fundamental underlying mechanisms is required to be able to retrace the tidal evolution of hot-Jupiter, and eventually understand their migration and formation.

But even with a perfect understanding of magnetic braking, and internal angular momentum transport \citep[we refer the reader to e.g.][for last advances]{Maeder2009,MathisTransport2013,Revilleetal2016}, this study would just be a first step in the theoretical effort to produce tidal models accurate enough to constrain migration scenarios. Indeed, in all scenarios of migration that do not involve interactions with the protoplanetary disk, the eccentricity of the proto-hot Jupiter must be excited to high values to allow its subsequent circularisation on a close-in orbit \citep{Dawson2018}. Our model here is not suitable for the treatment of eccentricity. In the classical equilibrium tide, it is easy to show that the circularisation of the orbit in a star-planet system is mainly produced by the dissipation of tides raised in the planet. Here we neglect those tides, so that our model would need to include them to also account for the evolution of the eccentricity. Since we are neglecting the interactions of tidal bulges between them, the orbital evolution due to the tides raised and dissipated in the planet can simply be obtained by adding the torque resulting from the response of the planet, to the point-mass-like tidal potential exerted by the star. The efficiency of tidal dissipation on the planet could also use the frequency-averaged formulation of \citet{Ogilvie2013}, assuming that the planet could be considered as a homogeneous body with a solid core \citep{Guenel2014}. Because this formulation is very sensitive to the core size, this raises the question of the mass of the solid core in gas-giants, for which there is still no definite observational evidence, even in the case of our own Jupiter. Finally, in the case of an eccentric orbit, the dynamical tide assumes an infinite number of terms, and non-dominant harmonics can be important through resonant contributions. In this case, the frequency-averaged dissipation would be needed for arbitrary value of $l$ and $m$, and the analytical formulation we have made explicit in Appendix can be used. The introduction of the eccentricity in the model is thus the next step to improve models of tidal evolution and constrain migration scenarios. 

To get a complete picture, it would also be necessary to take into account the dissipation of tidal gravity waves \citep{Zahn1970} \citep[or gravito-inertial waves for those strongly modified by rotation, e.g.][]{Ogilvie2007} in the radiative cores of low-mass stars through linear thermal diffusion \citep{Zahn1975,GoodmanDickson1998,Terquem1998,Ivanovetal2013,Chernovetal2013} or non-linear breaking mechanisms \citep{Barker2010,Barker2011}. This dissipation can be of the same order of magnitude that the one of tidal inertial waves in the external convective envelope \citep[we refer for instance the reader to Fig. 15 in][]{Ivanovetal2013}. Therefore, tidal gravity (gravito-inertial) waves can also play a key role to shape the orbital dynamics of short-period systems and the rotational evolution of their components \citep[see e.g. the examples studied by][]{Guillotetal2014,Chernovetal2017,Weinbergetal2017}. As in the case of tidal inertial waves, the dissipation of tidal gravito-inertial waves in the radiative core of low-mass stars depends on their mass, their age, and their rotation. This dependence, combined with observational constraints on the orbital state of short-period systems, can also allow us to constrain the mass and the evolutionary state of the host star \citep[][]{Chernovetal2017,Weinbergetal2017}. Besides, the effects of differential rotation, both in convective and in radiative zones, should also be taken into account, as it is expected that the range of frequencies at which tidal (gravito-)inertial waves may propagate is broader than in the case of solid-body rotation \citep{Baruteau2013,Gueneletal2016a,Mirouhetal2016} while corotation resonances may enhance their dissipation \citep{GoldreichNicholson1989,Gueneletal2016b}. Finally, one should also consider the effects of the elliptic instability \citep[e.g.][]{Cebronetal2013,Lebarsetal2015,BarkerObl2016} and of magnetic fields on tidal inertial and gravito-inertial waves \citep[][]{Wei2016,LinOgilvie2018,Wei2018}.

Efforts to confront observations of hot-Jupiters and migration theories require also better observational constraints. In a very near future, these studies will benefit from the observations of the ground-based near-IR SPIRou instrument \citep{Moutouetal2015} and of the space missions TESS \citep{Rickeretal2015} and CHEOPS \citep{Broegetal2013}. Currently, one of the most stringent limitation to test migration theories is the poor constrain we have on the age of most known hot-Jupiter systems. Besides, measuring the internal rotation profile of stars, possibly down to the core \citep[][]{Garcia2007,Becketal2012,Mosseretal2012,Deheuvelsetal2014,Deheuvelsetal2015,Benomar2015,Fossat2017,Gehanetal2018}, would help to better understand the mechanisms involved in the evolution of their angular momentum \citep[e.g.][and references therein]{Maeder2009,MathisTransport2013}. In this context, it is clear that a mission like PLATO \citep{Raueretal2014}, which will provide the complete characterisation of host stars using asteroseismology, will enable a breakthrough for dynamical studies, thanks to the precise and accurate determination of stellar radii and masses, differential rotation and, above all, ages.

\begin{acknowledgements}
The authors are grateful to the referee, Adrian Barker, for several valuable comments on the first version of the manuscript that helped improved the quality of the work. CD wishes to thank J.P. Marques, M. Janvier, T. Appourchaux, R. Cameron and E. Bolmont for helpful discussions. CD acknowledges funding from the ANR (Agence Nationale de la Recherche, France) program IDEE (ANR-12-BS05-0008) "Interaction Des \'Etoiles et des Exoplan\`etes", and from the German Space Agency (Deutsches Zentrum für Luftind Raumfahrt) under PLATO grant 50OO1501. SM acknowledges funding from European Research Council (ERC) under SPIRE grant (647383), and from CNES PLATO grant at CEA Saclay.
\end{acknowledgements}

 \bibliographystyle{aa} 
 \bibliography{DBZ} 

\begin{thebibliography}{94}
\expandafter\ifx\csname natexlab\endcsname\relax\def\natexlab#1{#1}\fi

\bibitem[{{Albrecht} {et~al.}(2012){Albrecht}, {Winn}, {Butler}, {Crane},
  {Shectman}, {Thompson}, {Hirano}, \& {Wittenmyer}}]{Albrecht2012}
{Albrecht}, S., {Winn}, J.~N., {Butler}, R.~P., {et~al.} 2012, \apj, 744, 189

\bibitem[{{Alexander}(1973)}]{Alexander1973}
{Alexander}, M.~E. 1973, \apss, 23, 459

\bibitem[{{Auclair-Desrotour} {et~al.}(2014){Auclair-Desrotour}, {Le
  Poncin-Lafitte}, \& {Mathis}}]{ADLPM2014}
{Auclair-Desrotour}, P., {Le Poncin-Lafitte}, C., \& {Mathis}, S. 2014, \aap,
  561, L7

\bibitem[{{Baraffe} {et~al.}(1998){Baraffe}, {Chabrier}, {Allard}, \&
  {Hauschildt}}]{Baraffe1998}
{Baraffe}, I., {Chabrier}, G., {Allard}, F., \& {Hauschildt}, P.~H. 1998, \aap,
  337, 403

\bibitem[{{Barker}(2011)}]{Barker2011}
{Barker}, A.~J. 2011, \mnras, 414, 1365

\bibitem[{{Barker}(2016)}]{BarkerObl2016}
{Barker}, A.~J. 2016, \mnras, 460, 2339

\bibitem[{Barker \& Ogilvie(2009)}]{Barker2009}
Barker, A.~J. \& Ogilvie, G.~I. 2009, \mnras, 395, 2268

\bibitem[{{Barker} \& {Ogilvie}(2010)}]{Barker2010}
{Barker}, A.~J. \& {Ogilvie}, G.~I. 2010, \mnras, 404, 1849

\bibitem[{{Baruteau} \& {Rieutord}(2013)}]{Baruteau2013}
{Baruteau}, C. \& {Rieutord}, M. 2013, Journal of Fluid Mechanics, 719, 47

\bibitem[{{Beck} {et~al.}(2012){Beck}, {Montalban}, {Kallinger}, {De Ridder},
  {Aerts}, {Garc{\'{\i}}a}, {Hekker}, {Dupret}, {Mosser}, {Eggenberger},
  {Stello}, {Elsworth}, {Frandsen}, {Carrier}, {Hillen}, {Gruberbauer},
  {Christensen-Dalsgaard}, {Miglio}, {Valentini}, {Bedding}, {Kjeldsen},
  {Girouard}, {Hall}, \& {Ibrahim}}]{Becketal2012}
{Beck}, P.~G., {Montalban}, J., {Kallinger}, T., {et~al.} 2012, \nat, 481, 55

\bibitem[{{Benomar} {et~al.}(2015){Benomar}, {Takata}, {Shibahashi},
  {Ceillier}, \& {Garc{\'{\i}}a}}]{Benomar2015}
{Benomar}, O., {Takata}, M., {Shibahashi}, H., {Ceillier}, T., \&
  {Garc{\'{\i}}a}, R.~A. 2015, \mnras, 452, 2654

\bibitem[{{Bolmont} {et~al.}(2017){Bolmont}, {Gallet}, {Mathis}, {Charbonnel},
  {Amard}, \& {Alibert}}]{Bolmont2017}
{Bolmont}, E., {Gallet}, F., {Mathis}, S., {et~al.} 2017, \aap, 604, A113

\bibitem[{{Bolmont} \& {Mathis}(2016)}]{Bolmont2016}
{Bolmont}, E. \& {Mathis}, S. 2016, Celestial Mechanics and Dynamical
  Astronomy, 126, 275

\bibitem[{{Bolmont} {et~al.}(2012){Bolmont}, {Raymond}, {Leconte}, \&
  {Matt}}]{Bolmont2012}
{Bolmont}, E., {Raymond}, S.~N., {Leconte}, J., \& {Matt}, S.~P. 2012, \aap,
  544, A124

\bibitem[{{Broeg} {et~al.}(2013){Broeg}, {Fortier}, {Ehrenreich}, {Alibert},
  {Baumjohann}, {Benz}, {Deleuil}, {Gillon}, {Ivanov}, {Liseau}, {Meyer},
  {Oloffson}, {Pagano}, {Piotto}, {Pollacco}, {Queloz}, {Ragazzoni}, {Renotte},
  {Steller}, \& {Thomas}}]{Broegetal2013}
{Broeg}, C., {Fortier}, A., {Ehrenreich}, D., {et~al.} 2013, in European
  Physical Journal Web of Conferences, Vol.~47, European Physical Journal Web
  of Conferences, 03005

\bibitem[{{C{\'e}bron} {et~al.}(2013){C{\'e}bron}, {Le Bars}, {Le Gal},
  {Moutou}, {Leconte}, \& {Sauret}}]{Cebronetal2013}
{C{\'e}bron}, D., {Le Bars}, M., {Le Gal}, P., {et~al.} 2013, \icarus, 226,
  1642

\bibitem[{{Charbonnel} {et~al.}(2013){Charbonnel}, {Decressin}, {Amard},
  {Palacios}, \& {Talon}}]{Charbonnel2013}
{Charbonnel}, C., {Decressin}, T., {Amard}, L., {Palacios}, A., \& {Talon}, S.
  2013, \aap, 554, A40

\bibitem[{{Chernov} {et~al.}(2017){Chernov}, {Ivanov}, \&
  {Papaloizou}}]{Chernovetal2017}
{Chernov}, S.~V., {Ivanov}, P.~B., \& {Papaloizou}, J.~C.~B. 2017, \mnras, 470,
  2054

\bibitem[{{Chernov} {et~al.}(2013){Chernov}, {Papaloizou}, \&
  {Ivanov}}]{Chernovetal2013}
{Chernov}, S.~V., {Papaloizou}, J.~C.~B., \& {Ivanov}, P.~B. 2013, \mnras, 434,
  1079

\bibitem[{{Cranmer} \& {Saar}(2011)}]{Cranmer2011}
{Cranmer}, S.~R. \& {Saar}, S.~H. 2011, \apj, 741, 54

\bibitem[{{Damiani} \& {Lanza}(2015)}]{Damiani2015}
{Damiani}, C. \& {Lanza}, A.~F. 2015, \aap, 574, A39

\bibitem[{{Dawson} \& {Johnson}(2018)}]{Dawson2018}
{Dawson}, R.~I. \& {Johnson}, J.~A. 2018, ArXiv e-prints
  [\eprint[arXiv]{1801.06117}]

\bibitem[{{Deheuvels} {et~al.}(2015){Deheuvels}, {Ballot}, {Beck}, {Mosser},
  {{\O}stensen}, {Garc{\'{\i}}a}, \& {Goupil}}]{Deheuvelsetal2015}
{Deheuvels}, S., {Ballot}, J., {Beck}, P.~G., {et~al.} 2015, \aap, 580, A96

\bibitem[{{Deheuvels} {et~al.}(2014){Deheuvels}, {Do{\u g}an}, {Goupil},
  {Appourchaux}, {Benomar}, {Bruntt}, {Campante}, {Casagrande}, {Ceillier},
  {Davies}, {De Cat}, {Fu}, {Garc{\'{\i}}a}, {Lobel}, {Mosser}, {Reese},
  {Regulo}, {Schou}, {Stahn}, {Thygesen}, {Yang}, {Chaplin},
  {Christensen-Dalsgaard}, {Eggenberger}, {Gizon}, {Mathis},
  {Molenda-{\.Z}akowicz}, \& {Pinsonneault}}]{Deheuvelsetal2014}
{Deheuvels}, S., {Do{\u g}an}, G., {Goupil}, M.~J., {et~al.} 2014, \aap, 564,
  A27

\bibitem[{{Efroimsky} \& {Makarov}(2013)}]{Efroimsky2013}
{Efroimsky}, M. \& {Makarov}, V.~V. 2013, \apj, 764, 26

\bibitem[{{Fabrycky} \& {Winn}(2009)}]{Fabrycky2009}
{Fabrycky}, D.~C. \& {Winn}, J.~N. 2009, \apj, 696, 1230

\bibitem[{{Fossat} {et~al.}(2017){Fossat}, {Boumier}, {Corbard}, {Provost},
  {Salabert}, {Schmider}, {Gabriel}, {Grec}, {Renaud}, {Robillot},
  {Roca-Cort{\'e}s}, {Turck-Chi{\`e}ze}, {Ulrich}, \& {Lazrek}}]{Fossat2017}
{Fossat}, E., {Boumier}, P., {Corbard}, T., {et~al.} 2017, \aap, 604, A40

\bibitem[{{Gallet} {et~al.}(2017){Gallet}, {Bolmont}, {Mathis}, {Charbonnel},
  \& {Amard}}]{Gallet2017}
{Gallet}, F., {Bolmont}, E., {Mathis}, S., {Charbonnel}, C., \& {Amard}, L.
  2017, \aap, 604, A112

\bibitem[{{Gallet} \& {Bouvier}(2013)}]{Gallet2013}
{Gallet}, F. \& {Bouvier}, J. 2013, \aap, 556, A36

\bibitem[{{Gallet} \& {Bouvier}(2015)}]{Gallet2015}
{Gallet}, F. \& {Bouvier}, J. 2015, \aap, 577, A98

\bibitem[{{Garc{\'{\i}}a} {et~al.}(2007){Garc{\'{\i}}a}, {Turck-Chi{\`e}ze},
  {Jim{\'e}nez-Reyes}, {Ballot}, {Pall{\'e}}, {Eff-Darwich}, {Mathur}, \&
  {Provost}}]{Garcia2007}
{Garc{\'{\i}}a}, R.~A., {Turck-Chi{\`e}ze}, S., {Jim{\'e}nez-Reyes}, S.~J.,
  {et~al.} 2007, Science, 316, 1591

\bibitem[{{Gehan} {et~al.}(2018){Gehan}, {Mosser}, {Michel}, {Samadi}, \&
  {Kallinger}}]{Gehanetal2018}
{Gehan}, C., {Mosser}, B., {Michel}, E., {Samadi}, R., \& {Kallinger}, T. 2018,
  ArXiv e-prints [\eprint[arXiv]{1802.04558}]

\bibitem[{{Goldreich} \& {Keeley}(1977)}]{Goldreich1977a}
{Goldreich}, P. \& {Keeley}, D.~A. 1977, \apj, 211, 934

\bibitem[{{Goldreich} \& {Nicholson}(1989)}]{GoldreichNicholson1989}
{Goldreich}, P. \& {Nicholson}, P.~D. 1989, \apj, 342, 1079

\bibitem[{{Goodman} \& {Dickson}(1998)}]{GoodmanDickson1998}
{Goodman}, J. \& {Dickson}, E.~S. 1998, \apj, 507, 938

\bibitem[{{Guenel} {et~al.}(2016{\natexlab{a}}){Guenel}, {Baruteau}, {Mathis},
  \& {Rieutord}}]{Gueneletal2016a}
{Guenel}, M., {Baruteau}, C., {Mathis}, S., \& {Rieutord}, M.
  2016{\natexlab{a}}, \aap, 589, A22

\bibitem[{{Guenel} {et~al.}(2016{\natexlab{b}}){Guenel}, {Mathis}, {Baruteau},
  \& {Rieutord}}]{Gueneletal2016b}
{Guenel}, M., {Mathis}, S., {Baruteau}, C., \& {Rieutord}, M.
  2016{\natexlab{b}}, ArXiv e-prints [\eprint[arXiv]{1612.05071}]

\bibitem[{{Guenel} {et~al.}(2014){Guenel}, {Mathis}, \& {Remus}}]{Guenel2014}
{Guenel}, M., {Mathis}, S., \& {Remus}, F. 2014, \aap, 566, L9

\bibitem[{{Guillot} {et~al.}(2014){Guillot}, {Lin}, {Morel}, {Havel}, \&
  {Parmentier}}]{Guillotetal2014}
{Guillot}, T., {Lin}, D.~N.~C., {Morel}, P., {Havel}, M., \& {Parmentier}, V.
  2014, in EAS Publications Series, Vol.~65, EAS Publications Series, 327--336

\bibitem[{{Hansen}(2010)}]{Hansen2010}
{Hansen}, B.~M.~S. 2010, \apj, 723, 285

\bibitem[{{Hansen}(2012)}]{Hansen2012}
{Hansen}, B.~M.~S. 2012, \apj, 757, 6

\bibitem[{{Ivanov} {et~al.}(2013){Ivanov}, {Papaloizou}, \&
  {Chernov}}]{Ivanovetal2013}
{Ivanov}, P.~B., {Papaloizou}, J.~C.~B., \& {Chernov}, S.~V. 2013, \mnras, 432,
  2339

\bibitem[{Kawaler(1988)}]{Kawaler1988}
Kawaler, S.~D. 1988, \apj, 333, 236

\bibitem[{{Kopal}(1959)}]{Kopal1959}
{Kopal}, Z. 1959, The international astrophysics series, Vol.~5, {Close binary
  systems} (Chapman {\&} Hall LTD.)

\bibitem[{{Lai}(2012)}]{Lai2012}
{Lai}, D. 2012, \mnras, 423, 486

\bibitem[{{Lainey}(2016)}]{Lainey2016}
{Lainey}, V. 2016, Celestial Mechanics and Dynamical Astronomy, 126, 145

\bibitem[{{Lanza} \& {Mathis}(2016)}]{Lanza2016}
{Lanza}, A.~F. \& {Mathis}, S. 2016, Celestial Mechanics and Dynamical
  Astronomy, 126, 249

\bibitem[{{Le Bars} {et~al.}(2015){Le Bars}, {C{\'e}bron}, \& {Le
  Gal}}]{Lebarsetal2015}
{Le Bars}, M., {C{\'e}bron}, D., \& {Le Gal}, P. 2015, Annual Review of Fluid
  Mechanics, 47, 163

\bibitem[{Leconte {et~al.}(2010)Leconte, Chabrier, Baraffe, \&
  Levrard}]{Leconte2010}
Leconte, J., Chabrier, G., Baraffe, I., \& Levrard, B. 2010, \aap, 516, A64

\bibitem[{{Lin} \& {Ogilvie}(2017)}]{LinOgilvie2017}
{Lin}, Y. \& {Ogilvie}, G.~I. 2017, \mnras, 468, 1387

\bibitem[{{Lin} \& {Ogilvie}(2018)}]{LinOgilvie2018}
{Lin}, Y. \& {Ogilvie}, G.~I. 2018, \mnras, 474, 1644

\bibitem[{{Love}(1911)}]{Love1911}
{Love}, A.~E.~H. 1911, {Some Problems of Geodynamics} (Cambridge University
  Press)

\bibitem[{{MacGregor} \& {Brenner}(1991)}]{MacGregor1991}
{MacGregor}, K.~B. \& {Brenner}, M. 1991, \apj, 376, 204

\bibitem[{{Maeder}(2009)}]{Maeder2009}
{Maeder}, A. 2009, {Physics, Formation and Evolution of Rotating Stars},
  Astronomy and Astrophysics Library (Springer Berlin Heidelberg)

\bibitem[{{Mathis}(2013)}]{MathisTransport2013}
{Mathis}, S. 2013, in Lecture Notes in Physics, Berlin Springer Verlag, Vol.
  865, Lecture Notes in Physics, Berlin Springer Verlag, ed. M.~{Goupil},
  K.~{Belkacem}, C.~{Neiner}, F.~{Ligni{\`e}res}, \& J.~J. {Green}, 23

\bibitem[{{Mathis}(2015{\natexlab{a}})}]{Mathis2015sf2a}
{Mathis}, S. 2015{\natexlab{a}}, in SF2A-2015: Proceedings of the Annual
  meeting of the French Society of Astronomy and Astrophysics. Eds.: F.
  Martins, S. Boissier, V. Buat, L. Cambr{\'e}sy, P. Petit, pp.401-405, ed.
  F.~{Martins}, S.~{Boissier}, V.~{Buat}, L.~{Cambr{\'e}sy}, \& P.~{Petit},
  401--405

\bibitem[{{Mathis}(2015{\natexlab{b}})}]{Mathis2015}
{Mathis}, S. 2015{\natexlab{b}}, \aap, 580, L3

\bibitem[{{Mathis} {et~al.}(2016){Mathis}, {Auclair-Desrotour}, {Guenel},
  {Gallet}, \& {Le Poncin-Lafitte}}]{Mathis2016}
{Mathis}, S., {Auclair-Desrotour}, P., {Guenel}, M., {Gallet}, F., \& {Le
  Poncin-Lafitte}, C. 2016, \aap, 592, A33

\bibitem[{{Mathis} \& {Le Poncin-Lafitte}(2009)}]{MLP2009}
{Mathis}, S. \& {Le Poncin-Lafitte}, C. 2009, \aap, 497, 889

\bibitem[{{Matt} {et~al.}(2015){Matt}, {Brun}, {Baraffe}, {Bouvier}, \&
  {Chabrier}}]{Matt2015}
{Matt}, S.~P., {Brun}, A.~S., {Baraffe}, I., {Bouvier}, J., \& {Chabrier}, G.
  2015, \apjl, 799, L23

\bibitem[{{McLaughlin}(1924)}]{McLaughlin1924}
{McLaughlin}, D.~B. 1924, \apj, 60

\bibitem[{{Mirouh} {et~al.}(2016){Mirouh}, {Baruteau}, {Rieutord}, \&
  {Ballot}}]{Mirouhetal2016}
{Mirouh}, G.~M., {Baruteau}, C., {Rieutord}, M., \& {Ballot}, J. 2016, Journal
  of Fluid Mechanics, 800, 213

\bibitem[{{Mosser} {et~al.}(2012){Mosser}, {Goupil}, {Belkacem}, {Marques},
  {Beck}, {Bloemen}, {De Ridder}, {Barban}, {Deheuvels}, {Elsworth}, {Hekker},
  {Kallinger}, {Ouazzani}, {Pinsonneault}, {Samadi}, {Stello}, {Garc{\'{\i}}a},
  {Klaus}, {Li}, {Mathur}, \& {Morris}}]{Mosseretal2012}
{Mosser}, B., {Goupil}, M.~J., {Belkacem}, K., {et~al.} 2012, \aap, 548, A10

\bibitem[{{Moutou} {et~al.}(2015){Moutou}, {Boisse}, {H{\'e}brard},
  {H{\'e}brard}, {Donati}, {Delfosse}, \& {Kouach}}]{Moutouetal2015}
{Moutou}, C., {Boisse}, I., {H{\'e}brard}, G., {et~al.} 2015, in SF2A-2015:
  Proceedings of the Annual meeting of the French Society of Astronomy and
  Astrophysics, ed. F.~{Martins}, S.~{Boissier}, V.~{Buat}, L.~{Cambr{\'e}sy},
  \& P.~{Petit}, 205--212

\bibitem[{{Murray} \& {Dermott}(1999)}]{Murray1999}
{Murray}, C.~D. \& {Dermott}, S.~F. 1999, {Solar system dynamics} (Cambridge
  University Press)

\bibitem[{{Nielsen} {et~al.}(2017){Nielsen}, {Schunker}, {Gizon}, {Schou}, \&
  {Ball}}]{Nielsen2017}
{Nielsen}, M.~B., {Schunker}, H., {Gizon}, L., {Schou}, J., \& {Ball}, W.~H.
  2017, \aap, 603, A6

\bibitem[{{Ogilvie}(2013)}]{Ogilvie2013}
{Ogilvie}, G.~I. 2013, \mnras, 429, 613

\bibitem[{{Ogilvie}(2014)}]{Ogilvie2014}
{Ogilvie}, G.~I. 2014, \araa, 52, 171

\bibitem[{{Ogilvie} \& {Lesur}(2012)}]{Ogilvie2012}
{Ogilvie}, G.~I. \& {Lesur}, G. 2012, \mnras, 422, 1975

\bibitem[{Ogilvie \& Lin(2004)}]{Ogilvie2004}
Ogilvie, G.~I. \& Lin, D. N.~C. 2004, \apj, 610, 477

\bibitem[{Ogilvie \& Lin(2007)}]{Ogilvie2007}
Ogilvie, G.~I. \& Lin, D. N.~C. 2007, \apj, 661, 1180

\bibitem[{{Penev} {et~al.}(2014){Penev}, {Zhang}, \& {Jackson}}]{Penev2014}
{Penev}, K., {Zhang}, M., \& {Jackson}, B. 2014, \pasp, 126, 553

\bibitem[{{Press} {et~al.}(1992){Press}, {Teukolsky}, {Vetterling}, \&
  {Flannery}}]{Press1992}
{Press}, W.~H., {Teukolsky}, S.~A., {Vetterling}, W.~T., \& {Flannery}, B.~P.
  1992, {Numerical recipes in FORTRAN. The art of scientific computing}, 2nd
  edn. (Cambridge University Press)

\bibitem[{{Rauer} {et~al.}(2014){Rauer}, {Catala}, {Aerts}, {Appourchaux},
  {Benz}, {Brandeker}, {Christensen-Dalsgaard}, {Deleuil}, {Gizon}, {Goupil},
  {G{\"u}del}, {Janot-Pacheco}, {Mas-Hesse}, {Pagano}, {Piotto}, {Pollacco},
  {Santos}, {Smith}, {Su{\'a}rez}, {Szab{\'o}}, {Udry}, {Adibekyan}, {Alibert},
  {Almenara}, {Amaro-Seoane}, {Eiff}, {Asplund}, {Antonello}, {Barnes},
  {Baudin}, {Belkacem}, {Bergemann}, {Bihain}, {Birch}, {Bonfils}, {Boisse},
  {Bonomo}, {Borsa}, {Brand{\~a}o}, {Brocato}, {Brun}, {Burleigh}, {Burston},
  {Cabrera}, {Cassisi}, {Chaplin}, {Charpinet}, {Chiappini}, {Church},
  {Csizmadia}, {Cunha}, {Damasso}, {Davies}, {Deeg}, {D{\'{\i}}az}, {Dreizler},
  {Dreyer}, {Eggenberger}, {Ehrenreich}, {Eigm{\"u}ller}, {Erikson}, {Farmer},
  {Feltzing}, {de Oliveira Fialho}, {Figueira}, {Forveille}, {Fridlund},
  {Garc{\'{\i}}a}, {Giommi}, {Giuffrida}, {Godolt}, {Gomes da Silva},
  {Granzer}, {Grenfell}, {Grotsch-Noels}, {G{\"u}nther}, {Haswell}, {Hatzes},
  {H{\'e}brard}, {Hekker}, {Helled}, {Heng}, {Jenkins}, {Johansen},
  {Khodachenko}, {Kislyakova}, {Kley}, {Kolb}, {Krivova}, {Kupka}, {Lammer},
  {Lanza}, {Lebreton}, {Magrin}, {Marcos-Arenal}, {Marrese}, {Marques},
  {Martins}, {Mathis}, {Mathur}, {Messina}, {Miglio}, {Montalban}, {Montalto},
  {Monteiro}, {Moradi}, {Moravveji}, {Mordasini}, {Morel}, {Mortier},
  {Nascimbeni}, {Nelson}, {Nielsen}, {Noack}, {Norton}, {Ofir}, {Oshagh},
  {Ouazzani}, {P{\'a}pics}, {Parro}, {Petit}, {Plez}, {Poretti}, {Quirrenbach},
  {Ragazzoni}, {Raimondo}, {Rainer}, {Reese}, {Redmer}, {Reffert},
  {Rojas-Ayala}, {Roxburgh}, {Salmon}, {Santerne}, {Schneider}, {Schou},
  {Schuh}, {Schunker}, {Silva-Valio}, {Silvotti}, {Skillen}, {Snellen}, {Sohl},
  {Sousa}, {Sozzetti}, {Stello}, {Strassmeier}, {{\v S}vanda}, {Szab{\'o}},
  {Tkachenko}, {Valencia}, {Van Grootel}, {Vauclair}, {Ventura}, {Wagner},
  {Walton}, {Weingrill}, {Werner}, {Wheatley}, \& {Zwintz}}]{Raueretal2014}
{Rauer}, H., {Catala}, C., {Aerts}, C., {et~al.} 2014, Experimental Astronomy,
  38, 249

\bibitem[{{Remus} {et~al.}(2012){Remus}, {Mathis}, \& {Zahn}}]{Remus2012}
{Remus}, F., {Mathis}, S., \& {Zahn}, J.-P. 2012, \aap, 544, A132

\bibitem[{Remus {et~al.}(2015)Remus, Mathis, Zahn, \& Lainey}]{Remus2015}
Remus, F., Mathis, S., Zahn, J.~P., \& Lainey, V. 2015, Astronomy {\&}
  Astrophysics, 573, A23

\bibitem[{{R{\'e}ville} {et~al.}(2016){R{\'e}ville}, {Folsom}, {Strugarek}, \&
  {Brun}}]{Revilleetal2016}
{R{\'e}ville}, V., {Folsom}, C.~P., {Strugarek}, A., \& {Brun}, A.~S. 2016,
  \apj, 832, 145

\bibitem[{{Ribas} {et~al.}(2014){Ribas}, {Mer{\'{\i}}n}, {Bouy}, \&
  {Maud}}]{Ribas2014}
{Ribas}, {\'A}., {Mer{\'{\i}}n}, B., {Bouy}, H., \& {Maud}, L.~T. 2014, \aap,
  561, A54

\bibitem[{{Ricker} {et~al.}(2015){Ricker}, {Winn}, {Vanderspek}, {Latham},
  {Bakos}, {Bean}, {Berta-Thompson}, {Brown}, {Buchhave}, {Butler}, {Butler},
  {Chaplin}, {Charbonneau}, {Christensen-Dalsgaard}, {Clampin}, {Deming},
  {Doty}, {De Lee}, {Dressing}, {Dunham}, {Endl}, {Fressin}, {Ge}, {Henning},
  {Holman}, {Howard}, {Ida}, {Jenkins}, {Jernigan}, {Johnson}, {Kaltenegger},
  {Kawai}, {Kjeldsen}, {Laughlin}, {Levine}, {Lin}, {Lissauer}, {MacQueen},
  {Marcy}, {McCullough}, {Morton}, {Narita}, {Paegert}, {Palle}, {Pepe},
  {Pepper}, {Quirrenbach}, {Rinehart}, {Sasselov}, {Sato}, {Seager},
  {Sozzetti}, {Stassun}, {Sullivan}, {Szentgyorgyi}, {Torres}, {Udry}, \&
  {Villasenor}}]{Rickeretal2015}
{Ricker}, G.~R., {Winn}, J.~N., {Vanderspek}, R., {et~al.} 2015, Journal of
  Astronomical Telescopes, Instruments, and Systems, 1, 014003

\bibitem[{{Rossiter}(1924)}]{Rossiter1924}
{Rossiter}, R.~A. 1924, \apj, 60

\bibitem[{{Spada} {et~al.}(2011){Spada}, {Lanzafame}, {Lanza}, {Messina}, \&
  {Collier Cameron}}]{Spada2011}
{Spada}, F., {Lanzafame}, A.~C., {Lanza}, A.~F., {Messina}, S., \& {Collier
  Cameron}, A. 2011, \mnras, 416, 447

\bibitem[{{Terquem} {et~al.}(1998){Terquem}, {Papaloizou}, {Nelson}, \&
  {Lin}}]{Terquem1998}
{Terquem}, C., {Papaloizou}, J.~C.~B., {Nelson}, R.~P., \& {Lin}, D.~N.~C.
  1998, \apj, 502, 788

\bibitem[{{Thompson} {et~al.}(2003){Thompson}, {Christensen-Dalsgaard},
  {Miesch}, \& {Toomre}}]{Thompson2003}
{Thompson}, M.~J., {Christensen-Dalsgaard}, J., {Miesch}, M.~S., \& {Toomre},
  J. 2003, \araa, 41, 599

\bibitem[{{Valsecchi} \& {Rasio}(2014)}]{Valsecchi2014}
{Valsecchi}, F. \& {Rasio}, F.~A. 2014, \apj, 786, 102

\bibitem[{{Wei}(2016)}]{Wei2016}
{Wei}, X. 2016, \apj, 828, 30

\bibitem[{{Wei}(2018)}]{Wei2018}
{Wei}, X. 2018, \apj, 854, 34

\bibitem[{{Weinberg} {et~al.}(2017){Weinberg}, {Sun}, {Arras}, \&
  {Essick}}]{Weinbergetal2017}
{Weinberg}, N.~N., {Sun}, M., {Arras}, P., \& {Essick}, R. 2017, \apjl, 849,
  L11

\bibitem[{{Winn} {et~al.}(2010){Winn}, {Fabrycky}, {Albrecht}, \&
  {Johnson}}]{Winn2010}
{Winn}, J.~N., {Fabrycky}, D., {Albrecht}, S., \& {Johnson}, J.~A. 2010, \apjl,
  718, L145

\bibitem[{{Winn} \& {Fabrycky}(2015)}]{Winn2015}
{Winn}, J.~N. \& {Fabrycky}, D.~C. 2015, \araa, 53, 409

\bibitem[{{Witte} \& {Savonije}(2002)}]{WS2002}
{Witte}, M.~G. \& {Savonije}, G.~J. 2002, \aap, 386, 222

\bibitem[{{Zahn}(1966)}]{Zahn1966a}
{Zahn}, J.~P. 1966, Annales d'Astrophysique, 29, 313

\bibitem[{{Zahn}(1970)}]{Zahn1970}
{Zahn}, J.~P. 1970, \aap, 4, 452

\bibitem[{{Zahn}(1975)}]{Zahn1975}
{Zahn}, J.-P. 1975, \aap, 41, 329

\bibitem[{{Zahn}(2008)}]{Zahn2008}
{Zahn}, J.-P. 2008, in EAS Publications Series, Vol.~29, EAS Publications
  Series, ed. M.-J. {Goupil} \& J.-P. {Zahn}, 67--90

\end{thebibliography}

\pagebreak
\newpage

\appendix
\section{Analytical integrated response for a piecewise-homogeneous fluid body}\label{hydrostat}
In \citet{Ogilvie2013}, the computation of the frequency-averaged dissipation $\int^{\infty}_{-\infty} {\rm Im}[k_l^m(\omega)] {\rm d}\omega/\omega$ for inertial waves in a piece-wise homogeneous body is given in Appendix B for the $l=m=2$ mode. It is the only mode required to compute the tidal evolution of coplanar circular systems. Here we need to compute this value for other values of $m$, namely $m=0$ and $m=1$. In this appendix, we recall the main steps leading to the formulation of $\int^{\infty}_{-\infty} {\rm Im}[k_l^m(\omega)] {\rm d}\omega/\omega$ as derived by \citet{Ogilvie2013}. We derive the formulation of $\int^{\infty}_{-\infty} {\rm Im}[k_l^m(\omega)] {\rm d}\omega/\omega$ for arbitrary values of $l$ and $m$, which is not provided explicitly in  \citet{Ogilvie2013} but is needed to study the obliquity tides.

\subsection{Low-frequency tidal forcing of a rotating barotropic fluid}
The star is considered as a steady axisymmetric body weakly perturbed by a tidal gravitational potential $\Psi$ (see Eq.~\ref{quadtide}). We focus on its convective zone that rotates with uniform angular velocity $\bm{\Omega}$. It is also considered to be a barotropic ideal fluid of pressure $p$ uniquely related to the density $\rho$. Finally, we assume isentropy so we can write ${\rm d} p /\rho = h$, $h$ being the specific enthalpy. In the reference frame rotating with the envelope, the linearised equations of motion, mass conservation and gravitational potential, are 
\begin{eqnarray}
\ddot{\bm{\xi}} + 2\bm{\Omega} \times \dot{\bm{\xi}} &=& - \nabla W, \label{eqmot}\\
h'+\Phi'+\Psi &=&W, \label{prespert}\\
\rho' &=& - \nabla \cdot (\rho \bm{\xi}),\\
\nabla^2 \Phi' &=& 4 \pi G \rho' \label{eqpoisson},
\end{eqnarray}
where the prime denotes an Eulerian perturbation, the dot denotes $\partial/\partial t $, $\bm{\xi}$ is the displacement. The external tidal potential is $\Psi$ and $\Phi'$ represent the internal self-gravitational potential perturbation. Thus $W$ is a reduced pressure perturbation. In the low-frequency limit, this quantity can be neglected to leading order and thus the right-hand side of Eq.~\eqref{prespert} can be set to 0. Even with this simplification, solving the system \eqref{eqmot}-\eqref{eqpoisson} is a task that requires further approximations if progress is to be made analytically.

Thus solutions are sought by decomposition : all quantities are supposed to be the sum of a non-wavelike part and a wavelike part. The non-wavelike part is an instantaneous hydrostatic response to the tidal potential. It is calculated by neglecting the Coriolis force, but involves the correct Eulerian perturbations of internal potential, pression and density induced by the tidal forcing. The wavelike part is the component of the response that is driven by the Coriolis force, and satisfies the anelastic approximation where acoustic waves are filtered out, and rigid boundary conditions on surfaces with both vacuum and the core. The non-wavelike part therefore satisfies the following equations:
\begin{eqnarray}
\ddot{\bm{\xi}}_{\rm nw} &=& - \nabla W_{\rm nw}, \label{eq1nw}\\
h'_{\rm nw}+\Phi'_{\rm nw}+\Psi &=&0, \label{eqprespertnm}\\
\rho'_{\rm nw} &=& - \nabla \cdot (\rho \bm{\xi}_{\rm nw}),\\
\nabla^2 \Phi' _{\rm nw}&=& 4 \pi G \rho'_{\rm nw}, \label{eq4nw}
\end{eqnarray}
together with the wavelike part that verifies
\begin{eqnarray}
\ddot{\bm{\xi}}_{\rm w} + 2\bm{\Omega} \times \dot{\bm{\xi}_{\rm w}} &=& - \nabla W_{\rm w} + \vect{f},\label{eqmotw}\\
\nabla \cdot (\rho \bm{\xi}_{\rm w})& =&0, \label{massconsw}\\
h'_{\rm w}&=&0,\\
\Phi'_{\rm w}&=&0, \\
\rho'_{\rm w}&=&0,
\end{eqnarray}
where $\vect{f} = - 2 \bm{\Omega} \times \dot{\bm{\xi}}_{\rm nw}$ is the Coriolis force per unit mass acting on the non-wavelike part which drives the wavelike-like part of the solution. The boundary conditions are also decomposed as
\begin{eqnarray}
\bm{\xi}_{{\rm nw},r} &=&0 \quad \text{at} \quad r=\alpha R_\star, \\
\bm{\xi}_{{\rm nw},r} &=&-\frac{\Phi' + \Psi}{g} \quad \text{at} \quad r=R_\star,\\
\bm{\xi}_{{\rm w},r} &=&0 \quad \text{at} \quad r=\alpha R_\star,\\
\bm{\xi}_{{\rm w},r} &=&0 \quad \text{at} \quad r=R_\star,
\end{eqnarray}
where $R_{\star}$ is the mean radius of the free surface of the star and $\alpha ={R_{\rm c}}/{R_\star}$ with $R_{\rm c}$ the mean radius of the radiative zone.

The system of ordinary differential equations for the wavelike and non-wavelike parts, together with their respective boundary conditions at interfaces, form our initial-value problem. This can be solved by means of Fourier (or Laplace) transforms methods. Let us consider the case where the tidal forces derives from a potential $\Psi = {\rm Re}\left[A_l(r/R_\star)^l Y_l^m(\theta, \phi) e^{-i\hat{\omega} t}\right]$, in the fluid frame with coordinates $(r, \theta, \phi)$ defined in Sec.\ref{sec:dynpot}, and $A_l$ an amplitude of appropriate dimensions. Taking the Fourier transform of Eqs.~\eqref{eqmotw} and \eqref{massconsw} is formally equivalent to assume that the forcing and the response depend harmonically in time through a common factor $e^{-i\hat{\omega} t}$.

\subsection{Impulsive forcing}
It is important to stress that even if the equations \eqref{eq1nw}-\eqref{eq4nw} contain time derivatives, they do in fact imply that the response is instantaneously related to the tidal potential, and the corresponding boundary conditions are instantaneous in time. Taking the Fourier transforms of those equations for an impulsive forcing thus brings back the problem to an hydrostatic setting with only spatial dependence . If we consider a tidal potential with the temporal dependence 
\begin{equation}
\Psi = \hat{\Psi}(r, \theta, \phi) H(t),
\end{equation}
where $H(t)$ is the Heaviside step function, we place ourself in the special case where the wave-like part of the response experiences an impulsive effective force that derives from this potential. In this case, the spatial structure of the tidal potential
\begin{equation}\label{eqtidpot}
\hat{\Psi} = {\rm Re} \left[A_l\left(\frac{r}{R_\star}\right)^l Y_l^m(\theta, \phi) \right]= {\rm Re} \left[\hat{\Psi_l}(r) Y_l^m(\theta, \phi) \right], 
\end{equation}
and the associated perturbed internal gravitational potential 
\begin{equation}
\hat{\Phi}'_{\rm nw} = {\rm Re} \left[\hat{\Phi}'_{{\rm nw},l}(r) Y_l^m(\theta, \phi) \right], 
\end{equation}
are the only components intervening in the Fourier-transformed problem. The Fourier transforms of Eq.~\eqref{eqprespertnm} and \eqref{eq4nw} combine to produce the inhomogeneous Helmholtz equation
\begin{equation}\label{eqHelmoz}
\nabla^2 \hat{\Phi}_{\rm nw}' + \frac{4 \pi G \rho}{v_s^2} (\hat{\Phi}_{\rm nw}' + \hat{\Psi}) = 0,
\end{equation}
where $v_s$ is the sound speed, which projects on the considered spherical harmonics
\begin{equation}
\frac{1}{r^2}\frac{{\rm d}}{{\rm d}r} \left(r^2 \frac{{\rm d} \hat{\Phi}'_{{\rm nw},l}}{{\rm d}r} \right) - \frac{l(l+1)}{r^2} \hat{\Phi'}_{{\rm nw},l}+ \frac{4 \pi G \rho}{v_s^2} (\hat{\Phi'}_{{\rm nw},l}+ \hat{\Psi}_l) = 0
\end{equation}
 in $R_{\rm c} < r< R_\star$. The known tidal potential $\hat{\Psi}_l$ is of the form $\hat{\Psi}_l = A_l\left(r/R_\star\right)^l Y_l^m(\theta, \phi)$ for the impulse problem and $\hat{\Phi'}_{{\rm nw},l}$ is the internal perturbed gravitational potential to be solved for. For the two-layered model, the solutions to this equation can be written under the form 
\begin{equation}\label{eqsolphi}
\hat{\Phi'}_{{\rm nw},l} =
\begin{cases}
 (B_1 + B_2)(r/R_\star)^l, \quad  0 < r < R_{\rm c}, \\
 B_1 (r/R_\star)^l + B_2 \alpha^{2l+1} (r/R_\star)^{-(l+1)},  \quad R_{\rm c} < r< R_\star,\\
(B_1 + B_2 \alpha^{2l+1}) (r/R_\star)^{-(l+1)},\quad  r>R_\star.
\end{cases}
\end{equation}
The coefficients must satisfy the matching conditions to ensure that the boundaries are equipotentials
\begin{eqnarray} \label{eqboundb}
(2l+1) B_2 &=& 3(1-f) (A_l +B_1+B_2)\\
(2l+1) B_1 &=& \frac{3f}{f+(1-f)\alpha^3} (A _l+B_1+B_2\alpha^{2l+1}),
\end{eqnarray}
with $f=\rho_{\rm e}/\rho_{\rm c}$ and $\alpha ={R_{\rm c}}/{R_\star}$, where $\rho_{\rm c}$ and $\rho_{\rm e}$ denote the density of the core and of the envelope respectively.
We give here the explicit algebraic solutions for $B_1$ and $B_2$ as a function of the stellar parameters, which are not made explicit in \citet{Ogilvie2013}
\begin{equation}\label{eqb1}
B_{1} = A_l  \frac{1+q_l}{p_l-q_l}
\end{equation}
\begin{equation}\label{eqb2}
B_{2} = A_l  \left(\frac{1+p_l}{p_l-q_l}\right)\left( \frac{1- f  }{\displaystyle{\frac{2l-2}{3}} +f } \right)
\end{equation}
where we have introduced
\begin{align}
p_l &=\frac{2l-2}{3}\left(1+\left(\frac{2l+1}{2l-2}\right)\left(\frac{1}{f}-1\right)\alpha^3\right)\label{eqpl}\\
q_l &= \frac{1- f }{\displaystyle{\frac{2l-2}{3}} + f } \alpha^{2l+1}.\label{eqql}
\end{align}
Next, we use Eq.~\eqref{eq1nw} to find the non-wavelike tidal displacement by using the change of variable $W_{\rm nw} = \ddot{X}$ so that its Fourier transform can be re-written as 
\begin{equation}
\hat{\bm{\xi}}_{\rm nw} = - \nabla \hat{X},
\end{equation}
where $\hat{X} = {\rm Re} [\hat{X}_l (r) Y_l^m(\theta, \phi)]$. It projects on the spherical harmonics $Y_l^m$ as 
\begin{equation}
\frac{1}{r^2}\frac{{\rm d}}{{\rm d}r} \left(r^2 \rho \frac{{\rm d} \hat{X}_{l}}{{\rm d}r} \right) - \frac{l(l+1)}{r^2} \rho \hat{X}_{l} = - \frac{\rho}{v_s^2} (\hat{\Phi'}_{{\rm nw},l}+ \hat{\Psi}_l) 
\end{equation}
for $R_{\rm c} < r< R_\star$. The solution satisfies Poisson's equation and can be written in the form
\begin{equation}\label{eqxnw}
\hat{X}_{l} =
\begin{cases}
C_1(r/R_\star)^l, \quad  0 < r < R_{\rm c}, \\
C_2 (r/R_\star)^l + C_3 \alpha^{2l+1} (r/R_\star)^{-(l+1)},  \quad R_{\rm c} < r< R_\star.
\end{cases}
\end{equation}
The regularity of $\hat{X}_l$ at each interface requires ${\rm d}\hat{X}_l/{\rm d}r = (\hat{\Phi'}_{{\rm nw},l}+ \hat{\Psi}_l)/g$. This implies the following relationships between the coefficients $C_1$, $C_2$, $C_3$, $B_1$ and $B_2$.
\begin{eqnarray}
l C_1 = l C_2 - (l+1) C_3 &=& \frac{3 (A_l + B_1 + B_2)}{4 \pi G \rho_{\rm c}},\\
l C_2 - (l+1) C_3 \alpha^{2l+1} &=& \frac{3 (A _l+ B_1 + B_2 \alpha^{2l+1})}{4 \pi G \rho_{\rm c} (f + (1-f)\alpha^3)}.
\end{eqnarray}
Again this can be written in terms of densities and radii for the two-layered model as
\begin{equation}\label{eqc1}
C_{1} = \frac{A_l}{4 \pi G l \rho_{\rm c}}  \left(\frac{2l+1}{\displaystyle{\frac{2l-2}{3}}+f}\right)\left( \frac{1+p_l}{p_l-q_l}\right),
\end{equation}
\begin{multline}\label{eqc2}
C_{2} = \frac{A_l}{4 \pi G l \rho_{\rm c}} \left(\frac{1}{f+(1-f)\alpha^3}\right) \\ \left(3 - \left(\frac{f}{\displaystyle{\frac{2l-2}{3}}+f} \right)\left(\frac{\alpha^{2l+1} + (2l+1)(1-\alpha^3)}{1-\alpha^{2l+1}}\right)\right) \left( \frac{1+p_l}{p_l-q_l}\right),
\end{multline}
\begin{multline}\label{eqc3}
C_{3} = \frac{A_l}{4 \pi G (l+1) \rho_{\rm c}} \left(\frac{3\alpha^{2l+1} - (2l+1)\alpha^3 + 2l-2}{1-\alpha^{2l+1}}\right) \\ \left(\frac{1-f}{\left(\displaystyle{\frac{2l-2}{3}}+f\right) \left(f+(1-f)\alpha^3\right)}\right)\left( \frac{1+p_l}{p_l-q_l}\right).
\end{multline}
This gives the explicit form for $\hat{X}$ that determines the non-wavelike displacement $\hat{\bm{\xi}}_{\rm nw}$. From $\hat{\bm{\xi}}_{\rm nw}$ we deduce the impulsive effective force $\hat{\bm{f}} = - 2 \bm{\Omega} \times \hat{\bm{\xi}}_{\rm nw}$ that allows us to solve for the wavelike part of the response.

Assuming that the fluid is at rest before the impulse, and considering that the Coriolis force does not act on the wavelike velocity during the impulse process (besides providing the effective restoring force from its action on the non-wavelike velocity), the wavelike velocity $\hat{\bm{u}}_{\rm w}$ is given by
\begin{equation}\label{eqimpmot}
\hat{\bm{u}}_{\rm w} = \hat{\bm{f}} - \nabla \hat{W}_{\rm w},
\end{equation}
where the pressure perturbation $\hat{W}_{\rm w}$ satisfies the anelastic approximation
\begin{equation}
\nabla \cdot (\rho \hat{\bm{u}}_{\rm w}) = 0
\end{equation}
and the boundary conditions $\hat{u}_{{\rm w},r}=0$. The wavelike velocity can be projected on to vector spherical harmonics, using standard methods used to describe stellar oscillations. We note $\hat{a}_l(r)$ and $\hat{b}_l(r)$ the coefficients representing the spheroidal part and $\hat{c}_{l\pm 1}(r)$ the toroidal part. The wavelike velocity is written as
\begin{multline}\label{eqprojuw}
\hat{\bm{u}}_{\rm w} = {\rm Re} [\bm{e}_r \hat{a}_l^m(r) Y_l^m + r^2 \hat{b}_l^m(r) \nabla Y_l^m \\
-r^2 \hat{c}_{l-1}^m(r)\ \bm{e}_r  \times \nabla Y_{l-1}^m - r^2 \hat{c}_{l+1}^m(r)\bm{e}_r  \times \nabla Y_{l+1}^m ].
\end{multline}

These spheroidal and toroidal components satisfy ordinary differential equations according to Eq.~\eqref{eqimpmot}, which involves $\hat{X}_l$ found previously from the non-wavelike part, and $\hat{W}_{\rm w}$ of the wavelike part. If we write $\hat{W}_{\rm w} = {\rm Re} [\hat{W}_l^m(r) Y_l^m(\theta, \phi)]$, it can be shown that $\hat{W}_l^m$ satisfies
\begin{equation}
\frac{1}{r^2}\frac{{\rm d}}{{\rm d}r} \left(r^2 \rho \frac{{\rm d} \hat{W}_{l}^m}{{\rm d}r} \right) - \frac{l(l+1)}{r^2} \rho \hat{W}_{l}^m = - \frac{2 i m \Omega}{r} \frac{{\rm d} \rho}{{\rm d}r} \hat{X}_l
\end{equation}
with boundary conditions ${\rm d}\hat{W}_l^m/{\rm d}r = -2im\Omega \hat{X}_l /r$ at $r=R_{\rm c}$ and $r=R_\star$. Consequently, $\hat{W}_l^m$ satisfies Poisson's equation and the solution is 
\begin{equation}\label{eqww}
\hat{W}_{l}^m =
\begin{cases}
-2 i m \Omega \frac{C_1}{l}(r/R_\star)^l, \quad  0 < r < R_{\rm c}, \\
-2 i m \Omega \left[\frac{C_2}{l} (r/R_\star)^l - \frac{C_3}{l+1} \alpha^{2l+1} (r/R_\star)^{-(l+1)}\right],  \quad R_{\rm c} < r< R_\star.
\end{cases}
\end{equation}

Using Eqs.~\eqref{eqxnw} and \eqref{eqww}, the spheroidal and toroidal parts of the initial velocity just after the impulse can be made explicit. This yields null values for the spheroidal component for all values of $l$, $\hat{a}_l^m=\hat{b}_l^m=0$, and a toroidal part given by

\begin{equation}\label{eqtor1}
\hat{c}_{l-1}^m =
\begin{cases}
-2 \Omega \tilde{q}_l^m r^{-2} (2l +1) C_1 (r/R_\star)^l, \quad  0 < r < R_{\rm c}, \\
-2 \Omega \tilde{q}_l^m r^{-2} (2l +1) C_2 (r/R_\star)^l, \quad  R_{\rm c} < r < R_{\star},
\end{cases}\end{equation}
and 
\begin{equation}\label{eqtor2}
\hat{c}_{l+1}^m =
\begin{cases}
0, \quad  0 < r < R_{\rm c}, \\
-2 \Omega \tilde{q}_{l+1}^m r^{-2} (2l +1) C_3 \alpha^{2l+1} (r/R_\star)^{-(l+1)}, \quad  R_{\rm c} < r < R_{\star},
\end{cases}\end{equation}
where $\tilde{q}_{l}^m$ is a coefficient arising from the coupling of spheroidal and toroidal velocity components by the Coriolis force, given in Eq.~\eqref{eqcoupling}. The energy transfer associated to the impulsive response is given by
\begin{equation}
\hat{E}=\frac{1}{2}\int \rho |\hat{\bm{u}}_{\rm w}|^2 {\rm d}V,
\end{equation}
which, thanks to the projection given in Eq.~\eqref{eqprojuw}, is simply found to be
\begin{equation}\label{eq:enerassum}
\hat{E}=\hat{E}_l^m+\hat{E}_{l-1}^m+ \hat{E}_{l+1}^m
\end{equation}
with 
\begin{eqnarray}
\hat{E}_l^m&=&0,\\
\hat{E}_{l-1}^m&=&\frac{1}{4} \int_{R_{\rm c}}^{R_\star} \rho r^2 [l(l-1)r^2 |\hat{c}_{l-1}^m|^2] dr, \label{eqenerlp1}\\
\hat{E}_{l+1}^m&=&\frac{1}{4} \int_{R_{\rm c}}^{R_\star} \rho r^2 [(l+1)(l+2)r^2 |\hat{c}_{l+1}^m|^2] dr. \label{eqenerlm1}
\end{eqnarray}
Thus inserting Eqs.~\eqref{eqtor1} and \eqref{eqtor2} into Eq.~\eqref{eqenerlp1} and \eqref{eqenerlm1}, we get 
\begin{eqnarray}\label{eqenerc}
\hat{E}_l^m&=&0,\\
\hat{E}_{l-1}^m&=&l (l-1) (2l+1) \left(\tilde{q}_l^m\right)^2  \Omega^2 R_\star\nonumber\\
&&[\rho_{\rm c} \alpha^{2l+1} |C_1|^2 + \rho_{\rm e} (1- \alpha^{2l+1}) |C_2|^2],\\
\hat{E}_{l+1}^m&=& (l+1)(l+2) (2l+1) \left(\tilde{q}_{l+1}^m\right)^2 \Omega^2 R_\star \nonumber\\
&&\rho_{\rm e} \alpha^{2l+1} (1-\alpha^{2l+1}) |C_3|^2 ,
\end{eqnarray}
which are readily expressed in terms of stellar quantities using Eqs.~\eqref{eqc1} -- \eqref{eqc3}:
\begin{multline}
\hat{E}_{l-1}^m=l (l-1) (2l+1) \left(\tilde{q}_l^m\right)^2 \Omega^2  R_\star  \left(\frac{A_l}{4 \pi G l \rho_{\rm c}}\right)^2 \left( \frac{1+p_l}{p_l-q_l}\right)^2 \\
\left[\rho_{\rm c} \alpha^{2l+1}  \left(\frac{2l+1}{\displaystyle{\frac{2l-2}{3}}+f}\right)^2 + \rho_{\rm e} \left(1- \alpha^{2l+1}\right) \left(\frac{1}{f+(1-f)\alpha^3}\right)^2 \right.\\ 
\left.\times \left(3 - \left(\frac{f}{\displaystyle{\frac{2l-2}{3}}+f} \right)\left(\frac{\alpha^{2l+1} + (2l+1)(1-\alpha^3)}{1-\alpha^{2l+1}}\right)\right)^2 \right],
\end{multline}

\begin{multline}
\hat{E}_{l+1}^m= (l+1)(l+2) (2l+1 \left(\tilde{q}_{l+1}^m\right)^2 \Omega^2  R_\star\\
\rho_{\rm e} \alpha^{2l+1} \left(\frac{A_l}{4 \pi G (l+1) \rho_{\rm c}}\right)^2 \left( \frac{1+p_l}{p_l-q_l}\right)^2 \\
\frac{\left(1-f \right)^2}{1-\alpha^{2l+1}} \left(\frac{3\alpha^{2l+1} - (2l+1)\alpha^3 + 2l-2}{\left(\displaystyle{\frac{2l-2}{3}}+f\right) \left(f+(1-f)\alpha^3\right)}\right)^2 .
\end{multline}

It can be shown that for a tidal potential of the form ${\Psi = {\rm Re}[A_l(r/R_\star)^l Y_l^m (\theta, \phi) e^{-i \omega t}]}$, the rate of energy transfer given by ${P=\int \rho \dot{\bm{\xi}} \cdot \left( - \nabla \Psi \right) {\rm d}V}$
is related to the potential Love numbers through
\begin{equation}
P = \frac{(2l+1) R_\star}{ 4 \pi G}\frac{1}{2}{|A_l|^2} \omega {\rm Im}[k_l^m(\omega)].
\end{equation}

By analogy, in the Fourier-transformed problem, again assuming that the force derives from a tidal potential of the form ${\Psi = {\rm Re}[A_l(r/R_\star)^l Y_l^m (\theta, \phi) e^{-i \omega t}]}$, we define the complex dimensionless response functions $K_l^m(\omega)$ as
\begin{equation}
K_l^m(\omega) \equiv \frac{4 \pi G}{|A_l|^2 R_\star (2l+1)}\int \rho \tilde{\bm{\xi}_w} \cdot \bar{\tilde{\bm{f}}}\ {\rm d}V.
\end{equation}
So that the total energy transferred to the wavelike tide is, using the same analogy,
\begin{equation}
\hat{E}= 2 \frac{(2l+1)R_\star}{4 \pi G}\int_{-\infty}^{\infty} \omega {\rm Im} \left[K_l^m(\omega)\right] \left| \tilde{\Psi}_l^m (\omega)\right|^2  \frac{{\rm d} \omega}{\omega}.
\end{equation}
In the case of an impulsive tidal potential, the energy transfer is simply
\begin{equation}
\hat{E}= 2 \frac{(2l+1)R_\star}{4 \pi G}\frac{A_l^2}{8\pi}\int_{-\infty}^{\infty} {\rm Im} \left[K_l^m(\omega)\right] \frac{{\rm d} \omega}{\omega},
\end{equation}
and using Eq.~\eqref{eq:enerassum}, we finally get that
\begin{equation}
\int_{-\infty}^{\infty} {\rm Im} \left[K_l^m(\omega)\right] \frac{{\rm d} \omega}{\omega} = \frac{\hat{E}_{l-1}^m+\hat{E}_{l+1}^m }{(2l+1) R_\star G}\left(\frac{4\pi G}{A_l}\right)^2, 
\end{equation}
so that
\begin{multline}
\int_{-\infty}^{\infty} {\rm Im} \left[K_l^m(\omega)\right] \frac{{\rm d} \omega}{\omega} = \frac{4}{3}\left(\frac{l-1}{l}\right)  \epsilon^2   \left( \frac{1+p_l}{p_l-q_l}\right)^2  \left( \frac{f}{f+(1-f)\alpha^3}\right)   \\
\left[\left(\tilde{q}_{l}^m\right)^2 \left(\frac{2l+1}{\displaystyle{\frac{2l-2}{3}}+f}\right)^2 \left(f+(1-f)\alpha^3\right)^2 \left(\frac{ \alpha^{2l+1} }{f} \right) \right. \\
+ \left(\tilde{q}_{l}^m\right)^2  \left(1- \alpha^{2l+1}\right) \left(3 - \left(\frac{f}{\displaystyle{\frac{2l-2}{3}}+f} \right)\left(\frac{\alpha^{2l+1} + (2l+1)(1-\alpha^3)}{1-\alpha^{2l+1}}\right)\right)^2 \\
+ \left(\tilde{q}_{l+1}^m\right)^2 \left(\frac{l}{l-1}\right)\left(\frac{l+2}{l+1}\right) \left(\frac{1-f} {\displaystyle{\frac{2l-2}{3}}+f}\right)^2 \left(\frac{\alpha^{2l+1}}{1-\alpha^{2l+1}}\right) \\
\Bigg. \left(3\alpha^{2l+1} - (2l+1)\alpha^3 + 2l-2 \right)^2 \Bigg],
\end{multline}
where $\epsilon$ is a dimensionless parameter measuring stellar rotation normalised to the breakup velocity
\begin{equation}
\epsilon = \Omega_\star \left(\frac{G M_\star}{R_\star}\right)^{-1/2}.
\end{equation}

Lastly, to the level of accuracy of our approximations, we have $ {\rm Im} \left[K_l^m(\omega)\right] =  {\rm Im} \left[k_l^m(\omega)\right]$. So in this way we get an expression of $\int^{\infty}_{-\infty} {\rm Im}[k_l^m(\omega)] {\rm d}\omega/\omega$ that depends only on the angular velocity and the masses and radii of the radiative core and the convective envelope, valid for arbitrary values of $l$ and $m$. 

\section{Static Love numbers}\label{statLove}
We stress here that the Helmholtz-like equation \eqref{eqHelmoz} for the non-wavelike part in the Fourier-transform problem, neglecting the effect of the Coriolis force, is exactly equivalent to the one that would be obtained by combining Newton's second law of action and Poisson's equation, for a non-rotating fluid in hydrostatic equilibrium experiencing the static perturbing gravitational potential from a point-like companion. The general form of the solutions given in Eq.~\eqref{eqsolphi} also describe the gravitational potential arising from the mass of the tidally distorted star in hydrostatic equilibrium at any point $\bm{r}$ in space.

Indeed, as it is well-known, in a frame of reference centred on the star, the gravitational potential produced by a point-like companion of mass $M_p$ placed at $\mathbold{r_p}=(r_p, \theta', \phi')$ with $r_p>R_\star$ can be expanded over the Legendre polynomials. The tidal potential $\Psi$ is simply the potential from which derives the variable part of the perturbing force, so 

\begin{equation}\label{tidepotstat}
\Psi(r, \theta,\phi) =  \frac{G M_p}{r_p}\sum_{l = 2}^{\infty} \left(\frac{r}{r_p}\right)^{l} P_l(\cos \gamma),
\end{equation}
where  $\gamma$ denotes the angular separation between the vectors $\mathbold{r}$ and  $\mathbold{r_p}$ \citep[e.g. ][]{Murray1999}. We have, by the addition theorem for spherical harmonics,
\begin{equation}
P_l(\cos \gamma) = \frac{4\pi}{2l+1}\sum_{m=-l}^{l} Y_{l}^{m}(\theta, \phi)\bar{Y}{_{l}^{m}}(\theta', \phi').
\end{equation}

In hydrostatic equilibrium, any surface of a fluid body characterised by equal density and equal pressure corresponds to a surface of equal total potential of forces acting upon the body.  Because of the tidal force, the star in hydrostatic equilibrium cannot assume a perfectly spherical shape. Following \citet{Kopal1959}, without any loss of generality, we can assume that on such a surface the radial distance $r$ can be written as 
\begin{equation}\label{eqshape}
r = r_{*} \left( 1 + \sum_{l=1}^{\infty} \sum_{m=-l}^{l}  S_{l}^{m}(r_{*})Y_l^m (\theta, \phi) \right),
\end{equation}
where the $Y^{m}_{l}(\theta, \phi)$ are the normalised spherical harmonics following the Condon-Shortley phase convention, and $S_{l}^{m}(r_{*})$ are complex functions of the mean radius $r_{*}$ that must verify
\begin{equation}
S_{l}^{-m} =(-1)^m \bar{S}_{l}^{m},
\end{equation}
for $r$ to be a real quantity. By virtue of the uniqueness of the potential function, and because the density must remain constant over an equipotential, the density can be regarded as a function of the single variable $r_*$, denoting the mean radius of the corresponding equipotential. Thus, for an arbitrary point $(r,\theta,\phi)$ in the interior of the star,  we can write the interior self-potential $U$ of this distorted configuration  as 
\begin{multline}\label{intpot}
U(r,\theta,\phi) = U_0 + \\
\sum_{l=1}^{\infty} \frac{4\pi G r^l}{2l+1} \sum_{m=-l}^{l} Y_{l}^{m}(\theta, \phi)  \int_{r_*}^{R_\star} \rho \frac{\partial (r_*^{2-l} S_{l}^{m}(r_{*}))}{\partial r_*}dr_*, 
\end{multline}
where $\rho$ is the density, $R_\star$ denotes the mean radius at the surface of the body, and we have introduced
\begin{equation}
U_0 =  4 \pi G \int_{r_*}^{R_\star} \rho r_* dr_*.
\end{equation}
In the same way, the exterior potential $V(r, \theta, \phi)$ can be expressed as
\begin{multline}\label{extpotv}
V(r,\theta,\phi) = \frac{V_0}{r} + \\
\sum_{l=2}^{\infty} \frac{4\pi G}{(2l+1)r^{l+1}} \sum_{m=-l}^{l} Y_{l}^{m}(\theta, \phi)  \int_{0}^{r_*} \rho \frac{\partial (r_*^{l+3} S_{l}^{m}(r_{*}))}{\partial r_*}dr_* 
\end{multline}
with 
\begin{equation}
V_0 =  4 \pi G \int_{0}^{r_*} \rho r_*^2 dr_*.
\end{equation}

Now let $\Phi$  be the total potential of forces acting on any arbitrary point. In hydrostatic equilibrium, the function $\Phi$ is to remain constant over any surface of equal pressure of density, over which $r$ verifies Eq.~\eqref{eqshape}. By binomial expansion of $r$, neglecting squares and higher order of $Y_{l}^{m}$, the shape of an equipotential is described by the spherical harmonic expansion with coefficients $S_l^m$ that are solution of the following equation
\begin{multline}\label{eqcond}
(2l+1) S_{l}^{m}(r_{*}) \int_{0}^{r_*} \rho r_*^2 dr_* - r_*^{l+1}   \int_{r_*}^{R_\star} \rho \frac{\partial (r_*^{2-l} S_{l}^{m}(r_*))}{\partial r_*}dr_*  \\
- \frac{1}{r_*^l} \int_{0}^{r_*} \rho \frac{\partial (r_*^{l+3} S_{l}^{m}(r_*))}{\partial r_*}dr_* =  M_p  \left(\frac{r_*}{r_p}\right)^{l+1} \bar{Y}{_{l}^{m}}(\theta', \phi'),
\end{multline}
for any $l\geq1$ and $-l\leq m\leq l$. This is the well-known Clairaut's equation. It implicitly specifies the spherical harmonics describing the form of equipotential level surfaces as distorted by an external force derived from the potential \eqref{tidepotstat}. Solving Clairaut's equation allows us to derive the explicit form of the radial distance $r_S$ describing the free surface of the perturbed body of mass $M_\star$ in terms of the mass $M_p$ of the perturber 
\begin{equation}\label{eqsurf}
r_S = R_{\star} \left( 1 + \sum_{l=1}^{\infty} \frac{M_p}{M_{\star}} \left(\frac{R_\star}{r_p}\right)^{l+1} \frac{2l +1}{ l+ \eta_l (R_\star)} P_l(\cos \gamma) \right),
\end{equation}
where $\eta_{l}(R_\star)$ is the surface value of the logarithmic derivative of the shape coefficients
\begin{equation}
\eta_l (r_*) = \frac{{r_*}}{S_{l}^{m}(r_*)}\frac{\partial S_{l}^{m}(r_*)}{\partial r_*}.
\end{equation}

The potential produced by this distorted configuration at any point $(r, \theta, \phi)$, with $r >> R_{\star}$ using the center of mass of the extended body as the origin, is generally written as
\begin{equation}\label{extpotdis}
\Phi(r,\theta,\phi) = \frac{G M_{\star}}{r} \left[1 + \sum_{l=2}^{\infty} k_l \frac{M_p}{M_\star} \left(\frac{R_\star}{r_p}\right)^{l+1} \left(\frac{R_\star}{r}\right)^{l}  P_l(\cos \gamma)   \right]
\end{equation}
where $k_l$ are called the static Love numbers \citep{Love1911}. Using equations Eqs.~\eqref{eqsurf} and \eqref{extpotv}, the static Love numbers are found to be
\begin{equation}
k_l = \frac{ l+1 - \eta_l (R_\star)}{ l+ \eta_l (R_\star)}.
\end{equation} 
Thus the Love numbers link the external potential to the internal structure through the surface value of the logarithmic derivative of the shape coefficients. 

For certain density distributions, Clairaut's equation admit simple solutions in a closed form and the Love number can take explicit forms. For example, we have the well-known result
\begin{equation}
k_l = \frac{3}{2l-2}
\end{equation}
if the density remains uniform throughout the interior of the body, and 
\begin{equation}
k_l = \frac{l+2}{l-1}
\end{equation}
if the whole mass of the configuration were confined to an infinitesimally thin surface shell. If the whole mass of the configuration is condensed at its centre, obviously, $ k_l = 0$ for all $l$ , and the exterior potential \eqref{extpotdis} is that of a point mass.
\\
Now in this paper, we consider a two-layer model that simplifies the actual density distribution within the star. To our knowledge, the expression of the Love numbers in this case is not easily available in recent literature so we provide it here for convenience, since it is a direct by-product of our derivations. Since we consider that $\rho_{\rm c}$ and $\rho_{\rm e}$ are uniform within their respective layers, Clairaut's equation, taken at the core boundary,  simplifies to
\begin{multline}
\left( \frac{2l-2}{3} + \frac{\rho_{\rm e}}{\rho_{\rm c}} \right) \rho_{\rm c}  R_{\rm c}^3  S_{l}^{m}(R_{\rm c}) - \left(\frac{R_{\rm c}}{R_\star}\right)^{l+1}   \rho_{\rm e} R_{\star}^{3} S_{l}^{m}(R_{\star})\\
 =  M_{\rm p}  \left(\frac{R_{\rm c}}{r_p}\right)^{l+1} \bar{Y}{_{l}^{m}}(\theta', \phi'),
\end{multline}
while it gives at the free surface
\begin{multline}
\left( \frac{2l+1}{3} \left(\frac{\rho_{\rm c}}{\rho_{\rm e}} -1 \right) \left(\frac{R_{\rm c}}{ R_{\star}}\right)^3  + \frac{2l-2}{3} \right)  \rho_{\rm e} R_{\star}^3S_{l}^{m}(R_{\star})  \\
+\left(\frac{\rho_{\rm e}}{\rho_{\rm c}} -  1 \right) \left(\frac{R_{\rm c}}{R_{\star}}\right)^{l} \rho_{\rm c} R_{\rm c}^{3} S_{l}^{m}(R_{\rm c}) 
 =  M_{\rm p}  \left(\frac{R_{\star}}{r_p}\right)^{l+1} \bar{Y}{_{l}^{m}}(\theta', \phi').
\end{multline}
Using those two equations to solve for $S_{l}^{m}(R_{\rm c})$ and $S_{l}^{m}(R_{\star})$ and letting $f=\rho_{\rm e}/\rho_{\rm c}$, we find
\begin{align}
S_{l}^{m}(R_{\star})& =   \frac{M_{\rm p}}{\rho_{\rm e} R_{\star}^3}  \left(\frac{R_{\star}}{r_p}\right)^{l+1} \frac{1+ q_l}{p_l - q_l}\  \bar{Y}{_{l}^{m}}(\theta', \phi'),\label{renv}\\
S_{l}^{m}(R_{\rm c})& =   \frac{M_{\rm p}}{\rho_{\rm c} R_{\rm c}^3}  \left(\frac{R_{\rm c}}{r_p}\right)^{l+1}\frac{1}{\frac{2l-2}{3} + f} \frac{1+ p_l}{p_l - q_l}\  \bar{Y}{_{l}^{m}}(\theta', \phi'), \label{rcore}
\end{align}
where $p_l$ and $q_l$ are defined in Eq.~\eqref{eqpl} and ~\eqref{eqql}, respectively. Those expressions are equivalent to the ones derived by \citet{Remus2015} for a two-layer planet. They considered the case of a purely elastic solid core, which is equivalent to the fluid case when setting the shear modulus to zero. For the two-layer model considered here, the potential at any point $(r, \theta, \phi)$ with $r >> R_{\star}$ produced by the distorted configuration can be then written in the form
\begin{multline}\label{extpot}
\Phi(r,\theta,\phi) =  \frac{ G M_{\star}}{r} +\sum_{l=2}^{\infty}  \sum_{m=-l}^{l} B_{1, {\rm eq}} \left(\frac{R_\star}{r}\right)^{l+1}  Y_{l}^{m}(\theta, \phi) \\
+\sum_{l=2}^{\infty}  \sum_{m=-l}^{l} B_{2, {\rm eq}} \left(\frac{R_{\rm c}}{R_\star}\right)^{2l+1}  \left(\frac{R_{\star}}{r}\right)^{l+1}    Y_{l}^{m}(\theta, \phi), \\
\end{multline}
where we have abbreviated
\begin{equation}\label{eqb1Heq}
B_{1,{\rm eq}} = \frac{4\pi }{(2l+1)}  \frac{G M_{\rm p}}{R_{\star}}  \left(\frac{R_{\star}}{r_p}\right)^{l+1}  \frac{1+q_l}{p_l-q_l}\bar{Y}{_{l}^{m}}(\theta', \phi') 
\end{equation}
\begin{equation}\label{eqb2Heq}
B_{2, {\rm eq}} = \frac{4\pi }{(2l+1)} \frac{G M_{\rm p}}{R_{\star}}  \left(\frac{R_{\star}}{r_p}\right)^{l+1}  \frac{1+p_l}{p_l-q_l}\frac{\left( 1- f \right) }{\left(\displaystyle{\frac{2l-2}{3}} +f \right)} \bar{Y}{_{l}^{m}}(\theta', \phi').
\end{equation}

Finally the potential produced by the distorted star Eq.\eqref{extpot} can also be rearranged to take the form of Eq.~\eqref{extpotdis} where the static Love number now has the explicit expression
\begin{equation}
k_l = \frac{ \left(2+p_l\right) q_l +1}{p_l-q_l}.
\end{equation}
This yields the expected $k_l = \frac{3}{2l-2}$ in the limit case of a homogenous body with $\rho_{\rm e} = \rho_{\rm c}$.

Note that the correspondence with Eqs.~\eqref{eqb1} and \eqref{eqb2} is easily made by considering that, in the hydrostatic case, the tidal potential expressed as Eq.~\eqref{tidepotstat} yields
\begin{equation}
A_l \equiv  \frac{4\pi }{(2l+1)}  \frac{G M_{\rm p}}{R_{\star}}  \left(\frac{R_{\star}}{r_p}\right)^{l+1} \bar{Y}{_{l}^{m}}(\theta', \phi'),
\end{equation}
with $A_l$ a constant since the perturber is considered here to be static.

In the dynamical case, if one neglects the Coriolis force, the reasoning presented here still holds, but the coordinates of the perturber now must assume a temporal dependence . Therefore, the non-wavelike part of the response (which is equivalent to the instantaneous adjustment of the star when neglecting the Coriolis force) can easily be solved by considering an adequate setting where the temporal dependence  of the coordinates of the perturber is removed and $A_l$ is indeed a constant. This is done in \citet{Ogilvie2013} by using a frame of reference that rotates with the star, considering a tidal forcing of impulsive nature, and taking the problem into the Fourier domain \citep[see also][]{Remus2012}. For the circular misaligned problem, the same method can be used. In this case, the radial coordinate of the companion is constant over an orbit with $r_p = a$. Its latitudinal coordinate $\theta'$ intervenes only through the spherical harmonics $\bar{Y}{_{l}^{m}}(\theta', \phi')$, which can be expressed as a function of the obliquity $\Theta$ by a simple rotation of the axis of the reference frame (hence the Wigner-d matrix coefficient in Eq.~\ref{quadtide}). Thus, the temporal dependence  only persists in the exponential term of $\bar{Y}{_{l}^{m}}(\theta', \phi')$, which contributes to the harmonic forcing, and defines the tidal frequency in the rotating frame.

\end{document}